\documentclass[a4paper,10pt]{article}
\pdfoutput=1 
\usepackage{ae,aecompl}
\usepackage{jheppub} 

\usepackage[T1]{fontenc}
\usepackage[none]{hyphenat}

\usepackage[italian,english]{babel}
\usepackage{hyperref}
\usepackage{changepage}
\usepackage{ifpdf}
\usepackage{subfigure}
\usepackage{amssymb}
\usepackage{amsfonts}
\usepackage{epsf}
\usepackage{rotating}
\usepackage{graphicx}
\usepackage{amsmath, mathtools}
\usepackage{fancyhdr}
\usepackage{lineno}
\usepackage{babel}
\usepackage{graphics}
\usepackage{pstricks}
\usepackage{color}
\usepackage{multirow}
\usepackage{slashed}
\usepackage{physics}
\usepackage{caption}
\usepackage{tabularx, array,booktabs, makecell}
\usepackage[toc,page]{appendix}
\usepackage{pifont}

\usepackage{tikz}
\usepackage{tikz-feynman}
\tikzfeynmanset{compat=1.0.0}

\newcommand{\mgFull}{\texttt{MadGraph5\_aMC@NLO v.3.4.2}}
\definecolor{pantoneCB}{rgb}{0.0588235, 0.298039, 0.505882}

\usepackage[mathscr]{euscript}
\let \eucal \mathscr
\usepackage{mathrsfs}

\hypersetup{
	colorlinks=true,
	linkcolor=purple,
	filecolor=magenta,
	urlcolor=[rgb]{0.0,0.42,0.59},
	citecolor=[rgb]{0.64,0.0,0.0},
}

\newcommand{\lsim}{\mathrel{\mathop{\kern 0pt \rlap
			{\raise.2ex\hbox{$<$}}}
		\lower.9ex\hbox{\kern-.190em $\sim$}}}
\newcommand{\gsim}{\mathrel{\mathop{\kern 0pt \rlap
			{\raise.2ex\hbox{$>$}}}
		\lower.9ex\hbox{\kern-.190em $\sim$}}}

\newcommand{\be}{\begin{equation}}
	\newcommand{\ee}{\end{equation}}
\newcommand{\bea}{\begin{eqnarray}}
	\newcommand{\eea}{\end{eqnarray}}

\newcommand{\half}{\frac{1}{2}}

\newcommand{\py}{\texttt{Pythia8.3} }

\newcommand{\lhs}{\lambda_{hs} }
\newcommand{\lst}{\lambda_{st} }
\newcommand{\lht}{\lambda_{ht} }

\newcommand{\aht}{A_{ht} }

\newcommand{\abi}{ab$^{-1}$ }

\newcolumntype{P}[1]{>{\centering\arraybackslash}p{#1}}
\newcolumntype{M}[1]{>{\centering\arraybackslash}m{#1}}

\DeclareMathAlphabet{\mathpzc}{OT1}{pzc}{m}{it}

\title{\boldmath Probing a scalar singlet-triplet extension of the Standard Model via VBF at the Muon Collider}

\author[a]{Priyotosh Bandyopadhyay,}
\author[a]{Snehashis Parashar}

\affiliation[a]{Indian Institute of Technology Hyderabad, Kandi,  Sangareddy-502285, Telangana, India}

\emailAdd{bpriyo@phy.iith.ac.in}
\emailAdd{ph20resch11006@iith.ac.in}

\preprint{IITH-PH-0001/24}

\abstract{In this article, we investigate the  $Y=0$ $SU(2)$ scalar triplet and $Z_2$-odd scalar singlet extension of the Standard Model (SM). Here, the triplet charged Higgs boson decays to $ZW^\pm$, breaking the custodial symmetry at the tree-level, proportional to the triplet vev, while the singlet provides the dark matter (DM) relic. The triplet neutral Higgs ($T^0$)  can decay fully invisibly owing to the triplet-singlet portal coupling $\lambda_{st}$. The  other SM Higgs portal couplings $\lambda_{ht}, \lambda_{hs}$ are constrained by the Higgs to di-photon observations, and the dark matter relic and direct searches as well as invisible Higgs decay bounds, respectively. For a cleaner signature, we indulge in a futuristic multi-TeV muon collider (MuC) to probe both the triplet scalars  ($T^\pm, T^0 $) via vector boson fusion with Forward muon tagging, at the centre-of-mass energies of 3 TeV and 10  TeV. The analysis is comprised of a traditional cut-based approach and a BDT classifier, where  the latter is more effective  for lower energies. With large missing energy contributions to the final states from combinations of DM mass and $\lst$, The 3 TeV MuC is projected to probe triplet scalar masses of 450 GeV with the BDT classifier. The 10 TeV MuC can pinpoint the custodial symmetry breaking $T^\pm \to ZW^\pm \to 3$-lepton decay up to 800 GeV of triplet scalar mass from cut-based analysis, with $\lst$ as low as 1.5 being adequate.}

\begin{document}
	\maketitle
	\flushbottom
	
\section{Introduction}
\label{sec:int}

The Standard Model (SM)  is a  successful gauge theory of the $SU(3)_C \times SU(2)_L \times U(1)_Y$ gauge group, adequately explaining the elementary particle interactions. The last keystone of the  SM is the Higgs boson, existence of which is a testimony of the electroweak symmetry breaking (EWSB) via at least one fundamental scalar, which was finally discovered at the Large Hadron Collider (LHC) with a mass of $\sim125.5$ GeV  \cite{CMS:2012qbp, ATLAS:2012yve}. However, existence of other fundamental scalars and their role in EWSB cannot be ruled from this discovery. On the contrary, there are ample reasons to believe that such extra scalar should exist.
Some of these reasons, stemming from apparent shortcomings of the SM, include the lack of a stable dark matter (DM) candidate \cite{Bertone:2004pz}, the metastability of the electroweak (EW) vacuum \cite{Degrassi:2012ry}, and the lack of explanation for tiny but non-zero neutrino masses \cite{Gonzalez-Garcia:2002bkq}, among others. These shortcomings motivate us to look beyond the Standard Model (BSM), by extending the existing scenario with new fields and/or symmetries. 


 A popular set of minimal extensions of the SM include the two-Higgs-doublet models (2HDM), with an additional scalar doublet augmenting the SM, with possible couplings to the SM fermions (for a review, see \cite{Branco:2011iw}). A discrete $Z_2$-symmetry, if imposed on the model, under which the extra doublet transforms as odd, can also yield the desired DM candidate \cite{Belyaev:2016lok}. The charged scalars coming from 2HDMs can couple to fermions, and hence have stringent constraints on their couplings from the SM quarks/leptonic sector measurements. In contrast, an $SU(2)_L$ scalar triplet extension with $Y=0$ can be desirable, as they do not have tree-level Yukawa couplings to the SM fermions \cite{ROSS1975135}. This also disallows the single production of charged scalars via fermionic channels. unlike the 2HDMs. If the neutral component of such a scalar triplet obtains a vacuum expectation value (vev), they participate in the EWSB and contribute to the mass of the $W^\pm$ boson, keeping the $Z$-boson unaffected \cite{Gunion:1989ci, Blank:1997qa, Forshaw:2001xq, Forshaw:2003kh, Chen:2006pb, Chankowski:2006hs, Chivukula:2007koj}. This results into the violation of the custodial symmetry of the SM at the tree level, which otherwise is a residual $SU(2)$ symmetry that preserves the $W^\pm$ and $Z$ mass relation via the $\rho$-parameter defined as \cite{Sikivie:1980hm, Georgi:1993mps}:
\begin{equation}
	\rho = \dfrac{M_{W^\pm}^2}{\cos^2\theta_W M_Z^2} = 1\,, \label{eq:rho}
\end{equation}
with $\theta_W$ being the Weinberg angle. A real scalar triplet of its nature can yield a pair of charged scalars $T^\pm$ and a neutral scalar $T^0$ alongside the existing SM Higgs boson $h$, after the EWSB and subsequent mixing. The tree-level violation of the custodial symmetry manifests itself as the non-vanishing $T^\pm Z W^\pm$ vertex, providing a unique decay signature that can be analysed at the colliders \cite{FileviezPerez:2008bj}. The triplet vev $v_t$ is however heavily restricted to $\lsim 3$ GeV from the precise experimental measurements of the $\rho$ parameter \cite{ParticleDataGroup:2022pth},  reducing the strength of this vertex, as well as allowing minimal mixing between the SM Higgs doublet and the triplet. The phenomenology of these exciting additional triplet scalars from a collider perspective is studied in abundance \cite{Chabab:2018ert, FileviezPerez:2022lxp, Ashanujjaman:2023etj, Butterworth:2023rnw}. The $Y=0$ triplet scalar is also studied as an extension of supersymmetric scenarios, with the custodial symmetry-violating nature embedded in a more elaborate spectrum of additional particles \cite{Bandyopadhyay:2013lca,Bandyopadhyay:2014tha,Bandyopadhyay:2014vma, Bandyopadhyay:2014raa}, where constraints from  $B\to X_s \gamma$ and other collider bounds were presented. A $Y=0$ complex $SU(2)$ triplet is also a viable extension, which carries a pure triplet pseudoscalar that can be a DM candidate without any $Z_2$ symmetry. It also contains two charged triplet-like Higgs bosons, with one having the custodial symmetry-breaking decay signature, and the other one, purely triplet, decaying into the DM and an off-shell $W^\pm$-boson \cite{Bandyopadhyay:2020otm}. Other complex triplet scalar realisations can come in the forms of the triplets with $Y=2$, prevalent in the type-II seesaw model with a neutrino mass motivation \cite{Konetschny:1977bn, Schechter:1980gr}. While enabling the custodial symmetry violation with a permissible triplet vev of $\lsim 1$ GeV, the type-II seesaw model comes with an additional doubly-charged scalar in its mass eigenstate spectrum, testable at the colliders \cite{Chun:2003ej, FileviezPerez:2008jbu, Melfo:2011nx, Han:2015hba, Bandyopadhyay:2020mnp, Ashanujjaman:2021txz}. A similar extension, known as the Georgi-Machacek (GM) model, extends the SM with a complex scalar triplet of $Y=2$, and a real scalar triplet having $Y=0$ \cite{Georgi:1985nv}. Such a structure allows a non-vanishing triplet vev and the consequent $T^\pm Z W^\pm$ vertex, while uniquely preserving the custodial symmetry at the tree-level \cite{Chanowitz:1985ug}. The GM model however is not very minimal, with a multitude of additional doubly-charged, singly-charged, and neutral scalars, heavily constrained by the existing LHC bounds \cite{CMS:2017fhs, CMS:2021wlt, ATLAS:2022pbd}. On the contrary, to realise a DM candidate with only a $Y=0$ real scalar triplet, one can imposes a discrete $Z_2$ symmetry under which the triplet remains odd. Subsequently, the neutral component does not obtain a non-zero vev, allowing it to be a stable DM candidate, with a TeV-scale mass range required to satisfy the observed relic density of the universe \cite{Araki:2010nak, Araki:2011hm, YaserAyazi:2014jby, Khan:2016sxm, Jangid:2020qgo}. The $v_t = 0$ nature of this inert triplet does not violate the custodial symmetry, denying us from the aforementioned $T^\pm Z W^\pm$ vertex. Such an inert triplet scenario is also studied from the perspective of having a first-order EW phase transition (FOPT) in the early universe, and its consequences \cite{Bandyopadhyay:2021ipw}. The resulting compressed spectrum of the inert triplet scalars lead to disappearing track signatures, probes of which are studied in both the LHC and a future muon collider \cite{Chiang:2020rcv, Bandyopadhyay:2024plc}. 

Weighing in on this overview, we establish that the most minimal framework to realise the custodial symmetry-violating decay of a triplet charged scalar, while also having a stable DM candidate, is to extend the SM simultaneously with a real triplet scalar of $Y=0$, as well as a scalar singlet $S$ which is odd under a discrete $Z_2$ symmetry. The triplet, being even under the $Z_2$ symmetry, obtains a non-zero vev and provides the $T^\pm \to ZW^\pm$ decay, while the singlet can satisfy the observed DM relic of the universe. Notably, an inert scalar singlet is commonly studied as the simplest BSM scenario with a DM candidate, without the introduction of another multiplet \cite{Gonderinger:2009jp, Guo:2010hq, Cline:2013gha, GAMBIT:2017gge}. Models that consider both the scalar singlet and the $Y=0$ scalar triplet as constituents of a multi-component dark sector also exist in literature \cite{Fischer:2013hwa, DuttaBanik:2020jrj}. Additionally, supersymmetric extensions of such a triplet and singlet scalar, with the possibility of a very light pseudoscalar ($a$), are extensively studied in refs  \cite{Bandyopadhyay:2015ifm,Bandyopadhyay:2015tva,Bandyopadhyay:2015oga,Bandyopadhyay:2017klv}, but without the $Z_2$-odd nature of the singlet. Apart from the aforementioned charged scalar decay to $ZW^\pm$, a decay into light pseudoscalar via $aW^\pm$ opens up, evading the bounds of light charged Higgs boson estimated from fermionic decay modes. The latter is also possible with only a singlet supersymmetric extension \cite{Bandyopadhyay:2015dio}. Our framework bridges the apparent gap by bringing the two ideas together: having a real scalar triplet with non-zero vev, alongside an inert scalar DM. Via a certain portal coupling, the neutral triplet scalar can decay invisibly into a pair of the DM singlets, which combined with the three-lepton signature of the $T^\pm \to ZW^\pm$ decay can help us probe the model at the colliders. 

While the small $v_t$ allows the triplet scalars to evade the existing LHC bounds by having small cross-sections, it also restricts us from producing a healthy number of signal events for us to analyse at the LHC. Thankfully, A future multi-TeV muon collider (MuC) can come to the rescue as a cleaner, more precise alternative \cite{Palmer:1996gs, Ankenbrandt:1999cta, AlAli:2021let, Accettura:2023ked}. With the salient feature of producing particles with the full energy of the beams as the centre-of-mass energy, combined with suppressed QCD-born backgrounds, the MuC stands as a beacon of hope for BSM searches \cite{Capdevilla:2020qel, Bandyopadhyay:2020mnp, Bandyopadhyay:2021pld, Sen:2021fha,  Asadi:2021gah, Huang:2021nkl, Choi:1999kn, Han:2020uak, Han:2021udl, Chiesa:2021qpr, Capdevilla:2021fmj, Jueid:2021avn, Homiller:2022iax, Han:2022edd, Han:2022ubw, Black:2022qlg, Jueid:2023zxx, Asadi:2023csb,Capdevilla:2024bwt, Ouazghour:2024twx, Ma:2024ayr}. At higher energies, the MuC essentially becomes a vector boson fusion (VBF) machine, suitable for producing triplet scalars via the large $TTVV$-type vertices \cite{Costantini:2020stv,Bandyopadhyay:2024plc}. The short term target of the MuC is to operate at a centre-of-mass energy of 3 TeV with 1 \abi of target luminosity, with which one can produce these triplet scalars with healthy cross-sections for a longer range of masses compared to the 14 TeV LHC \cite{MuonCollider:2022xlm}. The reduced hadronic background at the MuC can also provide a  more precise probe for the hadronically quiet multilepton signature that we are interested in. An upgrade of the MuC with a remarkable 10 TeV of centre-of-mass energy is also in the cards, with 10 \abi of target luminosity that can enhance the probe of this scenario significantly \cite{InternationalMuonCollider:2024jyv}. Another unique feature of VBF production processes at the MuC is the presence of Forward muons with high energy and large pseudorapidity, which can be identified at dedicated detectors to successfully trigger the VBF processes and harness the full potential of the MuC \cite{Ruhdorfer:2023uea, Forslund:2023reu, Bandyopadhyay:2024plc, Li:2024joa}. We constitute three different final states from the asymmetric VBF production of associated triplet charged Higgs and  triplet neutral  Higgs bosons  via $\mu^+ \mu^- \to T^\pm T^0 \mu^\mp \nu$, that include large missing transverse energy (MET) from fully invisible decay of the triplet neutral Higgs via  $T^0 \to SS$, alongside exactly one Forward muon, and the visible decay products of $T^\pm$, and perform a detailed analysis for three benchmark points of triplet masses.

At the 3 TeV MuC, the fiducial VBF production rates for our desired final states can still be comparatively lower than the SM backgrounds, where a regular cut-based analysis may not be enough to obtain at least 3$\sigma$ hint for the model. To enhance the significance of the signal, we employ a multivariate analysis using a boosted decision tree (BDT)-based classifier, using the \texttt{xgboost} package\cite{10.1145/2939672.2939785}. Application of such BDT classifiers in probing BSM signals that are more elusive are quite frequent in a plenty of works dealing with LHC searches (see recent works like \cite{Ashanujjaman:2022ofg, Bhaskar:2022ygp, Mukherjee:2023qjw, Bhattacherjee:2023kxw, Ghosh:2024boo} and the references therein), however, in the context of a future muon collider, BDT classifiers are also picking up pace \cite{Mekala:2023diu, Guo:2023jkz, Maharathy:2023dtp, Belfkir:2023vpo, Mekala:2023kzo, Andreetto:2024rra, Kim:2024jhx}. In our analysis, we highlight the clear advantage of utilizing the BDT classifier for the 3 TeV MuC, while also realising that the 10 TeV MuC is powerful enough to not require it for the purpose of this model.

The article is organized as follows: in \autoref{sec:mod}, we introduce the theoretical framework of the model, with the mass eigenstates, mixing, decay channels, and the bounds from various theoretical and experimental sources on the parameter space, culminating in the choice of three benchmark points that satisfy all of them. \autoref{sec:prodfs} contains details of the VBF production channel of the triplet scalars at the MuC, with a description of the three final states that we wish to analyse. We begin our analysis at the 3 TeV MuC, as detailed in \autoref{sec:an3}, implementing both a traditional cut-based approach, as well as the BDT classifier. In \autoref{sec:an10} we continue the analysis in the 10 TeV MuC for the benchmark points. As a collider-specific work requires, in \autoref{sec:reach} we present a benchmark-independent discovery projection of the model parameter space at both the 3 TeV and 10 TeV MuC. Finally, in \autoref{sec:conc} we summarize our results and conclude the work.

\section{The HTM+S Framework}
\label{sec:mod}

For this model, we extend the scalar sector of SM with a real $SU(2)$ triplet scalar carrying hypercharge $Y=0$ (usually called the Higgs triplet model or HTM), as well as a real scalar singlet $S$ that will eventually become the Dark Matter that we desire. The new scalar sector now contain the following multiplets: 
\begin{equation}	
	\Phi(1,2,1) =	\begin{pmatrix}
		\phi^+ \\ \phi^0
	\end{pmatrix},\quad \mathcal{T} (1,3,0)=\half \begin{pmatrix}
		T^0 & \sqrt{2}T^+ \\ \sqrt{2}T^- & -T^0 
	\end{pmatrix},\quad S(1,1,0),
\end{equation}
with the numbers in brackets representing the gauge charges under the SM gauge group of $SU(3)_C \times SU(2)_L \times U(1)_Y$. Additionally, we introduce a discrete $Z_2$ symmetry to the model, under which only $S$ transforms as odd, and rest of the SM fields as well as $\mathcal{T}$ transform as even. The tree-level scalar potential invariant under $Z_2$ and the SM gauge group is given by
\begin{align}
	\begin{split}
		V(\Phi,T,S) ={}& \mu_\Phi^2 \Phi^\dagger \Phi + \mu_T^2 Tr(T^\dagger T) + \mu_S^2 S^2+\lambda_h \abs{\Phi^\dagger \Phi}^2 + \lambda_t \abs{Tr(T^\dagger T)}\\
		{}&+\lambda_{s}S^4+\lambda_{ht} \Phi^\dagger \Phi Tr(T^\dagger T) + \lambda_{hs} \Phi^\dagger \Phi S^2 +\lambda_{st} Tr(T^\dagger T) S^2\\
		{}&+ A_{ht} \Phi^\dagger T \Phi 
	\end{split}
\end{align}
where $\mu_{\phi,T,S}$ are the bare mass terms, $\lambda_{h},\, \lambda_{t}, \, \lambda_{s}, \lht, \lhs, \lst $ are dimensionless quartic couplings, and $A_{ht}$ is a trilinear coupling with a mass dimension of one, which eventually facilitates mixing between the doublet and triplet scalars. The electroweak symmetry breaking (EWSB) for this model witnesses contribution from both the doublet and the triplet scalars, with their neutral components obtaining vacuum expectation values (vev) of $v_h$ and $v_t$, respectively. $S$ being $Z_2$-odd does not participate in EWSB, and becomes the DM candidate of our model. A point to be noted here is that, owing to the zero hypercharge, the triplet vev does not contribute to the mass of the $Z$-boson, while the $W^\pm$ bosons obtains some contribution as follows:

\begin{equation}
	M_{W^\pm}^2 = \frac{g_2^2}{4}(v_h^2 + 4v_t^2), \quad M_Z^2 = \frac{g_2^2+g_1^2}{4}v_h^2 \, .
\end{equation}

Now, as per \autoref{eq:rho}, the SM custodial symmetry preserves the EW $\rho$-parameter at the tree-level. However, owing to the contribution of $v_t$ to the $W^\pm$ mass, this parameter deviates from unity at tree-level in the HTM part of our framework:

\begin{equation}
	\rho = 1 + \dfrac{4v_t^2}{v_h^2}.
\end{equation}

Thus, the non-zero vev of the triplet is the custodial symmetry-breaking parameter. The experimental value of $\rho$ is determined to be $1.00038 \pm 0.00020$ \cite{ParticleDataGroup:2022pth}, which puts stringent constraint on the triplet vev, restricting it to $v_t \lsim 3$ GeV. The presence of such a small but non-zero $v_t$ also enables mixing between the SM Higgs doublet and the scalar triplet, leading to new mass eigenstates in our model, which will be discussed in the next subsection.

\subsection{Gauge mixing and scalar mass spectrum}

After EWSB, the charged and neutral gauge eigenstates of the doublet and triplet fields mix among themselves, resulting into physical mass eigenstates. In terms of the gauge basis, we can write the physical mass eigenstates with the help of $2\times2$ rotation matrices $\mathcal{R}_H$ and $\mathcal{R}_P$ as follows:
\begin{equation}\label{eq:htm_states}
	\begin{split}
		\begin{pmatrix}
			h \\ H^0
		\end{pmatrix} ={}& \mathcal{R}_H \begin{pmatrix}
			\phi^0 \\ T^0
		\end{pmatrix}\\
		\begin{pmatrix}
			G^\pm \\ H^\pm
		\end{pmatrix} ={}& \mathcal{R}_P \begin{pmatrix}
			\phi^\pm \\ T^\pm
		\end{pmatrix}\,.
	\end{split}
\end{equation}

Here, $h^0, H^0$ are CP-even neutral scalars, $H^\pm$ is a pair of charged scalars, and $G^\pm$ is the charged Goldstone boson. The charged sector mixing is given as \cite{Butterworth:2023rnw}:

\begin{align}
	G^\pm ={}& \cos \alpha_+ \phi^\pm + \sin\alpha_+ T^\pm \label{eq:htm_gp}\\
	H^\pm ={}& -\sin \alpha_+ \phi^\pm + \cos\alpha_+ T^\pm \label{eq:htm_tp}\, ,
\end{align}

where we have identified the rotation matrix as
\begin{equation}
	\mathcal{R}_P \equiv \begin{pmatrix}
		\cos\alpha_+ & \sin\alpha_+ \\
		-\sin\alpha_+ & \cos\alpha_+ 
	\end{pmatrix}, \quad \tan{2\alpha_+} = \frac{4v_h v_t}{4v_t^2 - v_h^2}
\end{equation}

Owing to the constraint from the $\rho$-parameter, we have $v_t \ll v_h$. As a result, the Goldstone boson $G^\pm$ is mostly doublet, while $H^\pm$ is mostly triplet ($T^\pm$). After EWSB, the mass matrix for the charged scalar eigenstates is obtained to be 

\begin{equation}\label{eq:htm_cm}
	\mathcal{M}_\pm^2 = 	\begin{pmatrix}
		A_{ht} v_t & \half A_{ht} v_h \\
		\half A_{ht} v_h & \frac{A_{ht} v_h^2}{4 v_t}
	\end{pmatrix}\,.
\end{equation}

One of the two eigenvalues of $\mathcal{M}_\pm^2$ is identically zero, corresponding to the massless Goldstone boson $G^\pm$ that is eaten by the $W^\pm$ to obtain its own mass. The non-zero eigenvalue gives the mass of the mostly triplet charged scalar as:
\begin{equation}
	M_{H^\pm}^2 = \frac{A_{ht} (v_h^2+4v_t^2)}{4 v_t} \label{eq:htm_mtp}
\end{equation}

Coming to the neutral sector, the physical mass eigenstates are written as\cite{Butterworth:2023rnw}:
\begin{align}
	h ={}& \cos \alpha_0 \phi^0 + \sin\alpha_0 T^0 \label{eq:htm_h0}\\
	H^0 ={}& -\sin \alpha_0 \phi^0 + \cos\alpha_0 T^0 \label{eq:htm_t0}\, ,
\end{align}

identifying the rotation matrix as 
\begin{equation}
	\mathcal{R}_H \equiv \begin{pmatrix}
		\cos\alpha_0 & \sin\alpha_0 \\
		-\sin\alpha_0 & \cos\alpha_0 
	\end{pmatrix}\,, \quad \tan{2\alpha_0} = \frac{4v_h v_t(\aht - 2 \lht v_t)}{v_h^2 \aht - 8v_h^2 v_t \lambda_{h} + 8v_t^3 \lambda_{t}}. \label{eq:htm_rn}
\end{equation}

The mass matrix of these neutral eigenstates is given by\cite{Chabab:2018ert}

\begin{equation}
	\mathcal{M}_0 ^2= 	\begin{pmatrix}
		2v_h^2 \lambda_h & -\half A_{ht} v_h  + v_h v_t \lambda_{ht}\\
		-\half A_{ht} v_h  + v_h v_t \lambda_{ht} & 2\left( \frac{A_{ht} v_h^2}{8v_t}+v_t^2 \lambda_t\right)
	\end{pmatrix} \,  = \begin{pmatrix}
		\mathcal{A} & \mathcal{B} \\
		\mathcal{B} & \mathcal{C}
	\end{pmatrix}\,. \label{htm_m0}
\end{equation}

Diagonalizing this mass matrix, we obtain the masses of the two neutral scalars as
\begin{align}
	M_{h}^2 = {}& \frac{1}{2}\left\{ (\mathcal{A}+\mathcal{C}) - \sqrt{4\mathcal{B}^2 + (\mathcal{A}-\mathcal{C})^2} \right\} \\
	M_{H^0}^2 ={}& \frac{1}{2}\left\{ (\mathcal{A}+\mathcal{C}) + \sqrt{4\mathcal{B}^2 + (\mathcal{A}-\mathcal{C})^2} \right\}
\end{align}


Once again, as a result of $v_t \ll v_h$, $H^0$ is mostly triplet ($T^0$), while $h$ remains mostly doublet. We identify $h$ as our SM-like Higgs boson of mass $\sim125.5$ GeV, and study a heavier mass spectrum of $H^\pm$ and $H^0$ accordingly for the model. Lastly, the scalar singlet DM mass receives contributions from both the doublet and the triplet vevs, as follows:

\begin{equation}
	M_{S}^2 = \mu_S^2 + \frac{v_h ^2 \lambda_{hs}}{2}+\frac{v_t ^2 \lambda_{st}}{2}.
\end{equation}

Owing to the >99\% triplet contribution to the mass eigenstates of $H^\pm$ and $H^0$, we will often refer to them merely as ``triplet scalars ($T^\pm, T^0$)"  throughout this article. Additionally, the smallness of the triplet vev also leads to very small tree-level mass splitting between $H^\pm$ and $H^0$, which allows us to consider $M_{H^\pm} \simeq M_{H^0}$ across the subsequent sections. After establishing the mass spectrum of the physical scalar eigenstates, we will have a look at their decay channels in the next subsection.

\subsection{Decays of the triplet scalars}
\label{sec:decay}

Different decay channels of triplet-like charged and neutral scalars are well studied in literature \cite{FileviezPerez:2008bj,Chabab:2018ert,Butterworth:2023rnw}, and the dominant decay modes for both of them are somewhat established, when the model is not further extended with a singlet scalar DM. In the HTM+S scenario, the inclusion of the DM does not affect the decay modes and branching ratios (BR) of $H^\pm$. The telltale sign of a triplet charged scalar with a non-zero vev lies in the $H^+ \to Z W^+$ decay, facilitated by the vertex 
\begin{equation}
g_{H^+ W^+ Z} = \frac{g_2}{2}\left(-v_h g_1\sin\theta_W \sin\alpha_+ + 2v_t g_2\cos\theta_W \cos\alpha_+ \right). \label{eq:hpzw}
\end{equation}
This vertex is a definitive proof of the custodial symmetry breaking in this model, as it vanishes when $v_t \to 0$. The other dominant decay modes of $H^\pm$ having a mass of 180 GeV or above are $H^+ \to t\bar{b}$ and $H^+ \to hW^+$, which are also present in other models as well, such as 2HDM extensions. They are facilitated by the following vertices:
\begin{align}
	g_{H^+ t \bar{b}} &= \frac{\sqrt{2}}{\sqrt{v_h^2 + 4 v_t^2}}(M_t + M_b)\sin\alpha_+ \,, \label{eq:hptb}\\
	g_{H^+ W^+ h} &= \frac{g_2}{2}\left(\cos\alpha_0 \sin\alpha_+ + \sin\alpha_0 \cos\alpha_+\right)\left(p_{H^+}^\mu + p_{h}^\mu\right). \label{eq:hphw}
\end{align}

\begin{figure}[h]
	\centering
	\subfigure[]{\includegraphics[width=0.4\linewidth]{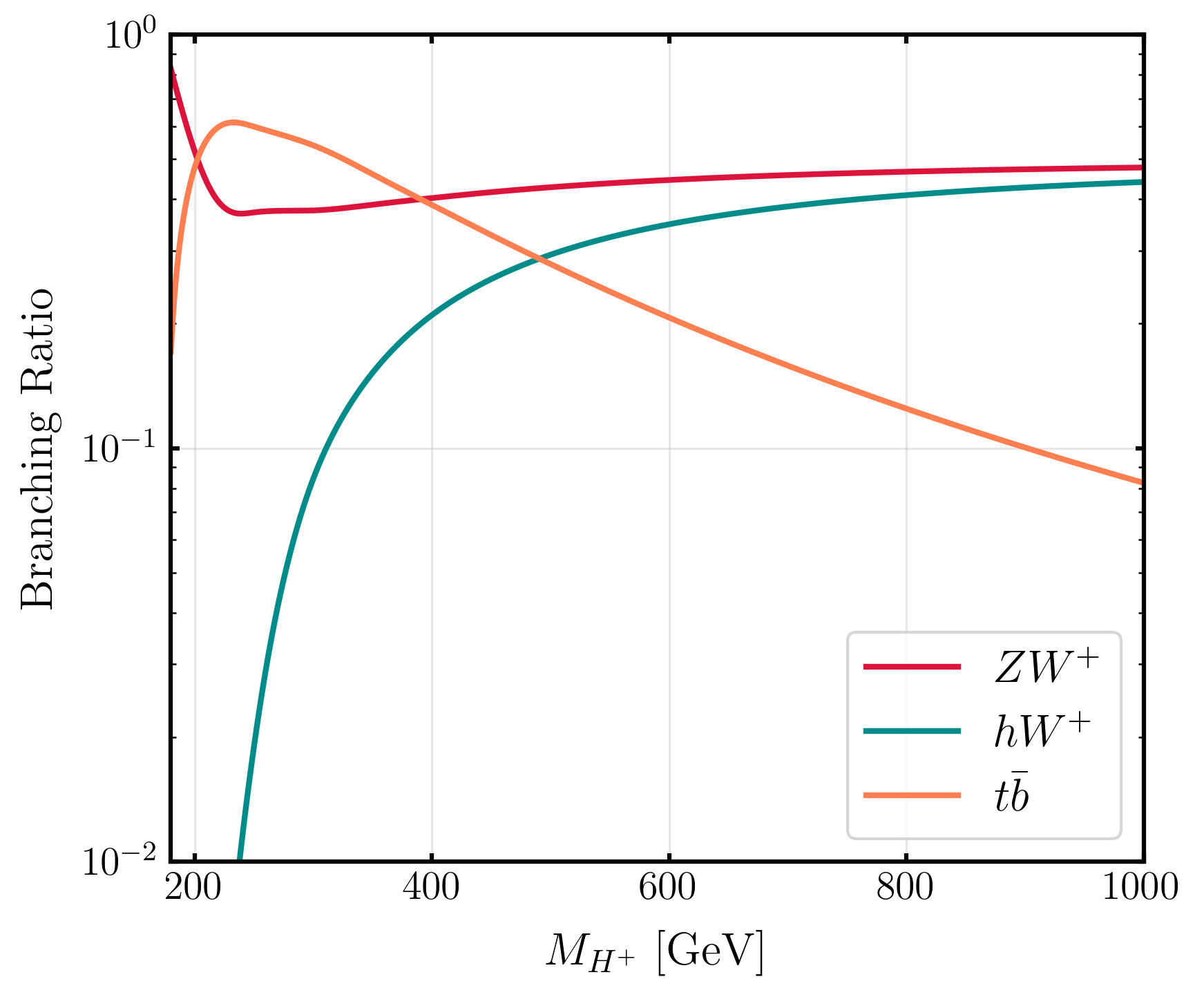}}
	\subfigure[]{\includegraphics[width=0.4\linewidth]{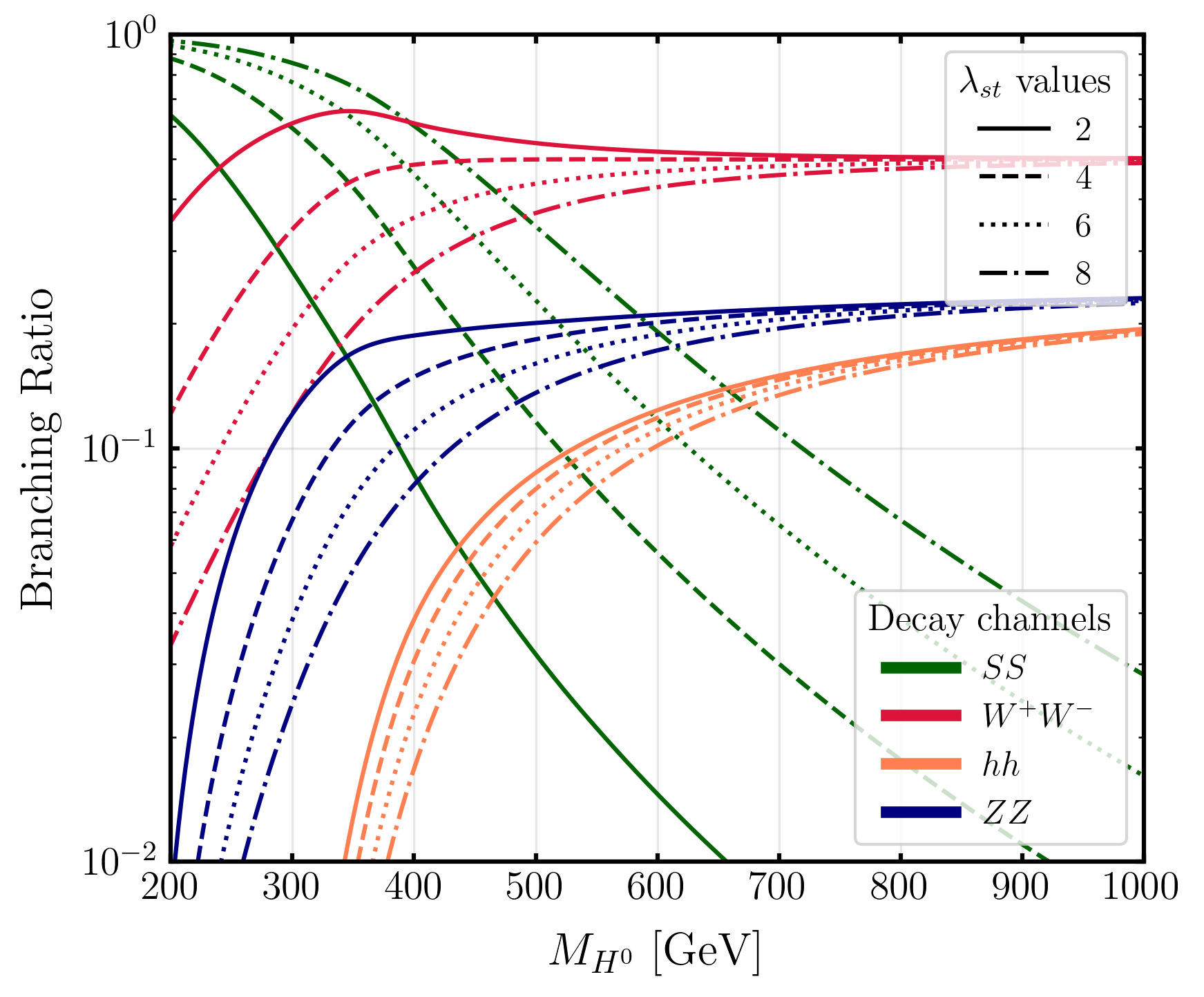}}
	\caption{(a) Charged scalar mass ($M_{H^+}$) vs branching ratios in three dominant decay modes $ZW^+$ (red), $hW^+$ (cyan), and $t\bar{b}$ (orange). (b) Heavy neutral scalar mass  ($M_{H^0}$) vs branching ratios in four dominant decay modes $W^+W^-$ (red), $hh$ (orange), $ZZ$ (navy), and $SS$ (green). The line styles correspond to different values of $\lambda_{st} = $ 2 (solid), 4 (dashed), 6 (dotted) and 8 (dash-dotted).}
	\label{fig:brplots}
\end{figure}

 Especially, the $t\bar{b}$ mode has contribution from only the tiny amount of doublet that remains in the $H^+$ mass eigenstate, represented by the $\sin\alpha_+$ in \autoref{eq:hptb}. The corresponding branching ratios for the $H^\pm$ as function of $M_{H^\pm}$ are displayed in \autoref{fig:brplots}(a). For the range of masses considered, the value of $v_t$ has been fixed at 3 GeV, varying only $A_{ht}$. To keep the $h$ mass $\sim 125.5$ GeV, the values of $\lambda_{h}, \lht$ are also varied accordingly, while fixing $\lambda_{t} = 0.2$. The branching ratios for the $ZW^+$, $h W^+$, and $t\bar{b}$ decay modes of $H^+$ are shown in red, teal, and orange curves respectively. The range for $M_{H^+}$ is chosen in such a way that all three of these decay modes are kinematically allowed. We notice that, for most of the mass range, $ZW^+$ mode remains dominant, consistently staying around $40\%-50\%$ throughout. However, between the triplet scalar mass values of 220 GeV to 380 GeV, the $t\bar{b}$ mode dominates, purely due to the combination of the parameters that influence the couplings and the respective partial decay widths, mentioned in \autoref{sec:appa}. The $t\bar{b}$ branching ratio falls down quickly as the $hW^+$ mode becomes kinematically available, whose own branching ratio reaches close to the $ZW^+$ mode as the charged scalar mass increases.

The significant difference in decay modes is observable in the $H^0$ decay, with the $\lambda_{st}$ parameter governing the $H^0 \to SS$ branching ratio. The other dominant decay channels of $H^0$ are into $W^+ W^-$, $ZZ$ and $hh$, which is also the case for a stand-alone triplet model. The vertices that enable these decays are as follows:
\begin{align}
	\begin{split}
	g_{H^0 SS} &= \lhs v_h \sin\alpha_0 - \lst v_t \cos\alpha_0,\\
	g_{H^0 W^+ W^-} &= \frac{g_2^2}{2}(4v_t \cos\alpha_0 - v_h\sin\alpha_0),\\
	g_{H^0 ZZ} &= \frac{g_1^2 + g_2^2}{2}v_h \sin\alpha_0,\\
	g_{H^0 hh} &= (6\lambda_{h} - 2\lht)v_h\cos\alpha_0^2\sin\alpha_0 + (-A_{ht}+2\lht v_t - 6\lambda_{t}
	 v_t)\cos\alpha_0\sin\alpha_0^2 \\ &{}+ (\frac{A_{ht}}{2} - \lht v_t)\cos\alpha_0^3 + \lht v_h \sin\alpha_0^3
	\end{split}
\end{align}

To showcase the influence of $\lst$ on the decay rate of $H^0$, in \autoref{fig:brplots}(b) we present the branching ratios of the four aforementioned dominant channels against $M_{H^0}$, keeping $v_t = 3$ GeV and also fixing the DM mass $M_S = 60$ GeV, for ease of comparison. In the plot, the green, red, orange and blue colours represent the decay channels of $SS,\,W^+W^-,\,hh$ and $ZZ$ respectively. We illustrate these branching ratios as a function of $M_{H^0}$ for four different choices of $\lst$ = 0.2, 0.4, 0.6 and 0.8, represented in the plot with solid, dashed, dotted, and dash-dot lines respectively. We see that, for lower masses of $H^0$, the $SS$ decay channel dominates over the other SM channels, which persists for a longer range when $\lst$ is larger. However, for $M_{H^0} \geq 500$ GeV, even for $\lst=8.0$ the $H^0 \to W^+ W^-$ mode becomes dominant. Above a triplet scalar mass of $\sim800$ GeV, the $SS$ mode carries the least branching ratio out of the four. In contrast, for $M_{H^0} \leq 250$ GeV, even for a low value of $\lst=2.0$, the $H^0\to SS$ mode dominates. For high values of $\lst$ in this lighter mass range, the branching ratio stays $\gsim 90\%$. The partial decay width expressions for these modes can also be found in \autoref{sec:appa}.

\begin{figure}[h]
	\centering
	\includegraphics[width=0.8\linewidth]{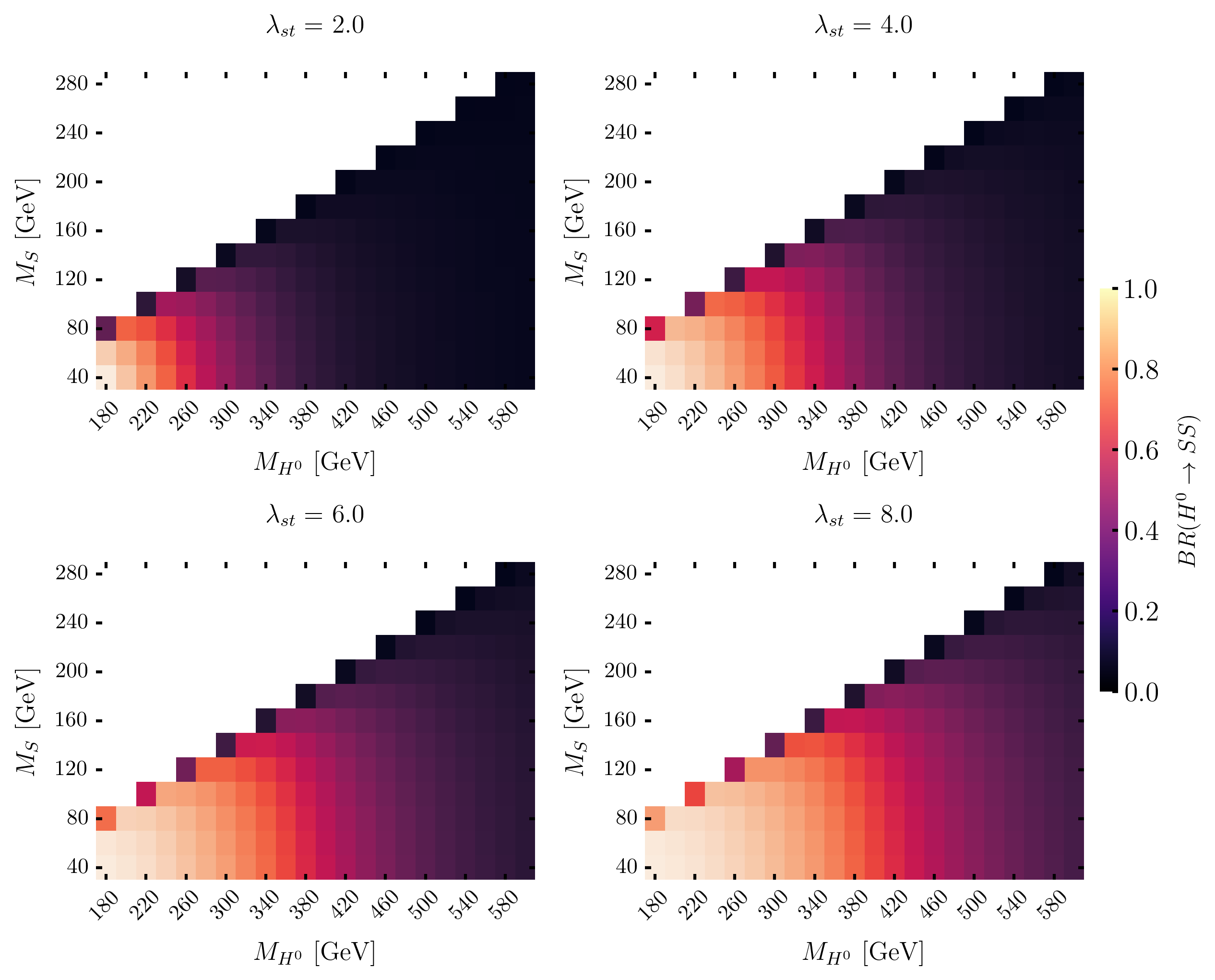}
	\caption{$M_H^0$ and $M_S$ dependence of the $H^0 \to SS$ branching ratio as 2-dimensional heatmaps, for four different $\lst$ values of 2.0, 4.0, 6.0, and 8.0 in consideration.}
	\label{fig:mhms}
\end{figure}

To obtain some more insight on the dependence of the $H^0 \to SS$ branching ratio on $M_H^0, \,M_S$ and $\lst$, in \autoref{fig:mhms} we choose to present heatmaps in the $M_H^0-M_S$ plane, with the colour bar over the corresponding value of $BR(H^0\to SS)$. The four panels correspond to four different $\lst$ values of 2.0, 4.0, 6.0, and 8.0, as written on top of each panel. We see that, for each value of $\lst$, the BR reduces for higher values of $M_{H^0}$ and $M_S$. While for $\lst = 2.0$ we do not get a $\gsim 20\%$ BR as we go beyond $M_{H^0} \geq 300$ GeV with any value of the $M_S$ considered, the situation is expectedly better as $\lst$ increases, with very optimistic BRs being obtainable for a sizeable portion of the $M_{H^0}-M_{S}$ plane with $\lst = 8.0$. Before we move towards the implications of all these decay channels at the colliders, we need to consider the constraints on the model, especially on the quartic couplings, from various theoretical and experimental bounds.

\subsection{Constraints on the quartic couplings}
\label{sec:quart}

While discussing the EWSB and its consequences, we obtained an upper limit on the triplet vev $v_t$ from the $\rho$-parameter. Additional constraints are imposed upon the quartic couplings of the model as well, which can be categorized as follows:

\begin{itemize}
	\item \underline{Bounded-from-below:} The scalar potential needs to be bounded from below, in order to ensure tree-level vacuum stability. The following conditions are imposed on the quartic couplings \cite{Butterworth:2023rnw, Robens:2015gla, Arhrib:2011uy}:
	\begin{equation}
		\{\lambda_{h}, \lambda_{t}, \lambda_{s} \} > 0; \quad \lambda_{hs} + 2\sqrt{\lambda_{h}\lambda_{s}} \geq 0; \quad  \lambda_{ht} + 2\sqrt{\lambda_{h}\lambda_{t}} \geq 0; \quad  \lambda_{st} + 2\sqrt{\lambda_{s}\lambda_{t}} \geq 0.
	\end{equation}
	
	\item \underline{Perturbativity:} The requirement of a perturbative theory generally demands $\abs{\lambda_i} \leq 4\pi$, where $\lambda_{i}$ stands for all the scalar quartic couplings \cite{Khan:2016sxm}. 
	
	\item \underline{$h \to \gamma \gamma$ measurement:} The presence of the charged scalar in the triplet implies its contribution to the $h\to\gamma\gamma$ branching ratio, imposing stringent limits on the $\lht$ coupling. Given the small $v_t$, the $H^\pm$ has >99\% contribution from the triplet, and hence the limits on the corresponding coupling from $h \to \gamma \gamma$ remains same as the one for an inert triplet model, as evaluated in ref. \cite{Bandyopadhyay:2024plc}. For our collider benchmarks, we will choose $\lambda_{ht}$ values that stringently keep the $h \to \gamma \gamma$ branching ratio within 1$\sigma$ of the recent ATLAS measurement \cite{ATLAS:2022tnm}. The $\lht$ coupling also plays a significant part in enabling the first-order electroweak phase transition\cite{Niemi:2018asa, Bandyopadhyay:2021ipw}, where for $M_{H^\pm} \geq 200$ GeV, one requires $\lht \geq 1.5$ to satisfy the FOPT conditions. This requirement stays in tension with the $h\to \gamma \gamma$ bound however, which allows such large values only for $M_{H^\pm} \gsim 350$ GeV \cite{Bandyopadhyay:2024plc}. Detailed discussion on FOPT conditions are beyond the scope of this work.
	
	\item \underline{$h\to$invisible decay:} In \autoref{fig:mhms}, we see that, no matter what $\lst$ we take, the highest $H^0 \to SS$ branching ratios are obtained when $M_S$ remains lower than 60 GeV. However, such DM masses kinematically allow the fully invisible $h \to SS$ decay, a heavily constrained mode from various experiments. The latest result from CMS \cite{CMS:2023sdw} provides a combined upper limit of $15\%$ on this branching ratio at a 95\% confidence level, with phenomenological studies like ref. \cite{Ngairangbam:2020ksz} showing that this can potentially be reduced to even more stringent values of $\lsim 7\%$. For the choice of $M_S < M_h/2$, the $\lst$ is hence restricted to be $\lsim 0.01$. This again is in contradiction with the requirement of FOPT for an inert scalar extension of the SM, which is feasible for $\lst \geq 1$ when $M_S$ = 50 GeV \cite{Bandyopadhyay:2021ipw}. The tension can however be easily ameliorated by taking $M_S > M_h/2$. For the purpose of our work, we will deal with the former choice of $M_S$ in our benchmark points, and hence we will exclusively work with $\lst < 0.01$. 
	
	\item \underline{DM relic and direct detection experiments:} As mentioned in the introduction, our primary motivation for adding the scalar singlet is to have a stable WIMP-like DM candidate in the model, without much complications. Due to the presence of both $\lhs$ and $\lst$ portal couplings, the annihilation rate of $SS \to SM\,SM$ in attaining the observed relic density of the universe is mediated by both the doublet and the triplet scalars. However, a majority of the contribution comes from $\lhs$ due to the SM Higgs boson being the lighter propagator, especially in the low-mass range of $S$, and hence we get further constraint on this portal coupling. $\mathcal{O}(1)$ values of $\lst$ are still permissible from the DM relic requirement. Even stronger constraints are obtained on these portal couplings from the DM direct detection experiments, with the LUX-ZEPLIN (LZ) data providing the most stringent of the lot \cite{LZ:2022ufs}. The doublet Higgs portal is again the preferred mode for $S$ to have spin-independent DM-nucleon scattering cross-sections, compressing the allowed range of $\lhs$ even lower. This is actually good for us, to successfully avoid the $h\to$invisible bounds. This constraint is also not very strong on $\lst$ as the triplet component of $H^0$ does not couple to quarks inside the nucleons. However, very high values of $\lst$ are still disfavoured by the LZ bound. 
	
\end{itemize}

With this discussion in mind, we establish a set of benchmark points in the next subsection, on which we will design our collider analysis strategy.

\subsection{Benchmarks for collider study}
\label{sec:bp}

\begin{table}[h]
	\renewcommand{\arraystretch}{1.2}
	\centering
	\begin{tabular}{|c|c|c|c|c|c|c|}
		\hline
		BP & $M_{H^+} \sim M_{H^0}$ [GeV] & $M_{S}$ [GeV] &$\lambda_{st}$ & $\lambda_{hs}$ & $\lambda_{ht}$ & $\Omega_{DM}^S$\\
		\hline
		BP1 &  200 & 58.8 & 7.5&0.00065&0.5&0.12\\	
		\hline
		BP2 & 350 & 58.5 & 6.5 & 0.0002 & 1.0&0.12\\
		\hline
		BP3 & 500 & 58.8  & 5.6 & 0.0001 & 1.0 &0.12\\
		\hline
	\end{tabular}
	\caption{Choices of benchmark points for the collider analysis.}
	\label{tab:bp}
\end{table}

Imposing the constraints discussed in the previous subsections, we finalize a set of benchmark points (BP), with three different masses of the triplet scalars, as detailed in \autoref{tab:bp}.
Across each benchmark, we have kept fixed values of $v_t = 3$ GeV, $\lambda_{s}$ = 0.2, and $\lambda_{t}$ = 0.2, while also ensuring that $M_h \sim 125.5$ GeV is held for the three BPs. For our choice of parameters, all the three benchmark masses of $H^\pm$ and $H^0$ are well within the limits of recent LHC searches for charged and heavy neutral scalars \cite{CMS:2019bnu, ATLAS:2020tlo, CMS:2021wlt, ATLAS:2021upq}. Utilizing \texttt{micrOMEGAs-5.3.41}, we ensure that the DM relic density $\Omega_{DM}^S$ matches the Planck data of observed relic density of the universe, $\Omega_{DM}^{Obs} = 0.1198 \pm 0.0012$ \cite{Planck:2018vyg}. To achieve both the correct relic density and a healthy $H^0 \to SS$ branching ratio, as well as a sense of uniformity in the DM part of the BPs, we keep $M_S \sim 59$ GeV for all three of the benchmarks. The DM portal couplings $\lambda_{st}$ and  $\lambda_{hs}$ are determined based on constraints from the relic, as well as the DM-nucleon spin-independent scattering cross-section bounds from the LZ experiment. The allowed values of $\lhs$ for all three BPs are low enough to stay within the permissible $h\to$invisible decay limit. The chosen $\lst$ values are adequate to yield strong branching ratios for $H^0 \to SS$, while simultaneously not violating the tree-level perturbativity and bounded-from-below limits, as well as the LZ bound for DM direct detection. Finally, the $\lambda_{ht}$ values, which do not have direct impact on the DM phenomenology, are chosen to stay within the allowed 1$\sigma$ limit of the ATLAS measurement of $h \to \gamma \gamma$ \cite{ATLAS:2022tnm, Bandyopadhyay:2024plc}. 

\begin{table}[h]
	\renewcommand{\arraystretch}{1.0}
\centering
\begin{tabular}{|c|c|c|c|}
	\hline
	\makecell{BP \\ {(Triplet mass)}} & Triplet scalar & Dominant decays & Branching Ratio (\%) \\
	\hline
	\multirow{4}{*}{\makecell{BP1 \\ (200 GeV)}} & \multirow{2}{*}{\makecell{$H^+$}} & $ZW^+$ & 51.8 \\
	&&$t\bar{b}$ & 48.2 \\
	\cline{2-4}
	& \multirow{2}{*}{\makecell{$H^0$}} & $SS$ & 93.7 \\
	&&$W^+ W^-$ & 6.2 \\
	\hline\hline
	\multirow{6}{*}{\makecell{BP2 \\ (350 GeV)}} & \multirow{3}{*}{\makecell{$H^+$}} & $t\bar{b}$ & 45.9 \\
	&&$ZW^+$ & 38.8 \\
	&&$hW^+$ & 15.3 \\
	\cline{2-4}
	& \multirow{3}{*}{\makecell{$H^0$}} & $SS$ & 66.3 \\
	&&$W^+ W^-$ & 26.1 \\
	&&$ZZ$ & 6.8 \\
	\hline\hline
	\multirow{8}{*}{\makecell{BP3 \\ (500 GeV)}} & \multirow{3}{*}{\makecell{$H^+$ }} & $ZW^+$ & 42.4 \\
	&&$hW^+$ & 30.0 \\
	&&$t\bar{b}$ & 27.6 \\
	\cline{2-4}
	& \multirow{5}{*}{\makecell{$H^0$ }} & $W^+ W^-$ & 43.2 \\
	&&$SS$ & 19.9 \\
	&&$ZZ$ & 16.3 \\
	&&$t\bar{t}$ & 10.8 \\
	&&$hh$ &9.7 \\
	\hline
\end{tabular}
\caption{Dominant decay modes and branching ratios for $H^+$ and $H^0$ in the three benchmark points.}
\label{tab:br}
\end{table}

In \autoref{tab:br} we mention the dominant decay channels and their respective branching ratios of $H^\pm$ and $H^0$ for each of the three BPs. The branching ratios are consistent with the trends shown in \autoref{fig:brplots}(a) and (b) for each case. Most notably, BP1 shows the highest branching ratios for both of our desired channels i.e. $H^\pm \to ZW^\pm$ and $H^0 \to SS$, with the latter being $\sim 94\%$, which is highly favourable for the analysis that we propose in the upcoming sections. For BP2, as per \autoref{fig:brplots}(a), the $H^+ \to t\bar{b}$ mode becomes dominant, while the $ZW^\pm$ channel still enjoys $\sim 39\%$ probability of decay. For $H^0$ in BP2, the $SS$ mode continues to dominate due to the high enough value of $\lst = 6.5$, with $\sim66\%$ probability. For BP3, the $ZW^\pm$ dominance is restored for $H^\pm$ with a decay probability of $\sim 42\%$. Whille $H^0 \to SS$ is not the dominant decay mode anymore, it still provides a healthy branching ratio of $\sim20\%$.

With all the details on the model structure, the relevant couplings and their constraints, and a set of well-defined and well-motivated benchmarks that respect the preliminary bounds from theoretical and experimental aspects, we are now equipped enough to begin the detailed collider study. The first step of the study is to identify the favourable production modes and corresponding final states, as entailed in the following section.

\section{VBF production and final states at a Muon Collider}
\label{sec:prodfs}

After establishing the benchmark points for the collider study, our task now is to find out the suitable mode of production that can yield discernible signatures of the model with healthy event rates across the range of the $H^\pm$ mass covered by the benchmarks, and possibly beyond. While the usual DY pair-production of $pp\to H^+ H^-$ can look tempting, we realize that this mode will not contain the DM singlet scalar as a decay product. At the colliders, the DM scalars can contribute to a sizeable missing transverse energy (MET), which is one of the primary discriminators that we would like to use for the model. This can be accounted for by exclusively studying the associated production of $H^\pm H^0$, which affords us the best of both worlds: one leg containing the charged triplet scalar $H^\pm$ which can lead to the multilepton signals that we want, and the other leg having the heavy neutral triplet scalar $H^0$ that has a sizeable probability of decaying completely into a pair of the DM singlet, $H^0 \to SS$, as exemplified in \autoref{tab:br}. The next step is to examine different ways and different colliders where this mode can be produced with respectable cross-sections. We establish the current LHC of $E_{CM} = 14$ TeV, the FCC-hh of $E_{CM} = 100$ TeV, and two MuC energies of 3 TeV and 10 TeV respectively, as the candidates. In the next subsection, we will see in detail how the MuC comes out as the most favourable environment for the purpose of our analysis.

\subsection{Production cross-sections}
\label{sec:cs}

\begin{figure}[h]
	\centering
	\includegraphics[width=0.5\linewidth]{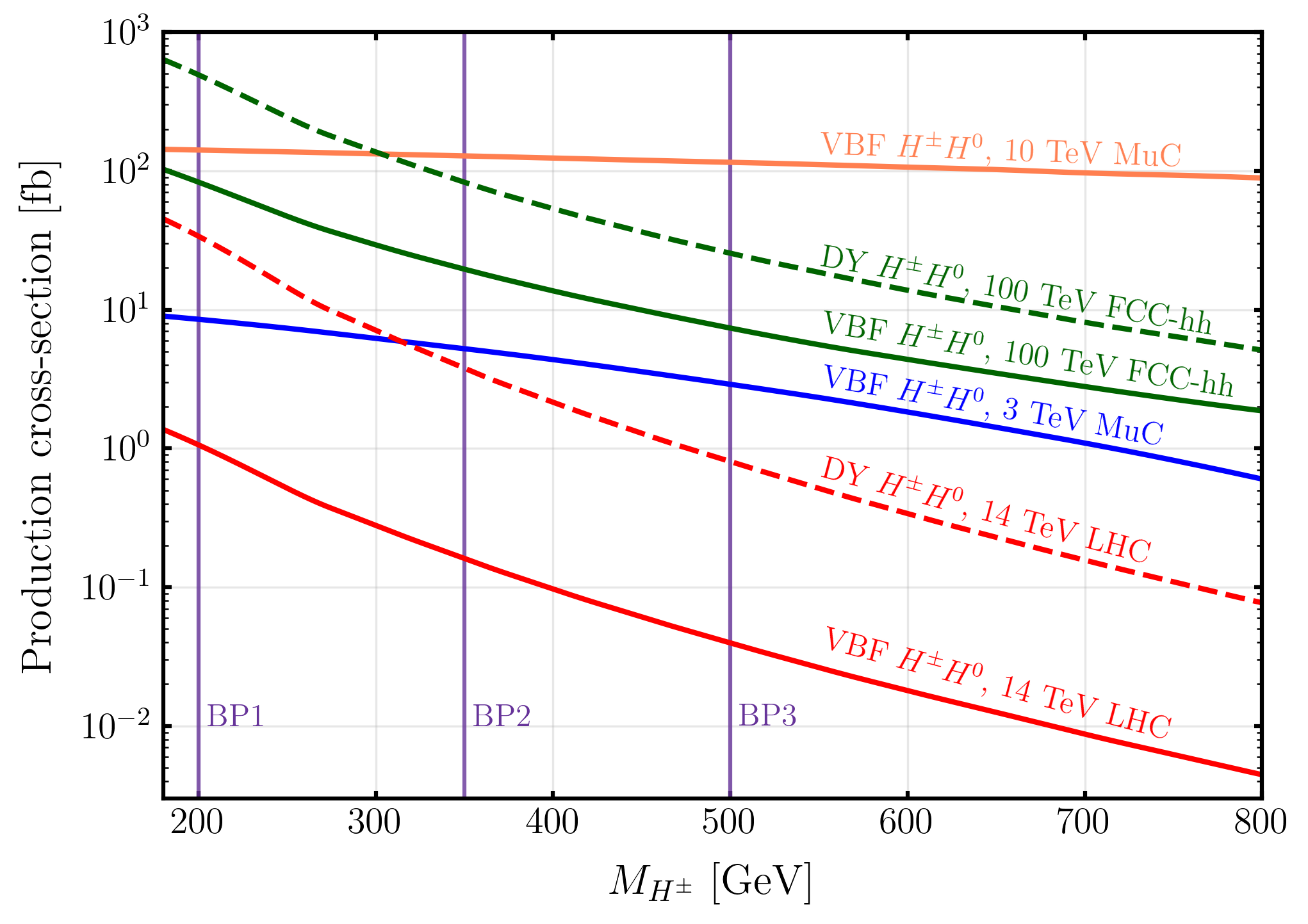}
	\caption{Production cross-sections of processes involving $H^\pm H^0$ in the final state, at the 14 TeV LHC (red), 100 TeV FCC-hh (green), 3 TeV MuC (blue), and 10 TeV MuC (orange). The VBF and DY processes are distinguished by solid and dashed lines, respectively. The vertical purple lines show our benchmark points.}
	\label{fig:csplot}
\end{figure}

In \autoref{fig:csplot} we showcase the production cross-sections for the processes that involve $H^\pm H^0$ in the final state, with respect to $M_{H^\pm} \simeq M_{H^0}$. The cross-sections are calculated at leading order (LO) using \mgFull \cite{Alwall:2014hca}, with \texttt{NNPDF30\_lo\_as\_0130} \cite{NNPDF:2014otw} as the choice of PDF for the $pp$ collision processes. At the hadronic colliders, both the Drell-Yan (DY) production $pp \to H^\pm H^0$ and the vector boson fusion (VBF) process of $pp \to H^\pm H^0 jj $ are feasible, whereas at the muon collider, only the VBF process of $\mu^+ \mu^- \to \mu^\pm \nu_\mu H^\mp H^0$ can be obtain, to ensure the conservation of charge. In \autoref{fig:csplot}, we distinguish between the colliders, with red for the 14 TeV LHC, green for the 100 TeV FCC-hh, blue for the 3 TeV MuC, and orange for the 10 TeV MuC, respectively. Additionally, we denote the VBF processes with solid lines, while the dotted lines represent the DY processes. Right away, one notices that the VBF processes at the MuC energies of 3 TeV and 10 TeV reigns superior over their hadron collider counterparts of 14 TeV LHC and 100 TeV FCC-hh respectively, consistent with the discussions in ref. \cite{Costantini:2020stv}. While the DY cross-sections at the LHC and FCC-hh start off above the MuC VBF rates, they decrease quickly as the $M_{H^\pm}$ crosses 300 GeV. The rates at which the MuC VBF cross-sections fall w.r.t the triplet scalar mass is less compared to those of the hadron collider processes, which in principle allows us to produce our desired final states for a longer range of masses. While for lower masses the DY production at hadron colliders look promising, in reality the overwhelming SM backgrounds will wash out any hints of the signal, especially at the 14 TeV LHC. VBF processes at the hadronic colliders usually carry less background, however, as the cross-sections fall rapidly for higher triplet scalar masses, it is not the most efficient channel out of all the ones being considered. In comparison, the VBF at MuC promises respectable production rates at both the 3 TeV and 10 TeV energies, and combined with the typically less background of a leptonic collider, it becomes apparent that this should be our collider and process of choice. The three benchmark points, from \autoref{tab:bp}, are shown as vertical purple lines, yielding VBF cross-sections of 8.7 fb, 5.4 fb, and 2.9 fb respectively for BP1, BP2 and BP3, at the 3 TeV MuC. With the increase in MuC energy of 10 TeV, these cross-sections for BP1-BP3 are enhanced to 141.4 fb, 128.3 fb, and 115.5 fb, respectively.

\begin{figure}[h]
	\centering
	\subfigure[]{\includegraphics[width=0.3\linewidth]{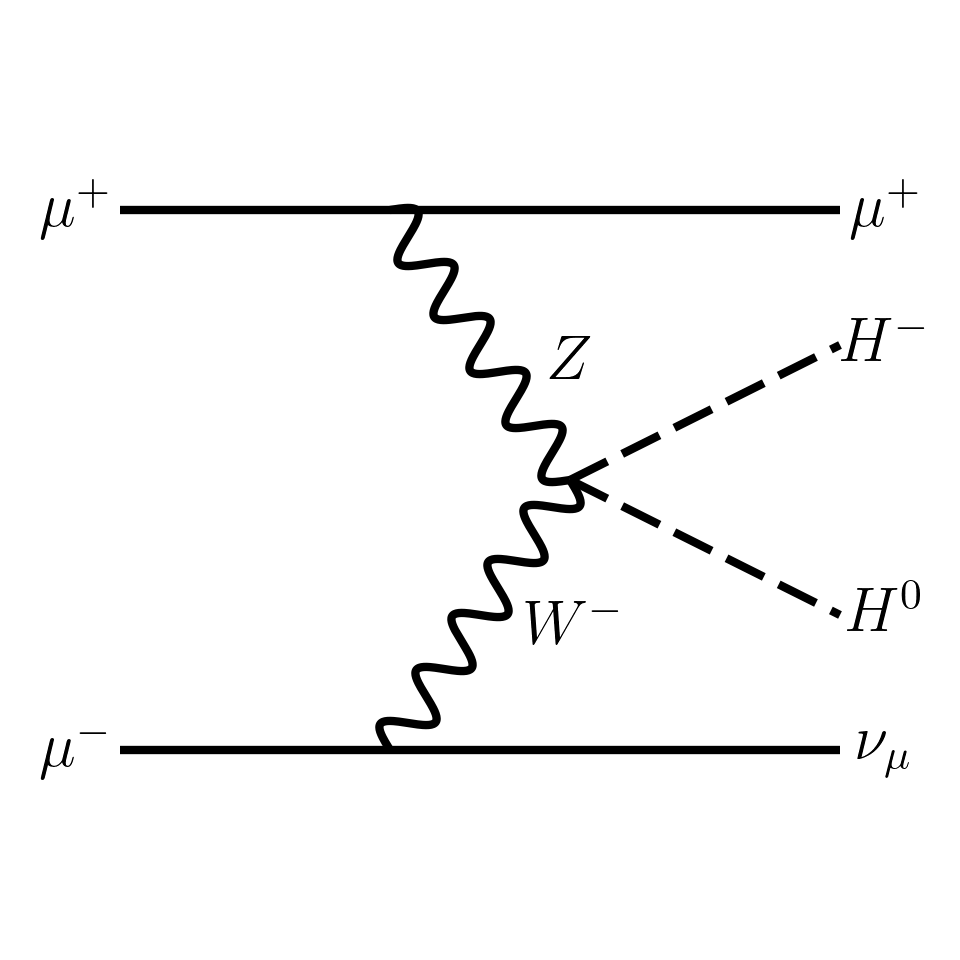}}
	\subfigure[]{\includegraphics[width=0.3\linewidth]{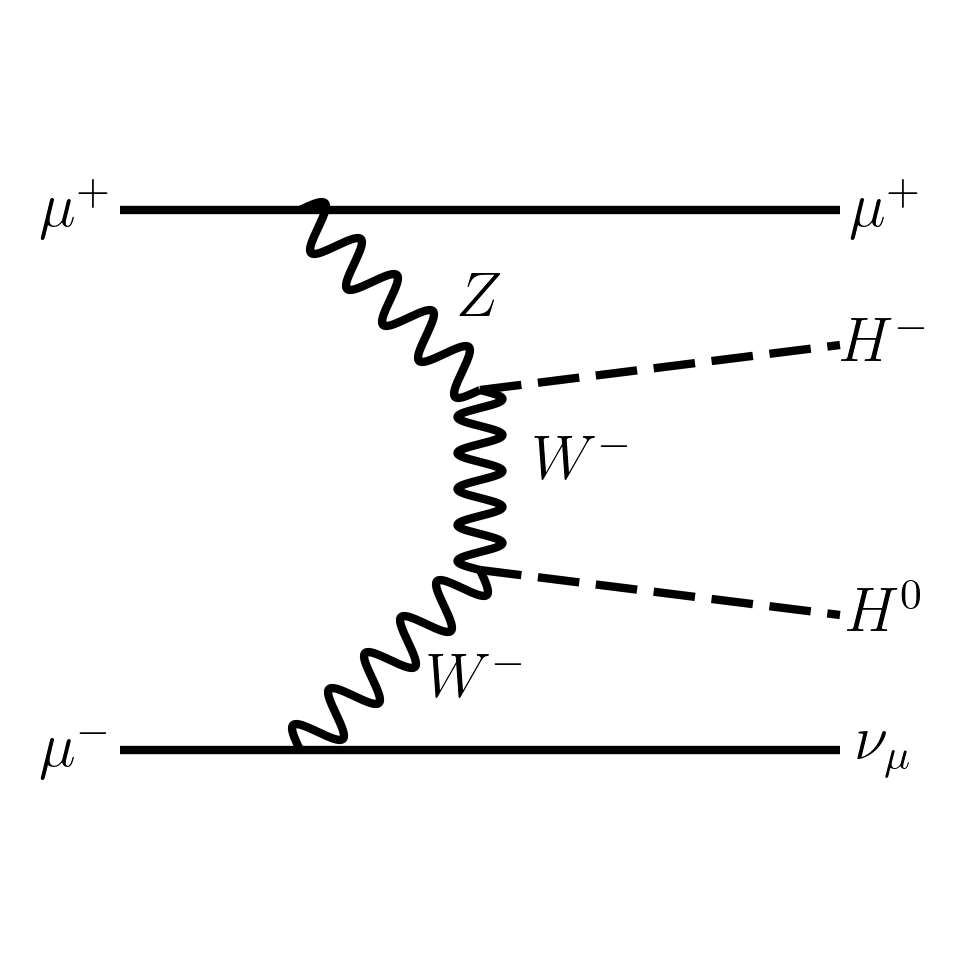}}
	\subfigure[]{\includegraphics[width=0.3\linewidth]{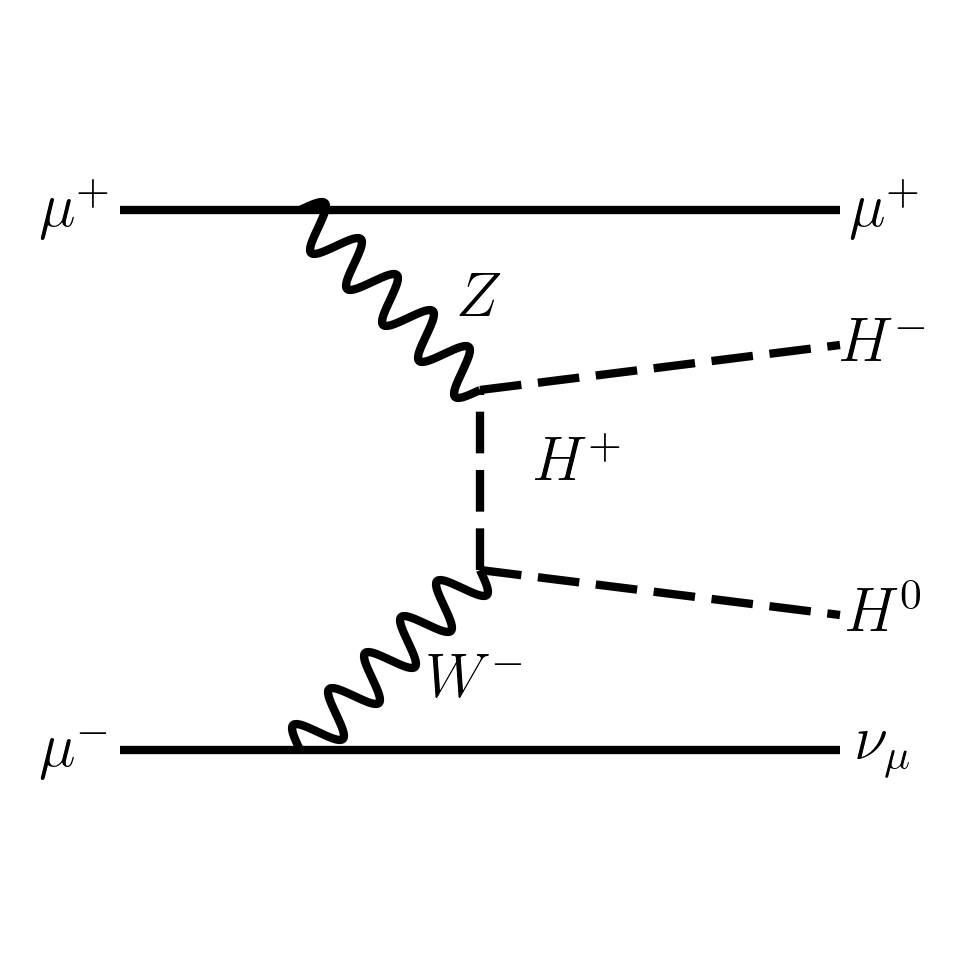}}
	\caption{Feynman diagrams contributing to the VBF production of $H^\pm H^0$ at a muon collider.}
	\label{fig:feynvbf}
\end{figure}

In \autoref{fig:feynvbf} we show the dominant Feynman diagrams that contribute to the VBF production of $H^\pm H^0$ at the muon collider, to have a look at some interesting features. Firstly, the four-point vertex of $H^\pm H^0 W^\mp Z$ in \autoref{fig:feynvbf}(a) contributes the most into the production of these triplet scalars \cite{Bandyopadhyay:2024plc}. This feature is absent in fermionic BSM models, and for doublet-like charged Higgs models such as 2HDM, this vertex is much smaller compared to the triplet case. Hence, for triplet scalars, such a production channel is always favourable to obtain a good number of events for analysis. The diagram in \autoref{fig:feynvbf}(b) has the $H^\pm Z W^\mp$ vertex, which is a feature of triplet extensions, facilitated by the presence of non-zero $v_t$. Lastly, we have the $t$-channel charged triplet mediated diagram in \autoref{fig:feynvbf}(c), which is present in most of the $SU(2)$ multiplet extensions of the SM, and usually contributes dominantly in fermionic models. Another salient feature of the VBF processes at the MuC is the presence of the spectator muons, which are typically produced with high pseudorapidity $\eta$, carrying a large portion of the beam energy as well. Known in contemporary literature as Forward muons, they can be used to trigger VBF processes at the MuC, successfully trimming down majority of the non-VBF SM background. It is important to mention here that, despite being relatively free of SM QCD backgrounds, the MuC is plagued by the so-called beam-induced background (BIB), which is a swarm of soft leptons, photons and pions in the high-$\eta$ region, coming from beam muons decaying in flight \cite{Collamati:2021sbv, Ally:2022rgk}. The current detector design proposal for the MuC includes two conical tungsten nozzles that cover the region of $\abs{\eta} \geq 2.5$, which absorb majority of the BIB particles, but restricts the main detector region to have a pseudorapidity coverage of $\abs{\eta} \leq 2.5$ \cite{MuonCollider:2022ded, InternationalMuonCollider:2024jyv}. However, the aforementioned Forward muons have high enough momenta to traverse through the nozzles, and can thus be found at dedicated detectors, which can be placed in the $\abs{\eta} \geq 2.5$ region \cite{Accettura:2023ked}. Based on this discussion, we can now closely inspect the possible final states that can establish a discovery potential of the model at the multi-TeV muon collider.

\subsection{Final states at the muon collider}
\label{sec:fs}

As discussed in the previous subsections, unique identification for the HTM+S scenario can be performed with the $\mu^\pm \nu_\mu H^\mp H^0$ production channel, resulting from $W^\pm Z$ fusion at a multi-TeV muon collider, looking for fully invisible decay of the $H^0 \to SS$. We can successfully trigger the process by demanding exactly one Forward muon at the dedicated detectors, which can not only help filter out most of the non-VBF SM backgrounds, but also VBF backgrounds emerging from $W^+W^-/ZZ$ fusions that yield either no Forward muons, or exactly two of them. This also prevents model backgrounds such as $\mu^+ \mu^- \to \mu^+ \mu^-  H^+ H^-$ from contaminating the signal as well. All things considered, the final states that we are going to analyse are described below:

\textbf{FS1: 3 leptons + MET + 1 Forward muon:}

When one looks for a triplet-like charged scalar with a non-zero vev, 3$\ell$ + MET is the go-to final state, which can emerge from the $H^\pm \to Z W^\pm$ decay. For our model, a large enough MET can also establish it as a signature for the dark sector decay of $H^0$. Keeping one Forward muon ($F\mu$) to trigger the event, the contributing decay chain is:
\begin{align*}
	\mu^\pm H^\mp H^0 \nu_\mu &{}\to \mu^\pm (Z W^\mp) (SS) \nu_\mu \\
	&{} \to \mu^\pm (\ell^+ \ell^- \ell^\mp \nu) (SS) \nu_\mu \to 3\ell + {\rm MET} + 1F\mu  \\
		\mu^\pm H^\mp H^0 \nu_\mu &{}\to \mu^\pm (h W^\mp) (SS) \nu_\mu \\
		&{}\to \mu^\pm (W^+ W^-) \ell^\mp \nu) (SS) \nu_\mu \\
		&{} \to \mu^\pm ((\ell^+ \nu) (\ell^- \bar{\nu}) \ell^\mp \nu) (SS) \nu_\mu \to 3\ell + {\rm MET} + 1F\mu 
\end{align*}
The $H^\pm \to h W^\pm$ decay channel can contribute for the BP2 and BP3, as seen from the branching ratios in \autoref{tab:br}, albeit with a much less probability compared to the $ZW^\pm$ decay channel. 

\textbf{FS2: 2 $b$-jets + 1 lepton + MET + 1 Forward Muon:}

While this signature is not unique to a triplet charged scalar, it can potentially yield a higher number of events compared to FS1, owing to combined branching ratios of $H^\pm \to h W^\pm$ and $H^\pm \to t\bar{b}$ being quite large, as shown in \autoref{tab:br}. A combined emergence of both FS1 and FS2 can strongly hint towards the triplet case nonetheless. The three decay chains that yield this signature are:
\begin{align*}
	\mu^\pm H^\mp H^0 \nu_\mu &{}\to \mu^\pm (h W^\mp) (SS) \nu_\mu \\
	&{}\to \mu^\pm ((b\bar{b}) \ell^\mp \nu) (SS) \nu_\mu \to 2b{\rm -jet}+1\ell + {\rm MET} + 1F\mu  \\
	\mu^\pm H^\mp H^0 \nu_\mu &{}\to \mu^\pm (Z W^\mp) (SS) \nu_\mu \\
	&{}\to \mu^\pm ((b\bar{b}) \ell^\mp \nu) (SS) \nu_\mu \to 2b{\rm -jet}+1\ell + {\rm MET} + 1F\mu  \\
	\mu^\pm H^\mp H^0 \nu_\mu &{}\to \mu^\pm (t\bar{b}) (SS) \nu_\mu \\
	&{}\to \mu^\pm ((W^\mp b) \bar{b}) (SS) \nu_\mu \\
	&{}\to \mu^\pm ((\ell^\mp \nu b) \bar{b}) (SS) \nu_\mu \to 2b{\rm -jet}+1\ell + {\rm MET} + 1F\mu  \\
\end{align*}
Here, for BP1 and BP2, the $H^+ \to t\bar{b}$ process will be dominant, while for BP3, the $H^+ \to hW^+$ is expected to contribute more. It is important to note here that, only the $\sim 1\%$ doublet contribution in the otherwise triplet-like charged scalar enables the $H^+ \to t\bar{b}$ decay. For heavy 2HDM charged scalars, this mode is usually how it they almost always decay. 

\textbf{FS3: 2 jets + 2 leptons + MET + 1 Forward muon:}

The two previous final states involve the contribution of the neutrino emerging from the leptonic decay of $W^\pm$ into the total MET. This also leaves the charged scalar decay not fully visible, hence restricting us from being able to reconstruct the invariant mass of $H^\pm$. However, FS3 involves fully visible decay products of $H^\pm$, enabling us to achieve this reconstruction. The primary decay chains leading to this final state are:
\begin{align*}
	\mu^\pm H^\mp H^0 \nu_\mu &{}\to \mu^\pm (Z W^\mp) (SS) \nu_\mu \\
	&{}\to \mu^\pm ((\ell^+ \ell^-)jj) (SS) \nu_\mu \to 2{j}+2\ell + {\rm MET} + 1F\mu  \\
	\mu^\pm H^\mp H^0 \nu_\mu &{}\to \mu^\pm (h W^\mp) (SS) \nu_\mu \\
	&{}\to \mu^\pm ((W^+ W^-) jj) (SS) \nu_\mu \\
	&{}\to \mu^\pm ((\ell^+ \nu)(\ell^- \bar{\nu}) jj) (SS) \nu_\mu \to 2{j}+2\ell + {\rm MET} + 1F\mu  \\
\end{align*}
Here, ``jet" ($j$) refers to non-$b$-tagged jets, as we  are interested in obtaining jets from fully hadronic decay of at least one resultant $W^\pm$ boson. Ensuring large enough MET to keep $H^0 \to SS$ as the primary decay channel for the neutral scalar, we can zero in on the charged scalar with the $M_{inv}^{jj\ell\ell}$ distribution.

With the final states established, we will now move to the primary objective of this article: to analyse the viability of these signals at a future muon collider. We will begin with the 3 TeV MuC energy, as detailed in the next section. 

\section{Analysis at a 3 TeV Muon Collider}
\label{sec:an3}

The 3 TeV muon collider is the short-term benchmark goal of the people involved in the research and development, as a part of the International Muon Collider Collaboration. We adopt their target luminosity of 1 \abi for our analysis, as well broadly considering the detector specifications and reconstruction algorithms recommended by them. We generate the events for each final state with their fiducial cross-section using \mgFull. Based on the final state composition, we allow the charged scalar to decay into that particular state at the gen-level, alongside ensuring the $H^0 \to SS$ decay to have the high-MET events that we desire. Because of this, the fiducial cross-sections are much lower than the inclusive cross-sections presented in \autoref{fig:csplot}. The SM backgrounds are also generated with the specific final states in mind, details of which can be found in the discussion for each final state that follows. After the generation, the events are showered and hadronized using \py \cite{Bierlich:2022pfr}, and the fast detector simulation is performed with \texttt{Delphes-3.5.0}\cite{deFavereau:2013fsa}, utilizing the \texttt{delphes\_card\_MuonColliderDet.tcl} card. We summarize the basic setup of the detector simulation as follows: 

\begin{itemize}
	\item Within the main detector, the charged leptons and jets are selected with $p_T \geq 20$ GeV, and $\abs{\eta} \leq 2.5$ because of the tungsten nozzles. The large $p_T$ cut is applied to reject leptons that may emerge from BIB at the calorimeters. 
	\item For leptonic colliders, the Valencia (VLC) jet clustering algorithm is deemed more favourable \cite{Boronat:2014hva, Boronat:2016tgd}. The detector card \texttt{delphes\_card\_MuonColliderDet.tcl}, which is written partly inspired by the CLIC detector recommendations \cite{Leogrande:2019qbe}, contains VLC jet modules with different jet radius $R$, starting from 0.2 to 1.5. For our analysis, when dealing with $b$-tagged jets i.e. FS2, we use $R=0.2$, to obtain higher multiplicity of such jets. For FS3 with only light jets, we use $R=0.3$. 
	\item For the muon collider card, similar to the CLIC card, the $b$-tagging is performed with three working points- loose, medium, and tight- that correspond to 90\%, 70\%, and 50\% $b$-tagging efficiencies. We choose to work with the loose working point for maximum efficiency, selecting the $b$-tagged jets by demanding the value of the index \texttt{Jet.BTag} $\geq 4$ for each $b$-jet \cite{Leogrande:2019qbe}.
	\item A very primitive idea of the Forward muon detector is implemented in the muon collider detector card, where muons between $2.5 \leq \abs{\eta} \leq 8.0$ can be detected with a 95\% efficiency. We put an additional cut of $p \geq 100$ GeV for the Forward muons, in order to avoid any contamination from softer BIB muons. 
\end{itemize}

In order to evaluate the significance of the observed signal events at each final state, we use two different metrics, defined as follows:

\begin{itemize}
	\item $\eucal{S}=S/\sqrt{S+B}$ , which is a commonly used approximation for Poissonian statistics.
	\item Approximate median significance \cite{Cowan:2010js}:
	\begin{equation}
		{\rm AMS} = \sqrt{2\left((S+B+B_r)\times \ln(1+\frac{S}{B+B_r})-S\right)}
	\end{equation}
	This is known as the Asimov approximation, and is useful when the signal and background numbers are small but of comparable order. $B_r$ is a small regulation term that keeps the AMS value from diverging and stabilizes the calculation\cite{10.5555/2996850.2996852, Choudhury:2024crp}\footnote{The regulation term $B_r$, typically chosen between 1-10, reduces the variance of the AMS when $B_r$ is too small. Specifically, if $B\to 0$, the logarithm in the expression becomes extreme. The numerical stability is ensured by the $B_r$ term, keeping variance of AMS low in a small search region, as explained in ref. \cite{10.5555/2996850.2996852}.}. For low statistics, AMS is a better approximation, whereas for $S \ll B$, one has AMS$\to \eucal{S}$. 
\end{itemize}

Establishing the fundamentals of our analysis strategy, we now proceed to elaborate on each of the final states under consideration.

\subsection{FS1: 3 leptons + MET + 1 Forward Muon}

As discussed in \autoref{sec:fs}, this final state is unique to the asymmetric VBF production of triplet-like scalars at the MuC, with a large MET value being contributed by the $H^0\to SS$ decay. While symmetric pair production such as $H^+ H^- \mu^+ \mu^-$ or $H^0 H^0 \mu^+ \mu^-$ may contribute minimally due to non-identification of some leptons in the main detector, demanding exactly one Forward muon can suppress them to a negligible amount, and hence we do not include them in the analysis. Additionally, demanding a hadronically quiet three-lepton final state also renders SM backgrounds involving multijets, top quarks, and Higgs bosons to be suppressed. However, leptonic decays of SM vector bosons can mimic this final state, and hence we consider the $\mu^\pm \nu_\mu VV$ and $\mu^\pm \nu_\mu VVV$ backgrounds here, where $V$ stands for $Z$ or $W^\pm$. For both the signal and the backgrounds, we simulate only the leptonic decay events at the generation level in \mgFull, which yields fiducial cross-sections, which are used in further analysis. Additionally, we only consider events with the $H^0 \to SS$ decay in the fiducial cross-section, to correctly asses the kinematics of this final state without losing too many events.

We start with a traditional cut-based analysis (CBA) at the 3 TeV MuC, where the raw cross-sections of the production process of our choice are of $\mathcal{O}(1)$ fb across the three BPs, as discussed in \autoref{sec:cs}. The demand for purely leptonic final states lead to even smaller fiducial cross-sections. Especially for BP3, this fiducial cross-section is $\sim 4 \times 10^{-3}$ fb at the 3 TeV MuC, which means that with 1 \abi of luminosity, one can observe a maximum of 4 events, which is not very ideal. Hence, we omit BP3 from the analysis of FS1.

\begin{figure}[h]
	\centering
	\includegraphics[width=0.9\linewidth]{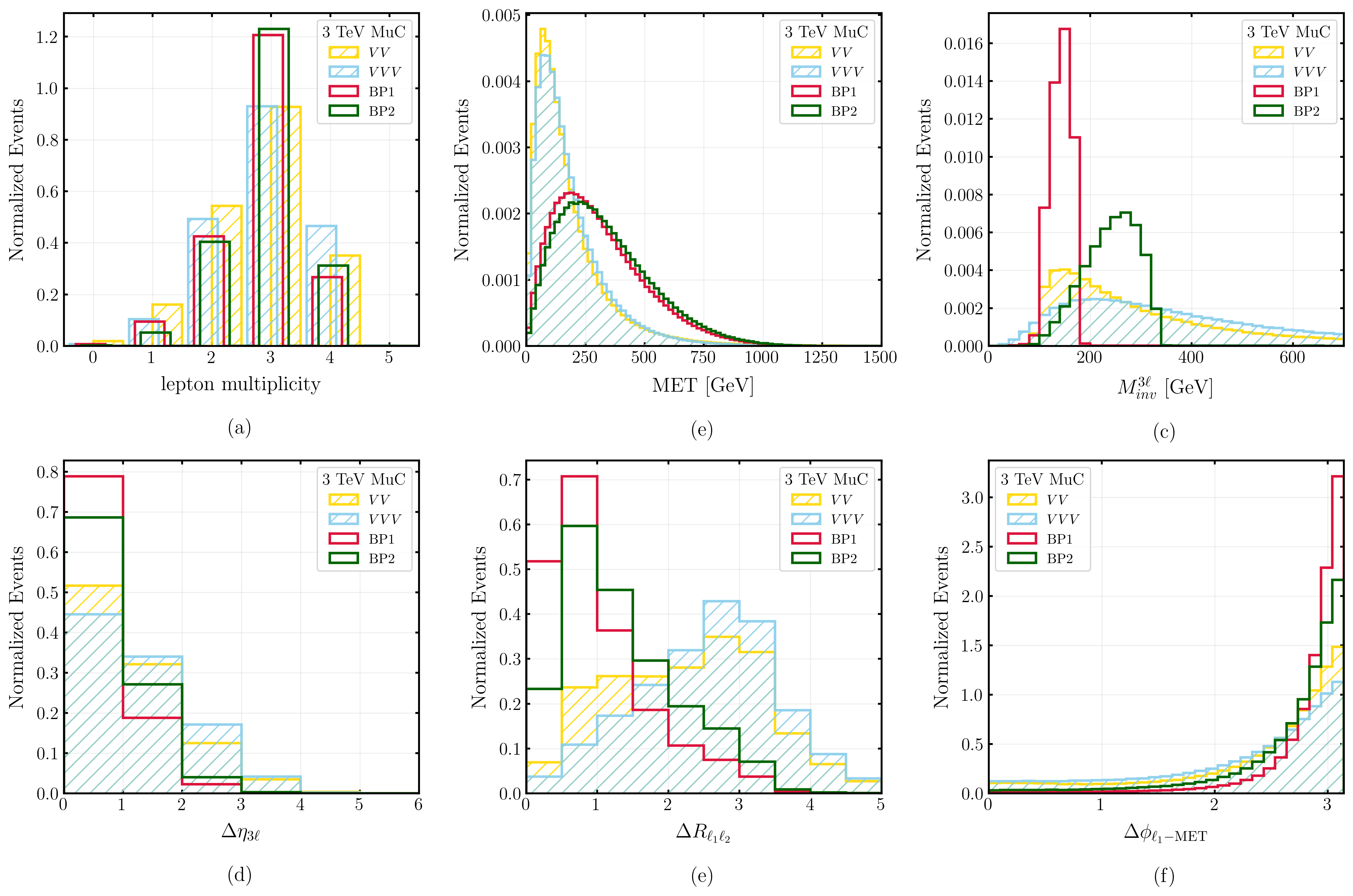}
	\caption{Distribution of discriminating kinematical variables for the $3\ell + {\rm MET} + 1\mu_F$ signal at a 3 TeV MuC. The backgrounds of $ZW$ and $ZZW$ are being taken with fully leptonic decays. The event histograms are normalized to one, for better visibility and comparison. The shaded histograms are for backgrounds, and the unshaded ones are for signals.}
	\label{fig:fs1kin}
\end{figure}

In \autoref{fig:fs1kin} we display the distributions of kinematical variables that can potentially be used in the cut-based analysis. In each plot, the yellow and light blue shaded histograms represent the $VV$ and $VVV$ backgrounds, while the unfilled red and green histograms stand for signal events from BP1 and BP2, respectively. The plots are normalized such that the area under each histogram is equal to one. The  \autoref{fig:fs1kin}(a) shows the multiplicity distribution of the charged leptons. While all the signals and backgrounds are generated with exactly three leptons in the final state, we see a maximum of four possible leptons, which can correspond to either the VBF spectator muon being within the main detector, or an energetic lepton from a radiated photon.  Another significant discriminator is the missing transverse energy, denoted as MET, which for the signal cases result dominantly from the singlet DM into which the $H^0$ is decaying. Evidently, from \autoref{fig:fs1kin}(e), the MET distributions for the signals peak around the corresponding $H^0$ masses, while the background distributions fall off for MET $\geq$ 200 GeV. Hence, even if one does not generate fiducial events with only the $H^0 \to SS$ decay, a cut of MET $\geq$ 200 GeV will essentially correspond to the fully invisible leg for the signal. \autoref{fig:fs1kin}(c) displays a crucial variable: the invariant mass of the three leptons. As the three leptons from the signal are produced from the resonance decay of the $H^+$, their invariant mass distribution does not cross the mass of the corresponding triplet scalar (200 GeV for BP1, 350 GeV for BP2). Whereas for the backgrounds, the non-resonant productions of the vector bosons allow the resulting leptons to have a wide distribution of the invariant mass. Taking advantage of the fact that the three leptons almost always emerge from the same leg i.e. the $H^+$ in the signal, we define the variable $\Delta\eta_{3\ell} = |\frac{\eta_{\ell_1} + \eta_{\ell_2}}{2}-  \eta_{\ell_3}|$, whose distribution is plotted in  \autoref{fig:fs1kin}(d). Expectedly, most of the signal events correspond to $\Delta\eta_{3\ell} \leq 1$, with a higher peak for BP1 due to more available boost resulting from the lower mass. The backgrounds show comparatively lower number of events in the same region. Another advantageous variable derived from the idea of having the three signal leptons close by is $\Delta R_{\ell_1 \ell_2}$, which is the separation of the two leading leptons in the $\eta-\phi$ plane. From \autoref{fig:fs1kin}(e), it is evident that the signal distribution of this separation peaks between 0.5 - 1.0, while the background peaks are observed around 2.5-3.0, signifying larger separation between the two leptons. \autoref{fig:fs1kin}(f) shows the distribution of the azimuthal angle separation between the leading lepton and the MET, denoted as $\Delta\phi_{\ell_1 - {\rm MET}}$. In case of the signal, the leading lepton and the MET directions are almost always back-to-back in the transverse plane, and hence for most of the signal events, their azimuthal separation is near $\pi$. The background distributions, in comparison, are flatter. Based on these kinematical observables, we can now now perform a traditional cut-based analysis (CBA) to optimize the signal significance. It is also important to know that, not all of these kinematical variables translate into optimal cuts, and hence we show the cut flow that yields the best possible signal significance. 

\begin{table}[h]
	\renewcommand{\arraystretch}{1.2}
	\centering
	\begin{tabular}{|c|c||c||c|c||c|c|}
		\hline 
	\multirow{2}{*}{BP}	& \multirow{2}{*}{Cut flow} & \multicolumn{3}{c||}{FS1 counts at 3 TeV MuC} &  \multicolumn{2}{c|}{{\makecell{Significance \\ at $ \int \mathcal{L} dt $ = 1 \abi}}} \\
		\cline{3-7}
		&& Signal& $VVV$ (BG) & $VV$ (BG) & $\eucal{S}$ & AMS \\
	    \hline\hline
		\multirow{5}{*}{BP1} & \makecell{$\sigma_{\rm fiducial} \times \int \mathcal{L} dt $  } &54.24&121.59&2168.36& -- &-- \\
		\cline{2-7}
		&S0:  $n_\ell =$ 3 + $n_{F\mu}$ = 1 &28.64&39.01&789.19&0.97$\sigma$&0.99$\sigma$\\
		\cline{2-7}
		&S1:  S0 +  MET $\geq$ 200 GeV &19.27&8.56&205.58&1.26$\sigma$&1.29$\sigma$\\
		\cline{2-7}
		&S2:  S1 +  $\Delta R_{\ell_1 \ell_2} \leq$ 2 &18.72&3.24&125.79&1.54$\sigma$&1.58$\sigma$\\
		\cline{2-7} 
		&S3: S2 + $M_{inv}^{3\ell} \leq$ 200 GeV &18.68&0.66&43.31&2.36$\sigma$&2.52$\sigma$\\
		\hline\hline
		\multirow{5}{*}{BP2} & \makecell{$\sigma_{\rm fiducial}  \times \int \mathcal{L} dt$ } &18.00&121.59&2168.36&--&--\\
		\cline{2-7}
		&S0:  $n_\ell =$ 3 + $n_{F\mu}$ = 1 &9.59&39.01&789.19&0.33$\sigma$&0.33$\sigma$\\
		\cline{2-7}
		&S1:  S0 +  MET $\geq$ 200 GeV &6.88&8.56&205.58&0.46$\sigma$&0.47$\sigma$\\
		\cline{2-7}
		&S2:  S1 + $\Delta R_{\ell_1 \ell_2} \leq$ 2   &6.26&3.24&125.8&0.54$\sigma$&0.54$\sigma$\\
		\cline{2-7} 
		&S3: S2 +  $M_{inv}^{3\ell} \leq$ 350 GeV &6.24&1.82&87.79&0.63$\sigma$&0.64$\sigma$\\
		\hline
	\end{tabular}
	\caption{Cut flow table of FS1 event counts for BP1 and BP2 at the 3 TeV MuC with 1 \abi luminosity, against the $VV$ and $VVV$ backgrounds. Signal significance is evaluated for events after each cut. }
	\label{tab:fs1_3tev}
\end{table}

\autoref{tab:fs1_3tev} shows the cut flow of event counts for the FS1, for BP1 and BP2 as well as the $VV$ and $VVV$ backgrounds, with the cuts over different kinematical variables imposed sequentially. The fiducial gen-level cross-sections for hadronically quiet three-lepton + invisible samples for the signals and the background processes are denoted as $\sigma_{\rm fiducial}$ in the same table. Right out of the box we notice that at the 3 TeV MuC with the target luminosity of 1 \abi, a maximum of $\sim$54 events can possibly be observed for BP1 in this final state, which is not very optimistic. Further cuts to reject the backgrounds unsurprisingly dilute this number. Here, S0 represents the pre-selection cut for the events, with three charged leptons each with $p_T \geq 20$ GeV in the main detector, as well as one Forward muon at the dedicated detectors. \autoref{fig:fs1kin}(a) clearly shows that despite always having three leptons at the gen-level, not all of them always pass the detector cuts. The $|\eta|$ coverage of 2.5 in the MuC detector sometimes fails to capture one or two of these leptons, more so in the  case of the backgrounds, where the production is more in the forward directions. The heavier mass of the triplet scalars mean that the leptons produced from them are relatively central, despite getting some leptons undetected. Nonetheless, with this pre-selection cut of S0, a mere 0.97$\sigma$ (0.99$\sigma$) significance values of $\eucal{S}$ (AMS) is shown by the BP1 signal. The next cut, S1, is imposed on the MET, with a demand of MET $\geq 200$ GeV to account for the purely dark decay of the $H^0$. This cut instantly filters out $75\% - 80\%$ of the backgrounds that pass the S0 cut, while only $\sim35\%$ of the signal suffers. The significance however does not seem much improvement, displaying $\eucal{S}$ (AMS) value of $1.26\sigma$ ($1.29\sigma$). This is followd by the S2 cut, applied on the separation between the two leading leptons, demanding $\Delta R_{\ell_1 \ell_2} \leq 2$ for optimum signal retention. A slight increase in the significance is observed here, with an $\eucal{S}$ (AMS) value of $1.54\sigma$ ($1.58\sigma$). Now, from the \autoref{fig:fs1kin}(c), we observe that the invariant mass of the three leptons is the most jarringly distinct feature for the signal. These three leptons must always come from the $H^\pm \to Z W^\pm \to 3\ell + \nu$ channel, and hence the edge of the invariant mass distribution must end at the $H^\pm$ resonance value, which is 200 GeV in this case. The vector bosons in the backgrounds are not produced on-shell, and hence the three-lepton invariant mass does not distinctively peak anywhere. The S3 cut of $M_{inv}^{3\ell} \leq$ 200 GeV thus reduce the backgrounds to $\sim 35\% (20\%)$ of the previous value of the $VV$ ($VVV$) case, while keeping the signal almost intact. This drastically enhances the significance to $2.36\sigma$ ($2.52\sigma$) in terms of $\eucal{S}$ (AMS).  For BP2, the much lesser fiducial cross-section means only a maximum of 18 events can be observed before the S0 cut. After the S3 cut (with an altered upper limit on the $M_{inv}^{3\ell} \leq$ 350 GeV), the $\eucal{S}$ (AMS) value for BP2 at the 3 TeV MuC with 1 \abi of luminosity comes out to be $0.63\sigma$ ($0.64\sigma$).

From the CBA, it is evident that, we need to look beyond this traditional approach, in order to enhance the distinction of the signal from the backgrounds. Thankfully, the recent advancements in the domain of multivariate analysis (MVA) using ML tools can come to our aid. Especially, we utilize the boosted decision tree (BDT) to create a classifier, which can be trained to distinguish between the signal and the background, using a set of low-level and high-level kinematical features. In the next subsection, we elaborate the analysis and results of this implementation.

%
%

\subsubsection{FS1 with a BDT}

In our attempt to enhance the signal significance for FS1 at a 3 TeV MuC with 1 \abi of integrated luminosity, we employ a BDT-based classifier using the \texttt{xgboost} \cite{10.1145/2939672.2939785} package (which we will refer to as XGB). Detailed explanations of the inner workings of BDTs, as well as different boosting algorithms including \texttt{xgboost}, can be found aplenty in contemporary literature such as ref. \cite{Cornell:2021gut, Choudhury:2024crp}, and hence we choose to omit these details from our work. For the first step of this analysis, after passing the pre-selection cut denoted as S0 in \autoref{tab:fs1_3tev}, we create feature dataframes for $\sim 5.5 \times 10^{5}$ signal events and $\sim 7 \times 10^{5}$ background events. For training the XGB classifier, we identify the following kinematical features:
\begin{itemize}
	\item Missing transverse energy (MET)
	\item $\eta-\phi$ plane distances of lepton pair combinations: $\Delta R_{\ell_1 \ell_2}, \Delta R_{\ell_1 \ell_3}, \Delta R_{\ell_2 \ell_3}$.
	\item $p_T$, $\eta$, and $\phi$ of the three leptons.
	\item $\eta$-separation between the lepton pair combinations, $\Delta \eta_{\ell_1 \ell_2}, \Delta \eta_{\ell_1 \ell_3}, \Delta \eta_{\ell_2 \ell_3}$.
	\item Average $\eta$ separation of the three leptons ($\Delta\eta_{3\ell} = \abs{\frac{\eta_{\ell_1} + \eta_{\ell_2}}{2} - \eta_{\ell_3}}$)
	\item Three-lepton invariant mass ($M_{inv}^{3\ell}$)
	\item $\eta$ of the Forward muons ($\eta_{F\mu}$)
	\item Azimuthal angle separation between the leading lepton and the MET, $\Delta\phi_{\ell_1{\rm MET}}$.
	\item Transverse mass of the leading lepton and the MET, $M_{T}^{\ell_1{\rm MET}}$. 
\end{itemize}

\begin{figure}[h]
	\centering
	\includegraphics[width=\linewidth]{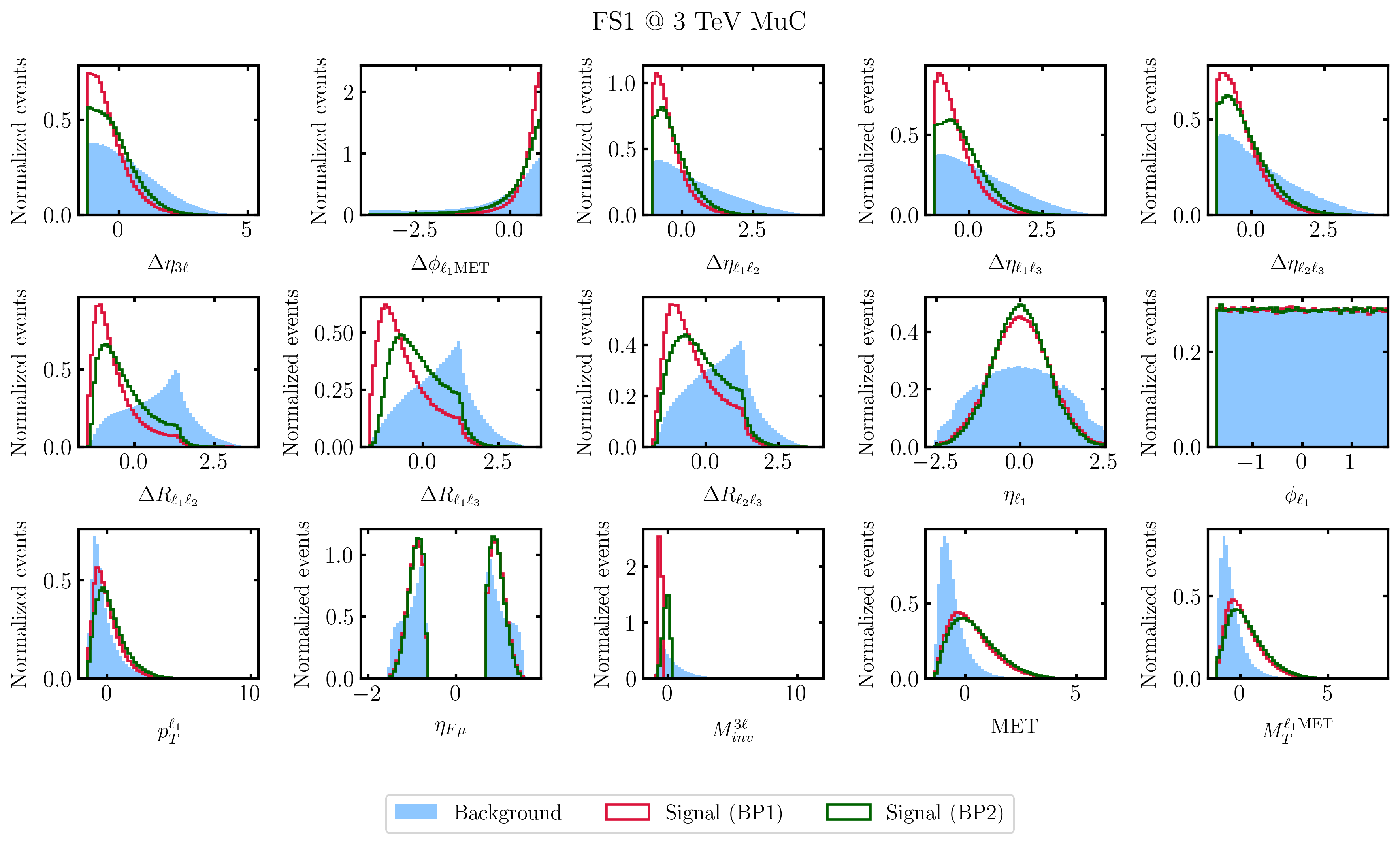}
	\caption{Feature distributions for the BP1 (red) and BP2 (green) signal and the combined $VV+VVV$ backgrounds (blue, filled), for \texttt{xgboost} training. The features are standardized to have mean =  0 and standard deviation = 1.}
	\label{fig:feat3tev}
\end{figure}

For unbiased training of an ML algorithm, it is important to scale the features such that all their distributions have mean = 0 and standard deviation = 1, also known as the standard scaling. In \autoref{fig:feat3tev}, we display the distributions of some of the standard-scaled features that are chosen to train the BDT on, for both the signal (red, unshaded) and background (blue, shaded). To train the XGB classifier, we split the combined signal and background dataframe into a training set with 60\% of the events, and the remaining 40\% into a test set. Half of the test set is used as a validation set to monitor overfitting during the classifier training. For best performance of the classifier training, we create the training set in such a way that it contains equal number of true signal and background events, which is known as balanced classes. It is imperative to have equal overall weights for the signal and background events in the training sample, so that the classifier can learn without bias towards one particular class. The two types of background channels are assigned weights according to their cross-section, while ensuring the sum of total weights are equal to those of the signal. The training set is then fitted with the XGB classifier with the following parameters:
\begin{verbatim}
	cls=xgb.XGBClassifier(base_score=0.5, booster='gbtree', colsample_bylevel=0.8,
	colsample_bynode=0.8, colsample_bytree=0.8, gamma=0.0,
	learning_rate=0.1, max_delta_step=0, max_depth=10, objective='binary:logistic',
	min_child_weight=5, n_estimators=1000, reg_alpha=1.0, reg_lambda=1.0, 
	scale_pos_weight=1, subsample=0.9, tree_method='exact')
\end{verbatim}
In the XGB classifier denoted as \texttt{cls}, the  \texttt{'n\_estimators'} parameter stands for the maximum number of decision trees, which is set to 1000 for better minimization of the loss function. Each tree corresponds to a round of 'boosting' the training, and for XGB the boosting method is known as gradient boosting (\texttt{'gbtree'}). The \texttt{'max\_depth'} parameter, set to 10, regulates the maximum depth of a tree. The \texttt{'learning\_rate'} and the \texttt{'n\_estimators'} parameters have a trade-off to keep the classifier from overfitting, and hence we keep the former as low as 0.1. The \texttt{'reg\_alpha'}, \texttt{'reg\_lambda'}, and \texttt{'gamma'} parameters also keep the model from overfitting, and the fractional sampling values of the \texttt{'colsample\_bylevel', 'colsample\_bynode', 'colsample\_bytree', 'subsample'} parameters keep a certain degree of randomness in the training, making the classifier better suited for a blind, general classification\footnote{Further details on each of the parameters can be found in \href{https://xgboost.readthedocs.io/en/latest/}{xgboost.readthedocs.io/en/latest/}}. While training, the model is validated at every boosting round by splitting half of the test set as a validation set, monitoring the minimization of the loss function. An early stopping parameter of 10 is employed to prevent overfitting, which terminates the boosting before reaching the \texttt{'n\_estimators'} value, if there is no improvement in the loss minimization over 10 consecutive rounds. In contemporary literature, \texttt{xgboost} has been found to be much more favourable over other BDT algorithms such as \texttt{AdaBoost} or \texttt{Random Forest}, due to its features such as gradient boosting, parallelization capabilities, enhanced speed of loss minimization, and multiple ways to regulate overtraining \cite{Jueid:2023fgo, Choudhury:2024crp}. 

\begin{figure}[h]
	\centering
		\hspace{-0.7cm}
		\subfigure[]{\includegraphics[width = 0.32\linewidth]{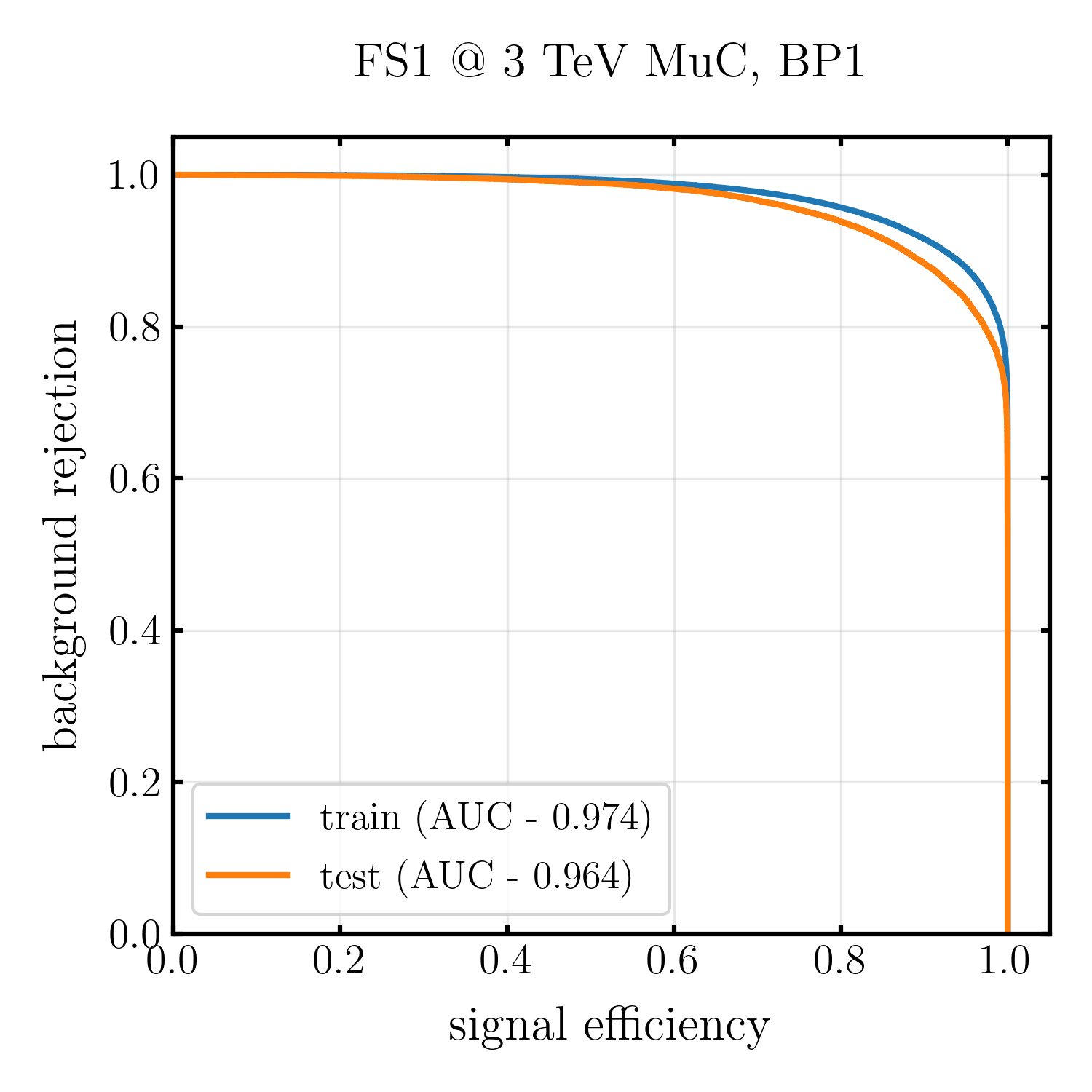}}
		\hspace{0.7cm}
		\subfigure[]{\includegraphics[width = 0.32\linewidth]{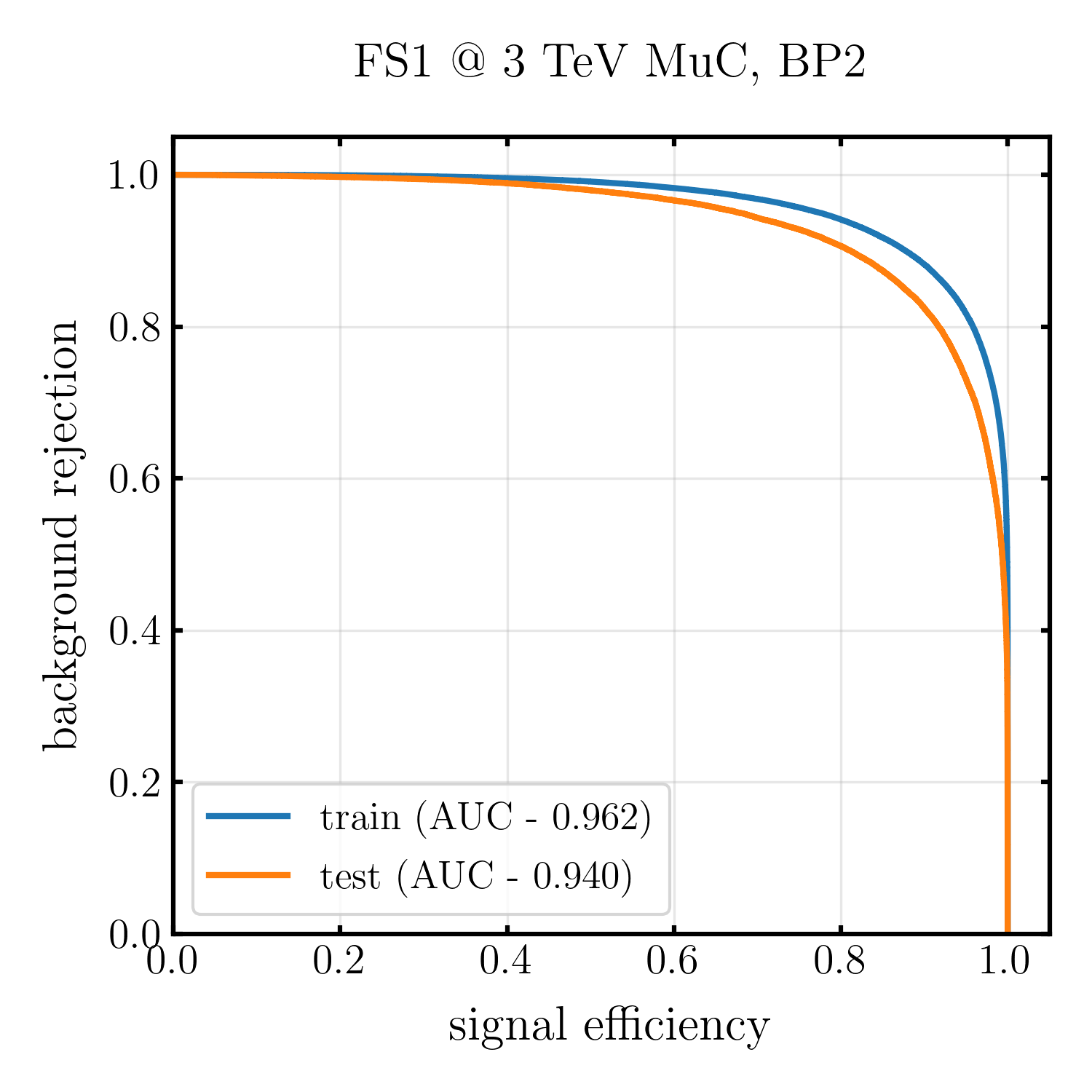}}
	\subfigure[]{\includegraphics[width = 0.37\linewidth]{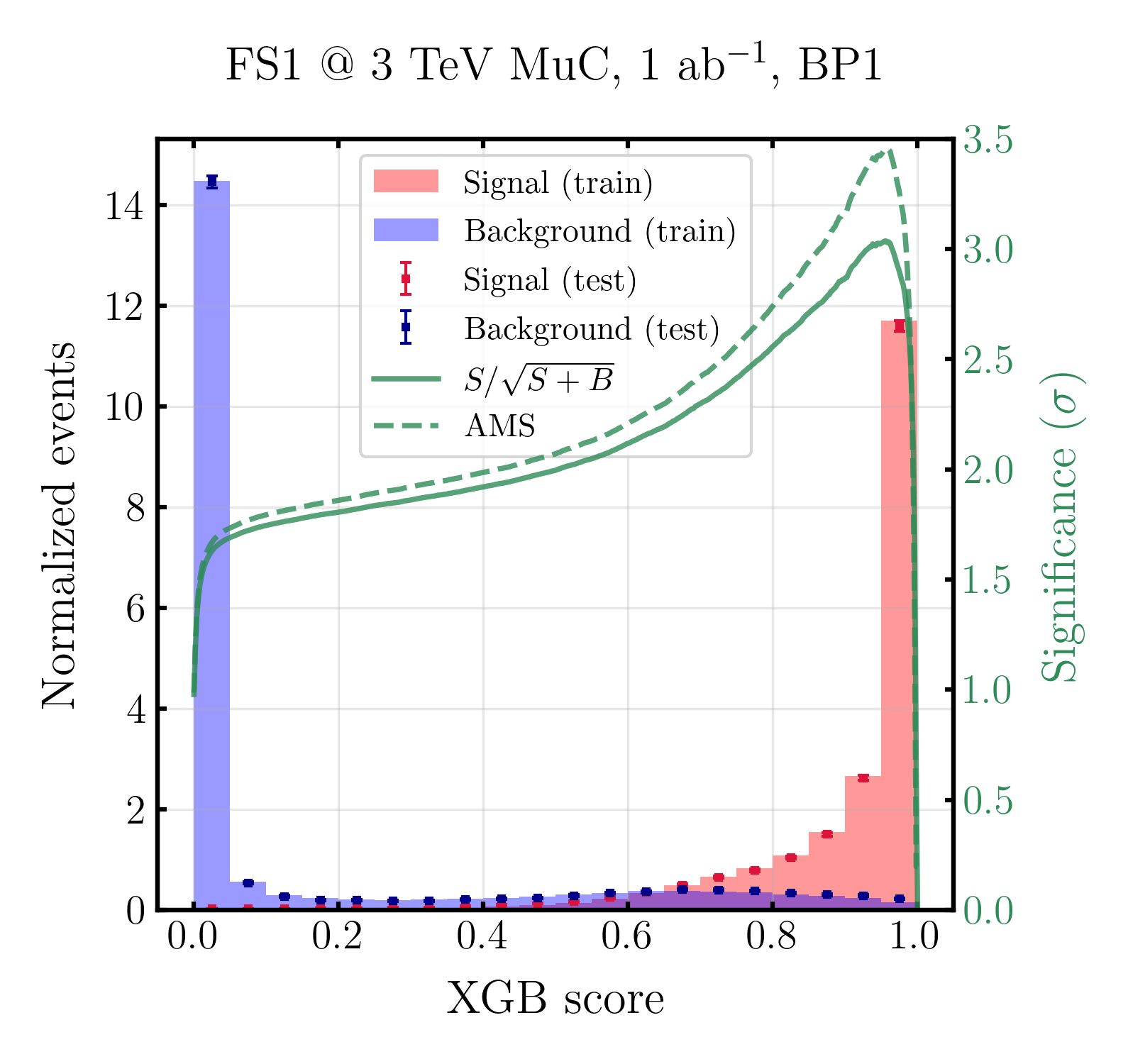}}
	\subfigure[]{\includegraphics[width = 0.37\linewidth]{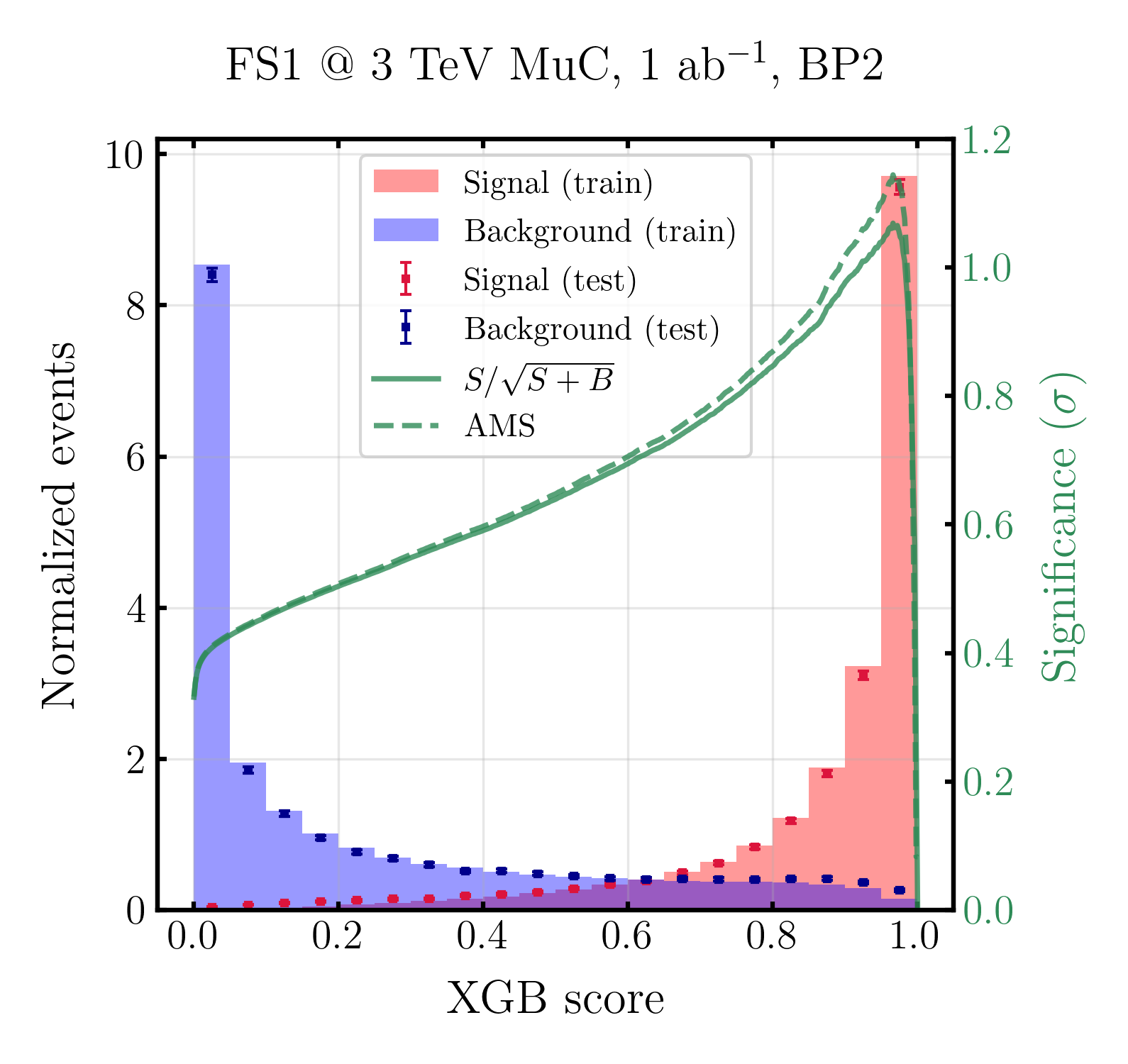}}
	\caption{Panels (a) and (b) show the ROC curves for BP1 and BP2 respectively, resulting from both the training data (blue) and the test set (orange), using features from FS1, with the AUC in brackets. Panels (c) and (d) show the signal (red) and background (blue) separation based on the XGB classifier score, for BP1 and BP2 respectively. The shaded histogram is for the training set, and the points with error bars are for the test set. The green curves represent the signal significance for the threshold cut on the XGB score. The y-axis on the right hand side with green ticks show the value of the significance in $\sigma$.}
	\label{fig:xgbres3tev_fs1}
\end{figure}

After the training of the XGB classifier, we now ask it to evaluate the probabilities of an event being signal-like or background-like, for both the train and test samples. Based on the predicted true positive rates (TPR) and false positive rates (FPR), we draw the receiver operating characteristic (ROC) curves shown in \autoref{fig:xgbres3tev_fs1}(a) and (b) for BP1 and BP2, respectively. The ROC curve is a measurement of the performance of the classifier, with the signal efficiency (TPR) on the x-axis, and the background rejection (1 - FPR) in the y-axis. The area under the curve (AUC) determines how efficiently the signal and background are separated. For our XGB classifier, in case of BP1, the train and test AUC comes out to be 0.974 and 0.964 respectively, which shows two things: great performance of the classifier, and little overfitting. For BP2, the AUC for both train and test sets drop to 0.962 and 0.940 respectively. The classifier score, or XGB score as denoted here, are probability thresholds set by the classifier to determine the number of true and false positives, in a range of 0 to 1. In \autoref{fig:xgbres3tev_fs1}(c) and (d) we show the signal and background event distributions as a function of the XGB score, for BP1 and BP2 respectively. Here, the shaded histograms represent the training set, and the points with the error bars denote the results from the test set. The red colours represent true signal events, while the true background events are shown in blue. For each value of the threshold, one can calculate the number of true signals and true background events above it, and subsequently calculate the signal significance with 1 \abi of luminosity using both the $\eucal{S} = S/\sqrt{S+B}$ method and the AMS method. The green curves in the \autoref{fig:xgbres3tev_fs1}(c) and (d) show the value of the signal significance for each of the XGB score thresholds, with the secondary y-axis in the right hand side denoting the value of the significance in units of $\sigma$, in green colour as well. The solid lines stand for the $S/\sqrt{S+B}$ values, while the dashed lines represent the AMS values. For BP1, we see that for a threshold of XGB score $\geq 0.956$, the maximum achievable signal significance is 3.04$\sigma$ (3.45$\sigma$) with $\eucal{S}$ (AMS), which is much better compared to the corresponding CBA outcomes of 2.36$\sigma$(2.52$\sigma$) from \autoref{tab:fs1_3tev}. While reaching a 5$\sigma$ discovery significance is still not possible due to the low fiducial cross-sections, a $>3\sigma$ significance can always hint at the existence of new physics, which we thrive for. For BP2 however, the small number of events mean that a maximum of 1.07$\sigma$ (1.14$\sigma$) significance can be obtained with XGB score $\geq 0.966$, using the $\eucal{S}$ (AMS) method. The cut flow for the XGB analysis is shown in \autoref{tab:fs1_3tev_bdt}. 

\begin{table}[h]
	\renewcommand{\arraystretch}{1.2}
	\centering
	\begin{tabular}{|c|c||c||c|c||c|c|}
		\hline 
		\multirow{2}{*}{BP}	& \multirow{2}{*}{Cut flow} & \multicolumn{3}{c||}{FS1 counts at 3 TeV MuC with BDT} & \multicolumn{2}{c|}{{\makecell{Significance \\ at $ \int \mathcal{L} dt $ = 1 \abi}}} \\
		\cline{3-7}
		&& Signal& $VVV$ (BG) & $VV$ (BG) &  $\eucal{S}$ & AMS \\
		\hline\hline
		\multirow{3}{*}{BP1} & \makecell{$\sigma_{\rm fiducial}$ (fb) } &54.24&121.59&2168.36 & --&--\\
		\cline{2-7}
		&S0:  $n_\ell =$ 3 + $n_{F\mu}$ = 1 &28.64&39.01&789.19&0.97$\sigma$&0.99$\sigma$\\
		\cline{2-7}
		&S5:  S0 +  XGB score $\geq 0.956$ &15.88&0.13&11.33&3.04$\sigma$&3.45$\sigma$\\
		\hline\hline
		\multirow{3}{*}{BP2} & \makecell{$\sigma_{\rm fiducial}$ (fb) } &18.00&121.59&2168.36& --&--\\
		\cline{2-7}
		&S0:  $n_\ell =$ 3 + $n_{F\mu}$ = 1 &9.59&39.01&789.19&0.33$\sigma$&0.33$\sigma$\\
		\cline{2-7}
		&S5:  S0 +  XGB score $\geq 0.966$ &3.77&0.14&8.55&1.07$\sigma$&1.14$\sigma$\\
		\hline
	\end{tabular}
	\caption{Cut flow table for FS1 event counts at the 3 TeV MuC with 1 \abi luminosity, for BP1 and BP2 against the $VV$ and $VVV$ backgrounds, with the XGB score threshold cut. Signal significance is evaluated for events after each cut.}
	\label{tab:fs1_3tev_bdt}
\end{table}

%
%
%

\subsection{FS2: 2 $b$-jets + 1 lepton + MET + 1 Forward Muon}

As mentioned in \autoref{sec:fs}, while this final state promises enhanced number of events due to higher decay probabilities and subsequent fiducial cross-sections, it is not essentially a direct probe of the custodial symmetry violating nature of the model. Nonetheless, FS1 and FS2 can be used in tandem to obtain an enhanced probe of this scenario. Here, the presence of two $b$-jets and exactly one Forward muon means that the dominant backgrounds are from the $\mu^+ \nu_\mu Z W^-$, $\mu^+ \nu_\mu h W^-$, and $\mu^+ \nu_\mu \bar{t}b$ VBF channels (and their conjugate processes), possessing large fiducial cross-sections. Processes like $\mu^+ \nu_\mu Z W^-$ can also contribute, possibly with larger MET values, but the fidual cross-sections are low, and hence can be neglected from the analysis. We generate the samples for signals in all three BPs, as well as these three backgrounds, with two $b$-quarks, one charged lepton, and missing energy, demanding $H^0\to SS$ decay as in the case of FS2.  

\begin{figure}[h]
	\centering
	\includegraphics[width=0.9\linewidth]{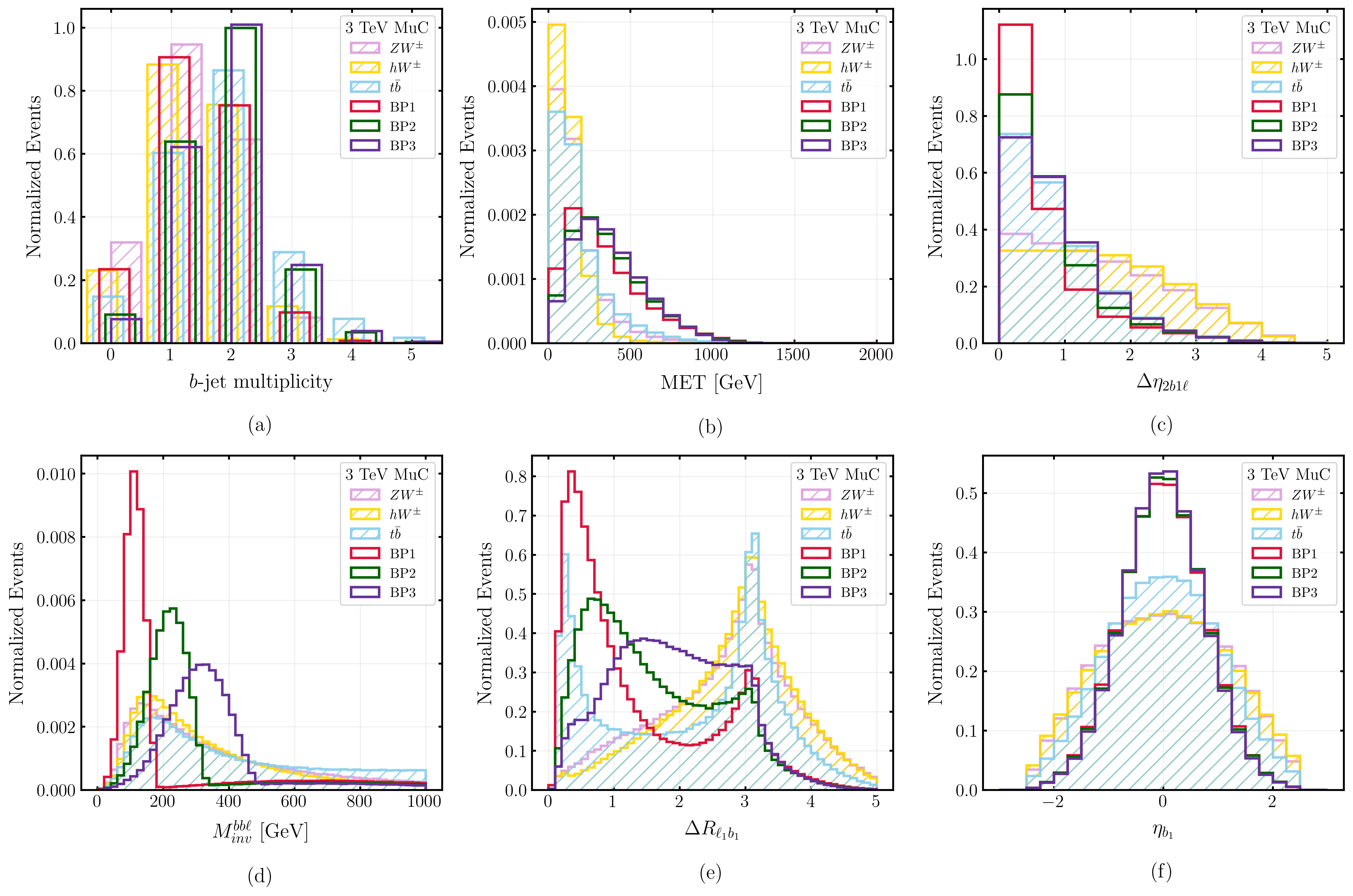}
	\caption{Distribution of discriminating kinematical variables for the $2 $b$-{\rm jet} + 3\ell + {\rm MET} + 1\mu_F$ signal at a 3 TeV MuC. The backgrounds of $ZW^\pm$, $hW^\pm$, and $t\bar{b}$ are being taken with the specific final state at gen-level. The event histograms are normalized to one, for better visibility and comparison. The shaded histograms are for backgrounds, and the unshaded ones are for signals.}
	\label{fig:fs2kin}
\end{figure}

\autoref{fig:fs2kin} displays the kinematical variables that can potentially distinguish the signal from the backgrounds, in a cut-based analysis. Starting with the $b$-jet multiplicity, \autoref{fig:fs2kin} shows that despite generating the events with exactly two $b$-quarks in the final state, both the signal and the backgrounds show a multiplicity distribution that does not always peak at 2 $b$-jets. especially for BP1, the multiplicity peaks at 1 due to the lower mass of $H^\pm$ allowing highly collimated and/or forward $b$-jets, making them difficult to be reconstructed as two separate ones. The same behaviour is observed for $ZW^\pm$ and $hW^\pm$ backgrounds. For BP2 and BP3 as well as the $t\bar{b}$ backgrounds, the relatively heavier masses and more central events allow the multiplicity to peak at 2. A few events are also observed with multiplicities of 3-4, due to mistagging, as well as for keeping a small jet radius of 0.2. The MET distributions in \autoref{fig:fs2kin}(b) follow almost the same pattern as FS1, with $\geq 200$ GeV peaks for the signal cases. From \autoref{fig:fs2kin}(c), the pseudorapidity separation between the two $b$-jets and the lepton, defined as $\Delta\eta_{2b1\ell}$, peaks within $\leq 0.5$ for all three signals, with the peak height decreasing as we go up in mass, due to less collimation between the three objects. While $ZW^\pm$ and $hW^\pm$ backgrounds have flat distributions, the $t\bar{b}$ shows almost the same trend as BP3. Now, similar to the three-lepton invariant mass in FS1, \autoref{fig:fs2kin}(d) shows similar behaviour for the invariant mass distribution of the two $b$-jets and the leptons (defined as $M_{inv}^{bb\ell}$), with the signal distributions having sharp edges at their respective benchmark charged Higgs boson masses. However, compared to the purely leptonic case, there is a more visible tail in each signal distribution, due to the jet resolution being less precise than leptons. An interesting variable is the $\Delta R_{\ell_1 b_1}$, the separation between the leading $b$-jet and the lepton in the $\eta-\phi$ plane, as shown in \autoref{fig:fs2kin}(e). For BP1 and BP2, while for most cases this separation peaks at $\leq 1$, a small secondary peak is observed around $\Delta R_{\ell_1 b_1} \sim 3$, corresponding to $H^+ \to t\bar{b}$ decays where the immediate $b$-jet from the $H^+$ becomes the leading jet. This branching ratio is smaller for BP3 and hence the behaviour is suppressed, however, no sharp peak is observed due to the higher mass of $H^+$ allowing less collimation. The $t\bar{b}$ background shows similar behaviour as BP1, with both the small- and large-value peaks having comparable heights. In contrast, the $ZW^\pm$ and $hW^\pm$ backgrounds peak at large values only. Putting optimal cuts on these variables, we can proceed towards the CBA as before.

\begin{table}[h]
	\renewcommand{\arraystretch}{1.2}
	\centering
	\begin{tabular}{|c|c||c||c|c|c||c|c|}
		\hline 
		\multirow{2}{*}{BP}	& \multirow{2}{*}{Cut flow} & \multicolumn{4}{c||}{FS2 counts at 3 TeV MuC} &   \multicolumn{2}{c|}{{\makecell{Significance \\ at $ \int \mathcal{L} dt $ = 1 \abi}}} \\
		\cline{3-8}
		&& Signal& \makecell{$ZW$ \\ (BG)} & \makecell{$hW$ \\ (BG)} &  \makecell{$t\bar{b}$ \\ (BG)} & $\eucal{S}$ & AMS\\
		\hline\hline
		\multirow{6}{*}{BP1} & \makecell{$\sigma_{\rm fiducial} \times \int \mathcal{L} dt $  } &488.04&4681.01&4091.50&2758.05& -- & --\\
		\cline{2-8}
		&S0:  $n_b = 2 + n_\ell =$ 1 + $n_{F\mu}$ = 1 &159.47&1243.73&1204.29&1272.13&2.56$\sigma$&2.59$\sigma$\\
		\cline{2-8}
		&S1:  S0 + MET $\geq$ 200 GeV &122.21&502.19&290.84&522.50&3.22$\sigma$&3.31$\sigma$\\
		\cline{2-8}
		&S2:  S1 + $\Delta R_{\ell_1 b_1} \leq$ 2.5  &98.75&229.59&115.64&283.84&3.66$\sigma$&3.86$\sigma$\\
		\cline{2-8} 
		&S3: S2 + $\Delta\eta_{2b1\ell} \leq 1$&90.32&138.24&63.94&227.02&3.96$\sigma$&4.20$\sigma$\\
		\cline{2-8}
		&S4:  S3 + $M_{inv}^{bb\ell} \leq$ 200 GeV&84.87&82.10&42.39&49.81&5.27$\sigma$&5.92$\sigma$\\
		\hline\hline
		\multirow{6}{*}{BP2} & \makecell{$\sigma_{\rm fiducial}  \times \int \mathcal{L} dt$ } &183.37&4681.01&4091.50&2758.05&--&--\\
		\cline{2-8}
		&S0:  $n_b = 2 + n_\ell =$ 1 + $n_{F\mu}$ = 1 &91.41&1243.73&1204.29&1272.13&1.48$\sigma$&1.49$\sigma$\\
		\cline{2-8}
		&S1:  S0 + MET $\geq$ 200 GeV &73.55&502.19&290.84&522.50&1.97$\sigma$&2.01$\sigma$\\
		\cline{2-8}
		&S2:  S1 + $\Delta R_{\ell_1 b_1} \leq$ 2.5  &60.14&229.59&115.64&283.84&2.29$\sigma$&2.35$\sigma$\\
		\cline{2-8} 
		&S3: S2 + $\Delta\eta_{2b1\ell} \leq 1$&49.74&138.24&63.94&227.02&2.27$\sigma$&2.34$\sigma$\\
		\cline{2-8}
		&S4:  S3 +  $M_{inv}^{bb\ell} \leq$ 350 GeV &46.83&117.76&58.14&89.69&2.65$\sigma$&2.77$\sigma$\\
		\hline\hline
		\multirow{6}{*}{BP3} & \makecell{$\sigma_{\rm fiducial}  \times \int \mathcal{L} dt$ } &28.51&4681.01&4091.50&2758.05&--&--\\
		\cline{2-8}
		&S0:  $n_b = 2 + n_\ell =$ 1 + $n_{F\mu}$ = 1  &14.71&1243.73&1204.29&1272.13&0.24$\sigma$&0.24$\sigma$\\
		\cline{2-8}
		&S1:  S0 + MET $\geq$ 200 GeV &11.98&502.19&290.84&522.50&0.33$\sigma$&0.33$\sigma$\\
		\cline{2-8}
		&S2:  S1 + $\Delta R_{\ell_1 b_1} \leq$ 2.5  &9.32&229.59&115.64&283.84&0.37$\sigma$&0.37$\sigma$\\
		\cline{2-8} 
		&S3: S2 + $\Delta\eta_{2b1\ell} \leq 1$&6.84&138.24&63.94&227.02&0.33$\sigma$&0.33$\sigma$\\
		\cline{2-8}
		&S4:  S3 +   $M_{inv}^{bb\ell} \leq$ 500 GeV &6.52&129.56&61.94&114.42&0.37$\sigma$&0.37$\sigma$\\
		\hline
	\end{tabular}
	\caption{Cut flow table of FS2 event counts for BP1-BP3 at the 3 TeV MuC with 1 \abi luminosity, against the backgrounds. Signal significance is evaluated for events after each cut.}
	\label{tab:fs2_3tev}
\end{table}

\autoref{tab:fs2_3tev} shows the cut flow for the FS2 analysis at a 3 TeV MuC, with event numbers evaluated at a luminosity of 1 \abi. The primary selection cut, S0, comprises of exactly two $b$-jets and one charged lepton each with $p_T \geq 20$ GeV, along with exactly one Forward muon. The next cut, S1, is similar to FS1, demanding MET $\geq 200$ GeV, which successfully filters out more than half the background. The S2 and S3 cuts are applied in order to obtain events with lower $\Delta R_{\ell_1 b_1}$ and $\eta_{2b1\ell}$, to help filter out more background events while keeping the signal counts respectably consistent. The S4 cut differs for each BP, demanding the $M_{inv}^{bb\ell}$ to be less than the benchmark masses. For all three BPs, this cut proves to be the most significant, similar to the FS1 case with the three-lepton invariant mass.  With the dominant contributions from the $H^+ \to t\bar{b}$ mode, as well as the $H^+ \to ZW^+$ mode, BP1 here has enough events to reach a maximum of 5.27$\sigma$ (5.92$\sigma$) value of $\eucal{S}$ (AMS) after the S4 cut. For BP2, the same number comes out to be 2.65$\sigma$ (2.77$\sigma$), which is still a better performance compared to FS1. BP3 however, with low fiducial cross-section to begin with, barely reaches a 0.4$\sigma$ significance after the S4 cut. To enhance the sensitivity for the model in this final state, we can apply another XGB classifier, with a new set of features.

\subsubsection{FS2 with a BDT}

While \autoref{tab:fs2_3tev} shows that for BP1, a 5$\sigma$ significance can already be achieved with the cut-based analysis, we wish to still demonstrate how the sensitivity can be benefited with the inclusion of an XGB classifier, especially for BP2 and BP3 with lower statistics. We identify the following features to train the classifier.

\begin{itemize}
	\item Missing transverse energy (MET). 
	\item $\eta-\phi$ plane separation between $b$-jet and lepton combinations: $\Delta R_{b_1\ell}$,  $\Delta R_{b_2\ell}$.
	\item $\eta$-separation between $b$-jet and lepton combinations: $\Delta \eta_{b_1\ell}$,  $\Delta \eta_{b_2\ell}$.
	\item $\eta$-separation between the two $b$-jets and the lepton, defined as $\Delta\eta_{2b\ell} = \abs{\frac{\eta_{b_1} + \eta_{b_2}}{2} - \eta_{\ell_1}}$.
	\item Azimuthal angle separation between each of the the $b$-jets and the lepton, and the MET direction: $\Delta\phi_{b_1{\rm MET}}$, $\Delta\phi_{b_2{\rm MET}}$, $\Delta\phi_{\ell{\rm MET}}$.
	\item  $p_T$, $\eta$, $\phi$ of $b$-jets and leptons.
	\item Di-$b$-jet invariant mass: $M_{inv}^{bb}$.
	\item Invariant mass of the two $b$-jets and the lepton: $M_{inv}^{bb\ell}$.
	\item Transverse mass of the MET with each of the two $b$-jets and the lepton: $M_T^{b_1{\rm MET}}$, $M_T^{b_1{\rm MET}}$, $M_T^{\ell{\rm MET}}$.

\end{itemize}

\begin{figure}[h]
	\centering
	\includegraphics[width=\linewidth]{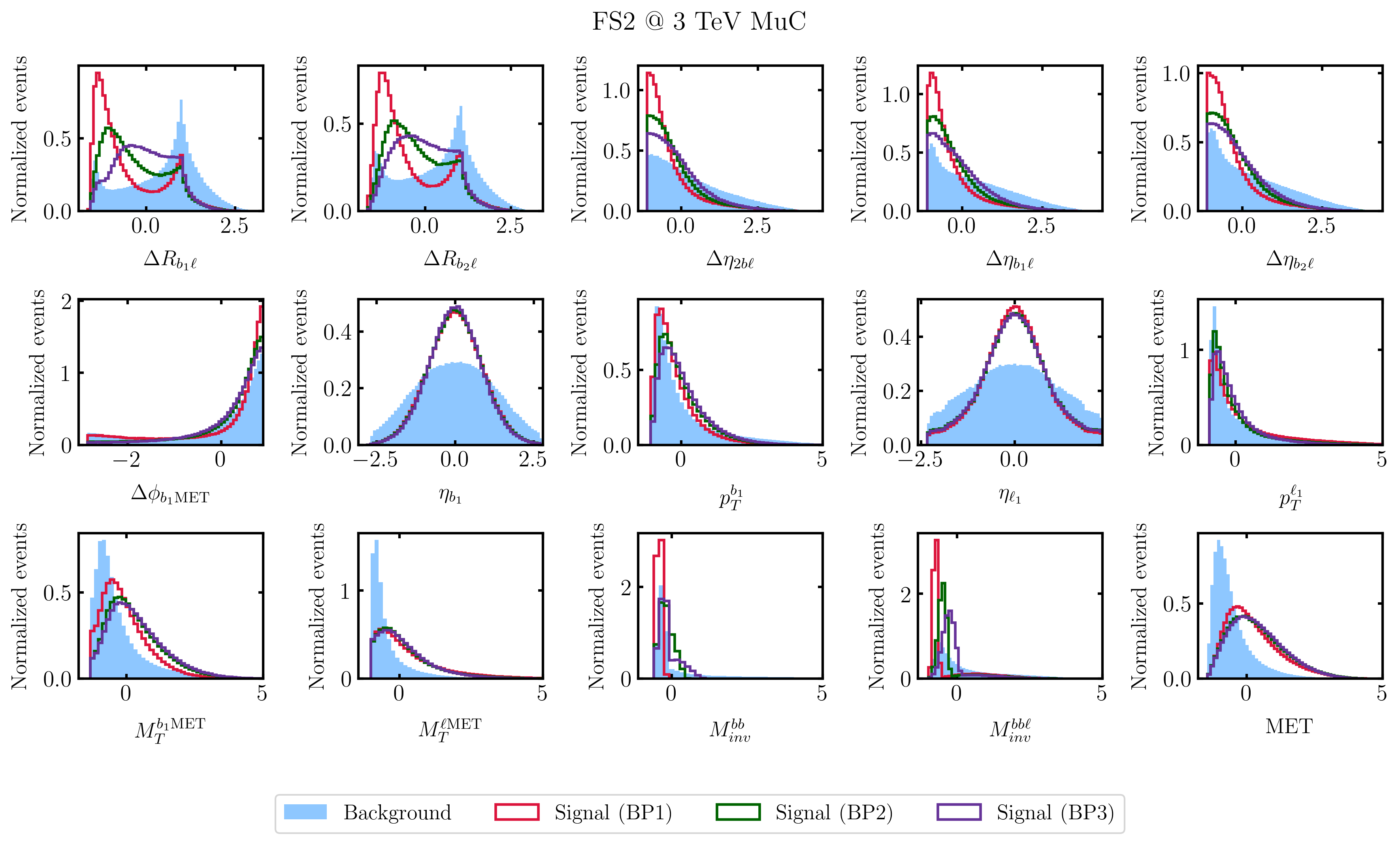}
	\caption{Feature distributions for the three signals and the combined $ZW+hW+t\bar{b}$ backgrounds, for \texttt{xgboost} training with FS2. The features are standardized to have mean =  0 and standard deviation = 1.}
	\label{fig:feat3tev_fs2}
\end{figure}

The features are again scaled to have mean = 0 and standard deviation = 1, for unbiased treatment by the classifier. In \autoref{fig:feat3tev_fs2} we display the distribution of the scaled features for the combined backgrounds (blue, filled histogram), along with the three BPs (red, green, and purple solid histograms), to showcase their separations.

After a similar train-test-validation splitting as FS1, we train the XGB classifier with the following parameters:

\begin{verbatim}
	cls=xgb.XGBClassifier(base_score=0.5, booster='gbtree', colsample_bylevel=0.8,
	colsample_bynode=0.8, colsample_bytree=0.8, gamma=0.0,
	learning_rate=0.08, max_delta_step=0, max_depth=10, objective='binary:logistic',
	min_child_weight=10, n_estimators=1000, reg_alpha=3.0, reg_lambda=2.0, 
	scale_pos_weight=1, subsample=0.9, tree_method='exact')
\end{verbatim}




\begin{figure}[h]
	\centering
	\subfigure[]{\includegraphics[width=0.31\linewidth]{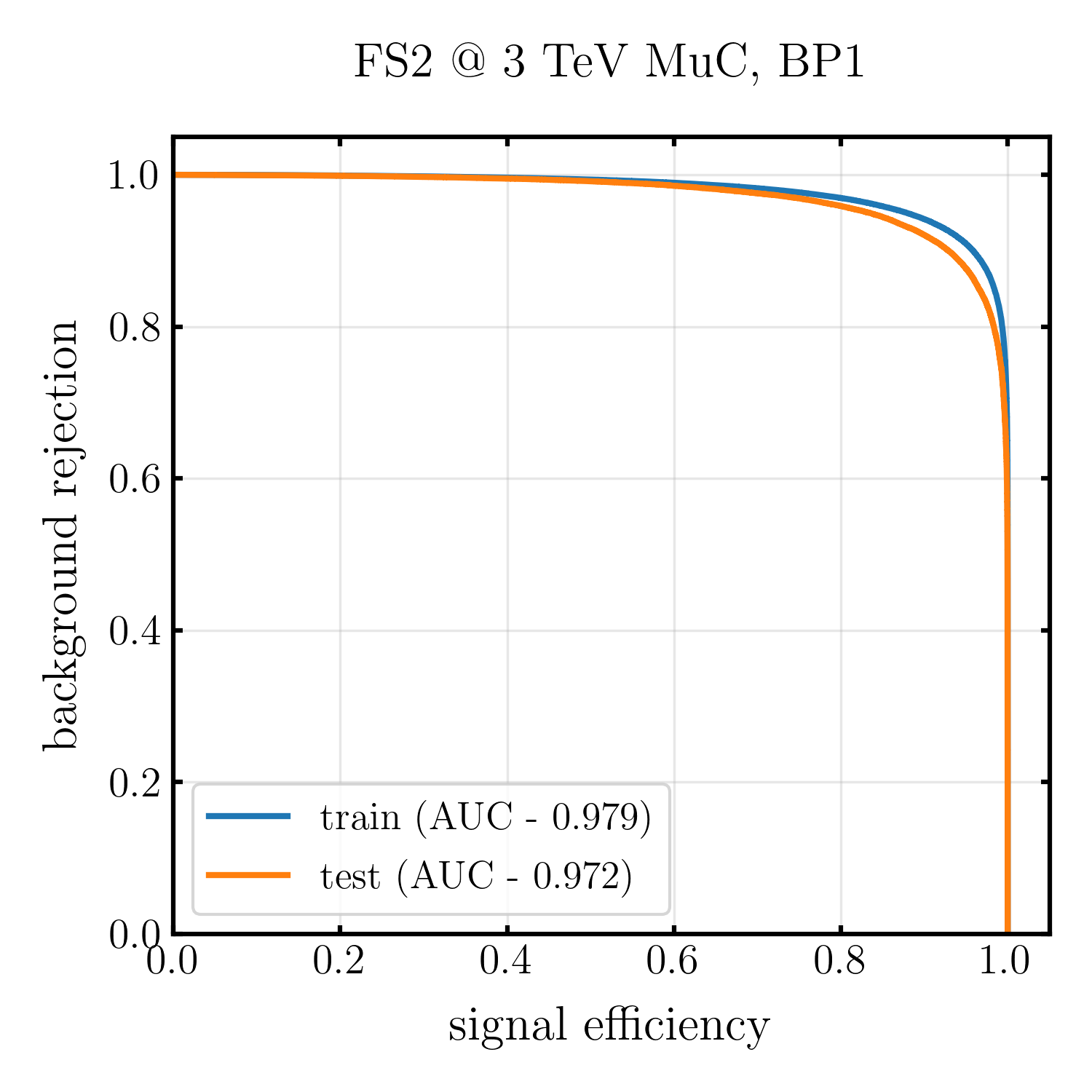}}
	\subfigure[]{\includegraphics[width=0.31\linewidth]{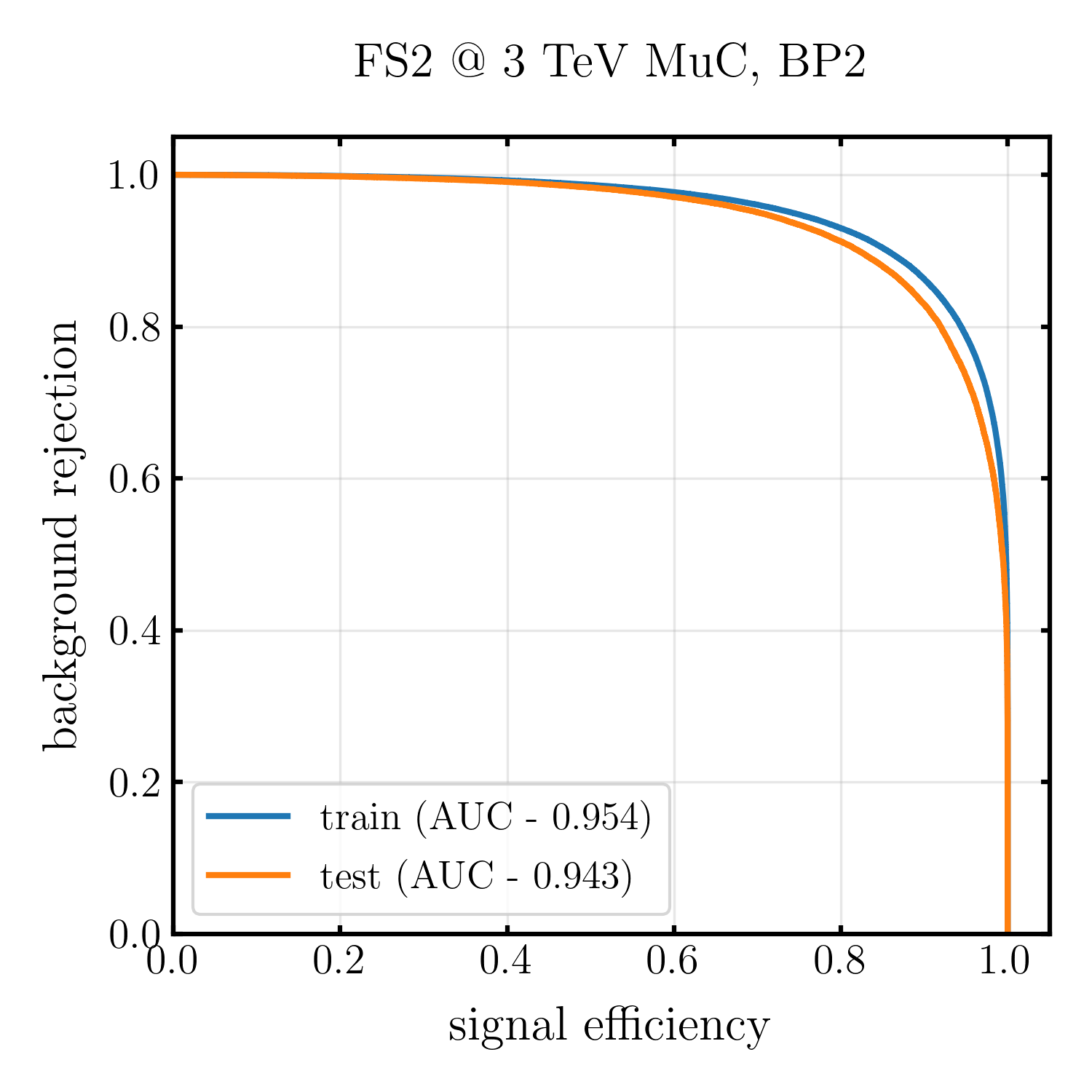}}
	\subfigure[]{\includegraphics[width=0.31\linewidth]{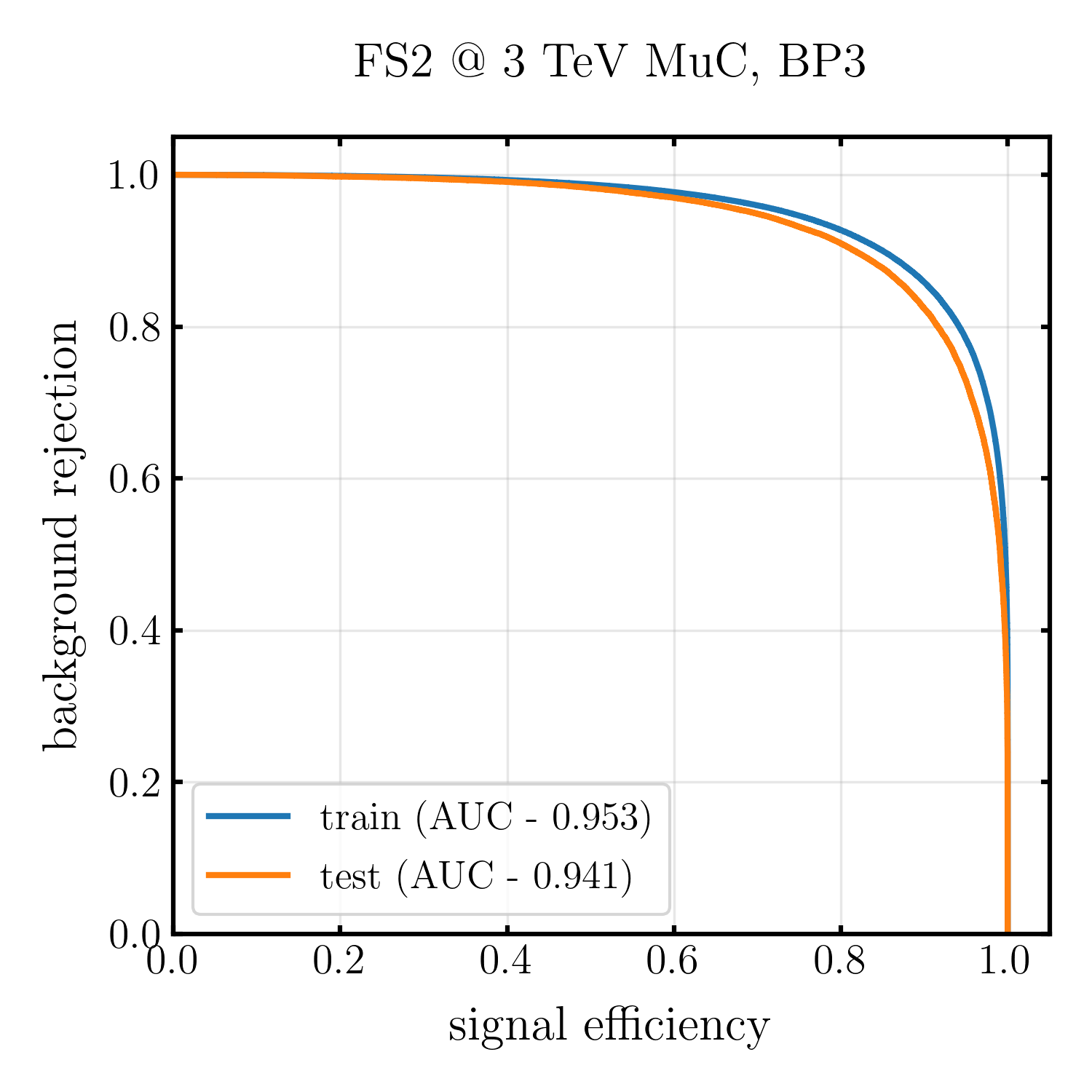}}
	\subfigure[]{\includegraphics[width=0.32\linewidth]{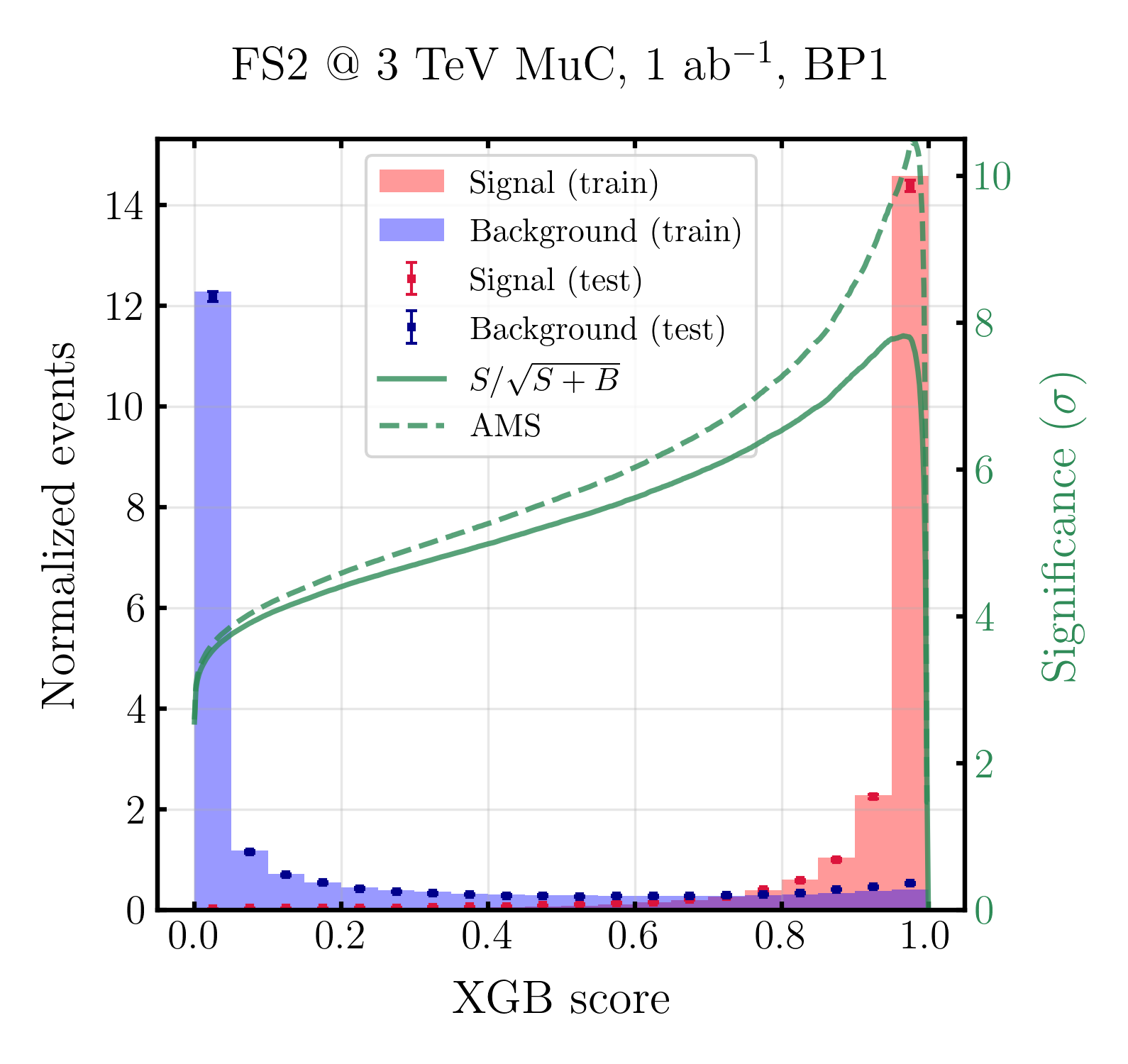}}
	\subfigure[]{\includegraphics[width=0.32\linewidth]{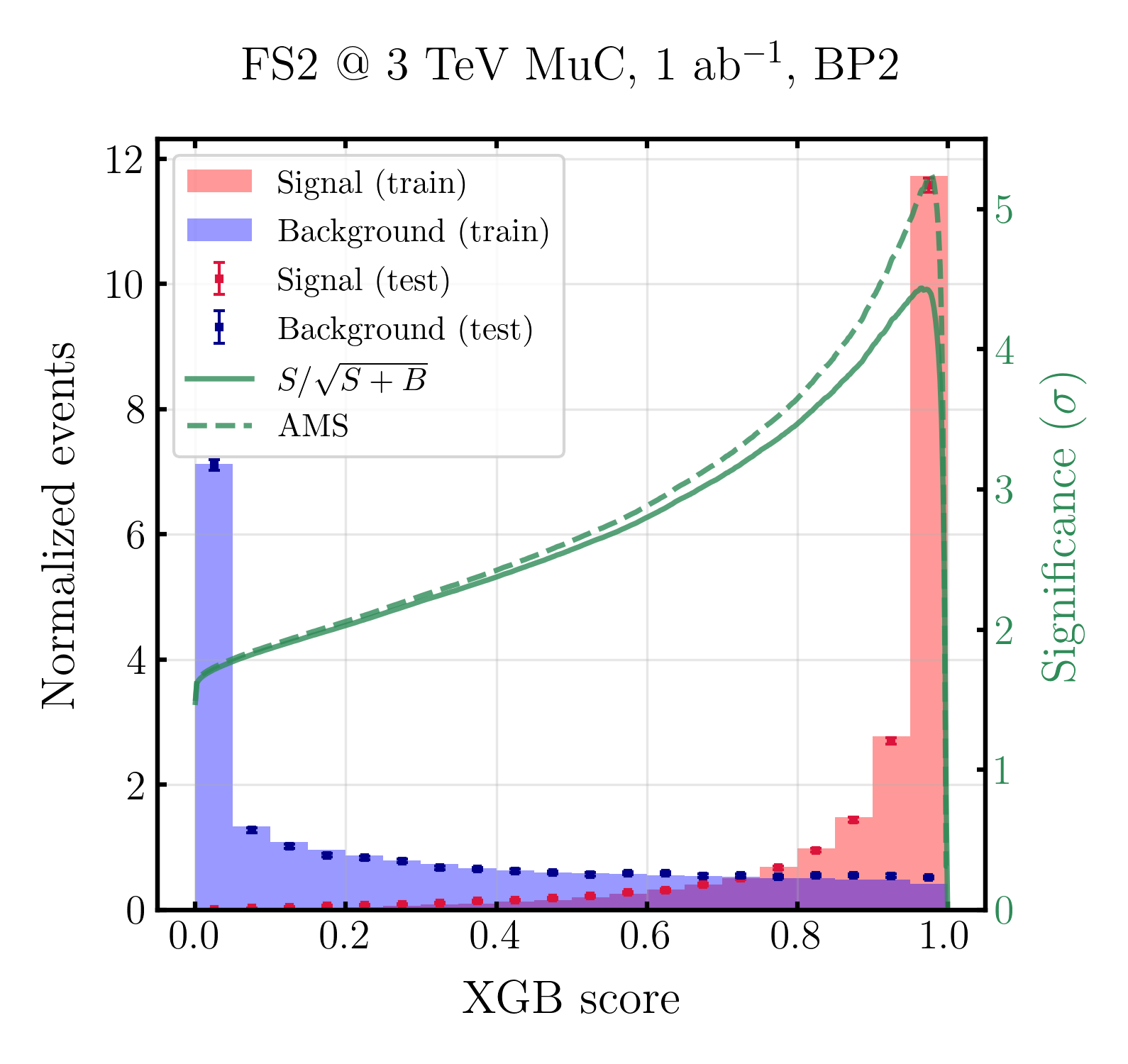}}
	\subfigure[]{\includegraphics[width=0.32\linewidth]{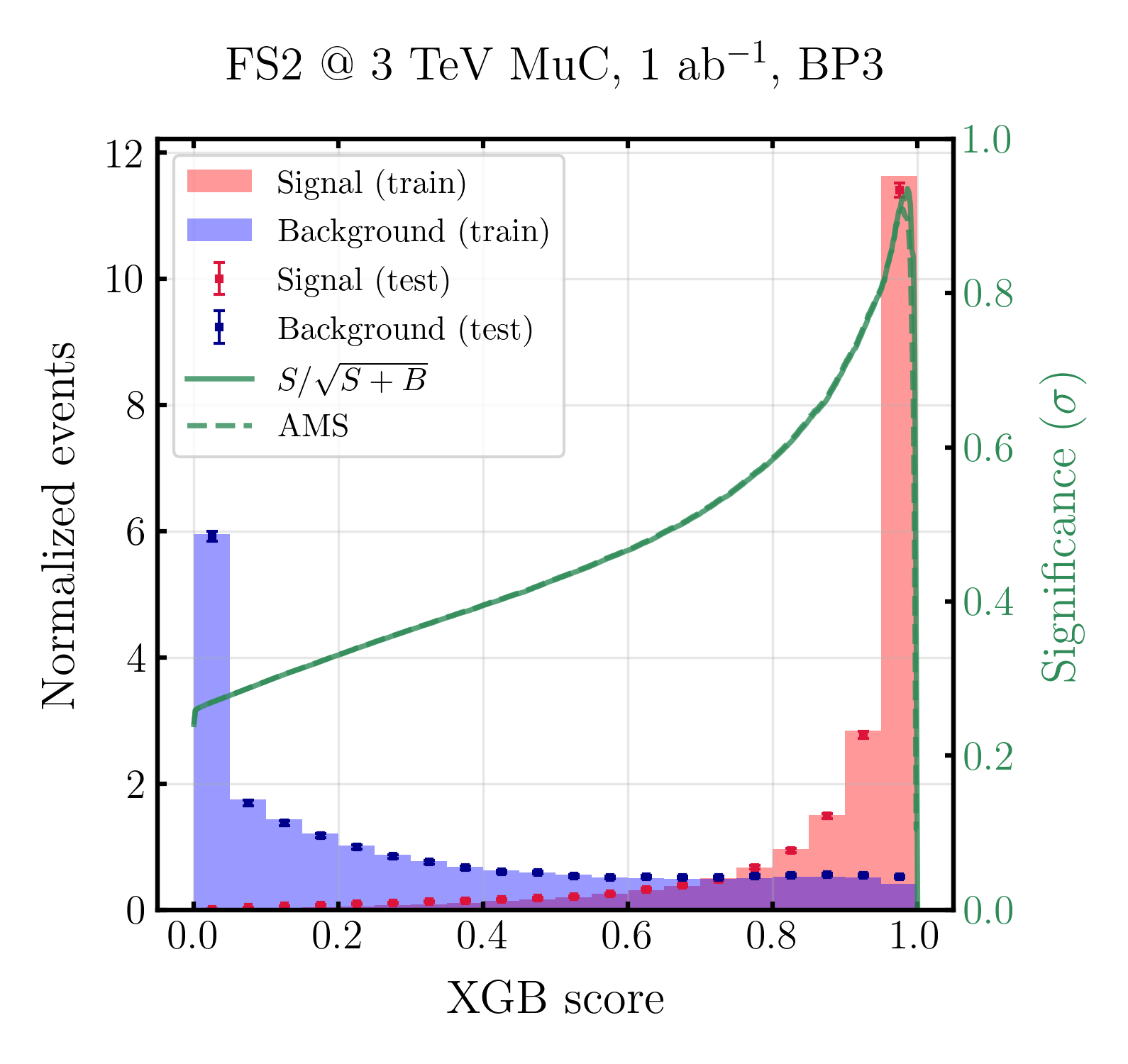}}
	\caption{Panels (a), (b), (c) show the ROC curves for BP1, BP2, BP3 respectively, resulting from both the training data (blue) and the test set (orange), using features from FS2, with the AUC in brackets. Panels (d), (e), (f) show the signal (red) and background (blue) separation based on the XGB classifier score, for BP1, BP2, BP3 respectively. The shaded histogram is for the training set, and the points with error bars are for the test set. The green curves represent the signal significance for the threshold cut on the XGB score. The y-axis on the right hand side with green ticks show the value of the significance in $\sigma$.}
	\label{fig:xgbres3tev_fs2}
\end{figure}

Once again, we evaluate the performance of the XGB classifier based on the ROC AUC metric, similar to FS1. In \autoref{fig:xgbres3tev_fs2}(a), (b) and (c), we display the ROC curves for each of the three benchmark points, with the training curve represented in blue, and the test curve shown in orange. Right away, we notice slightly better performance for both BP1 and BP2 in this FS2, compared to their FS1 counterpart. The features associated with FS2 have better discernibility compared to FS1, and hence the test AUC increase to 0.972 (0.943) for BP1 (BP2). BP3 shows a comparatively low AUC of 0.941 for the test sample. The corresponding signal-to-background separations based on the XGB score are plotted for BP1-BP3 in \autoref{fig:xgbres3tev_fs2}(d), (e) and (f) respectively. The true positives (signal) and true negatives (background) from the training set are displayed in red and blue filled histograms, respectively. The same for the test set are shown in red and blue points, with respective 1$\sigma$ error bars. The number of events are normalized to have total histogram area = 1, for better comparability. In each of the three plots, the overlaid green curves represent the achievable signal significance at 1 \abi luminosity with the final state, for each value of the XGB score threshold, evaluated using the test dataset. The secondary $y$-axis on the right hand side of the panels have the range for the signal significance ($\sigma$), in green text. The significance is again calculated with the $\eucal{S}=S/\sqrt{S+B}$ method (solid green curve) as well as the AMS method (dashed green curve).

\begin{table}[h]
	\renewcommand{\arraystretch}{1.2}
	\centering
	\begin{tabular}{|c|c||c||c|c|c||c|c|}
		\hline 
		\multirow{2}{*}{BP}	& \multirow{2}{*}{Cut flow} & \multicolumn{4}{c||}{FS2 counts at 3 TeV MuC} &   \multicolumn{2}{c|}{{\makecell{Significance \\ at $ \int \mathcal{L} dt $ = 1 \abi}}} \\
		\cline{3-8}
		&& Signal& \makecell{$ZW$ \\ (BG)} & \makecell{$hW$ \\ (BG)} &  \makecell{$t\bar{b}$ \\ (BG)} & $\eucal{S}$ & AMS\\
		\hline\hline
		\multirow{3}{*}{BP1} & \makecell{$\sigma_{\rm fiducial} \times \int \mathcal{L} dt $  } &487.80&4681.01&4091.50&2758.05& -- &--\\
		\cline{2-8}
		&S0:  $n_b = 2 + n_\ell =$ 1 + $n_{F\mu}$ = 1 &159.16&1243.73&1204.29&1272.13&2.56$\sigma$&2.59$\sigma$\\
		\cline{2-8}
		&S5:  S0 +  XGB score $\geq$ 0.970 &100.78&36.59&6.77&20.50&7.85$\sigma$&10.22$\sigma$\\
		\hline\hline
		\multirow{3}{*}{BP2} & \makecell{$\sigma_{\rm fiducial}  \times \int \mathcal{L} dt$ } &183.37&4681.01&4091.50&2758.05&--&--\\
		\cline{2-8}
		&S0:  $n_b = 2 + n_\ell =$ 1 + $n_{F\mu}$ = 1 &91.41&1243.73&1204.29&1272.13&1.48$\sigma$&1.49$\sigma$\\
		\cline{2-8}
		&S5:  S0 +  XGB score $\geq$ 0.970 &42.67&18.74&8.17&21.82&4.46$\sigma$&5.23$\sigma$\\
		\hline\hline
		\multirow{3}{*}{BP3} & \makecell{$\sigma_{\rm fiducial}  \times \int \mathcal{L} dt$ } &28.51&4681.01&4091.50&2758.05&--&--\\
		\cline{2-8}
		&S0:  $n_b = 2 + n_\ell =$ 1 + $n_{F\mu}$ = 1  &14.71&1243.73&1204.29&1272.13&0.24$\sigma$&0.24$\sigma$\\
		\cline{2-8}
		&S5:  S0 +  XGB score $\geq$ 0.986 &4.44&7.38&4.23&6.44&0.94$\sigma$&0.90$\sigma$\\
		\hline
	\end{tabular}
	\caption{Cut flow table of FS2 event counts for BP1-BP3 at the 3 TeV MuC with 1 \abi luminosity, against the backgrounds. Signal significance is evaluated for events after each cut.}
	\label{tab:fs2_3tev_bdt}
\end{table}

Based on the signal-to-background separation w.r.t the XGB classifier scores, we identify optimal cuts on the XGB score for each BP, and present them in \autoref{tab:fs2_3tev_bdt}, in the cut-flow format. After the pre-selection cut, we apply the S5 cut of XGB score , which is put at the values of 0.976, 0.982, and 0.988 for BP1, BP2, and BP3 respectively. Immediately, we notice the stark enhancement in the signal significance compared to the CBA in \autoref{tab:fs2_3tev}, with $\eucal{S}$ values of 7.85$\sigma$, 4.46$\sigma$, and 0.94$\sigma$ respectively, for BP1-BP3.  For the same cuts, the AMS score for the three BPs are 10.22$\sigma$, 5.23$\sigma$. and 0.90$\sigma$, which means that using the classifier, a $5\sigma$ discovery of the charged triplet scalar at the 350 GeV benchmark mass for BP2 can also be established.

%
%
%

\subsection{FS3: 2 jets + 2 leptons + MET + 1 Forward Muon}

It is also imperative to look for a final state that has fully visible decay products for the charged triplet scalar, in order to establish a mass probe for the model. As mentioned in \autoref{sec:fs}, we can either achieve it from the $H^+ \to ZW^+ / hW^+$ channels using two b-jets and two light jets, or we can demand the $Z$-boson to decay leptonically, essentially yielding FS3. While the generator-level number of evens can be expected to be more in the $2b+2j$ case, the $b$-tagging efficiencies and the presence of top-initiated backgrounds can make it more difficult to observe. Hence, we analyse the $2\ell + 2j$ signal with a large MET content accounting for the $H^0 \to SS$ decay. As we are specifically targetting light ($uds$) jets, we can successfully get rid of backgrounds coming from top quark processes with a $b$-jet veto. The backgrounds that contribute predominantly are from $\mu^\pm \nu VV$ and $\mu^\pm \nu VVV$ processes.

\begin{figure}[h]
	\centering
	\includegraphics[width=0.9\linewidth]{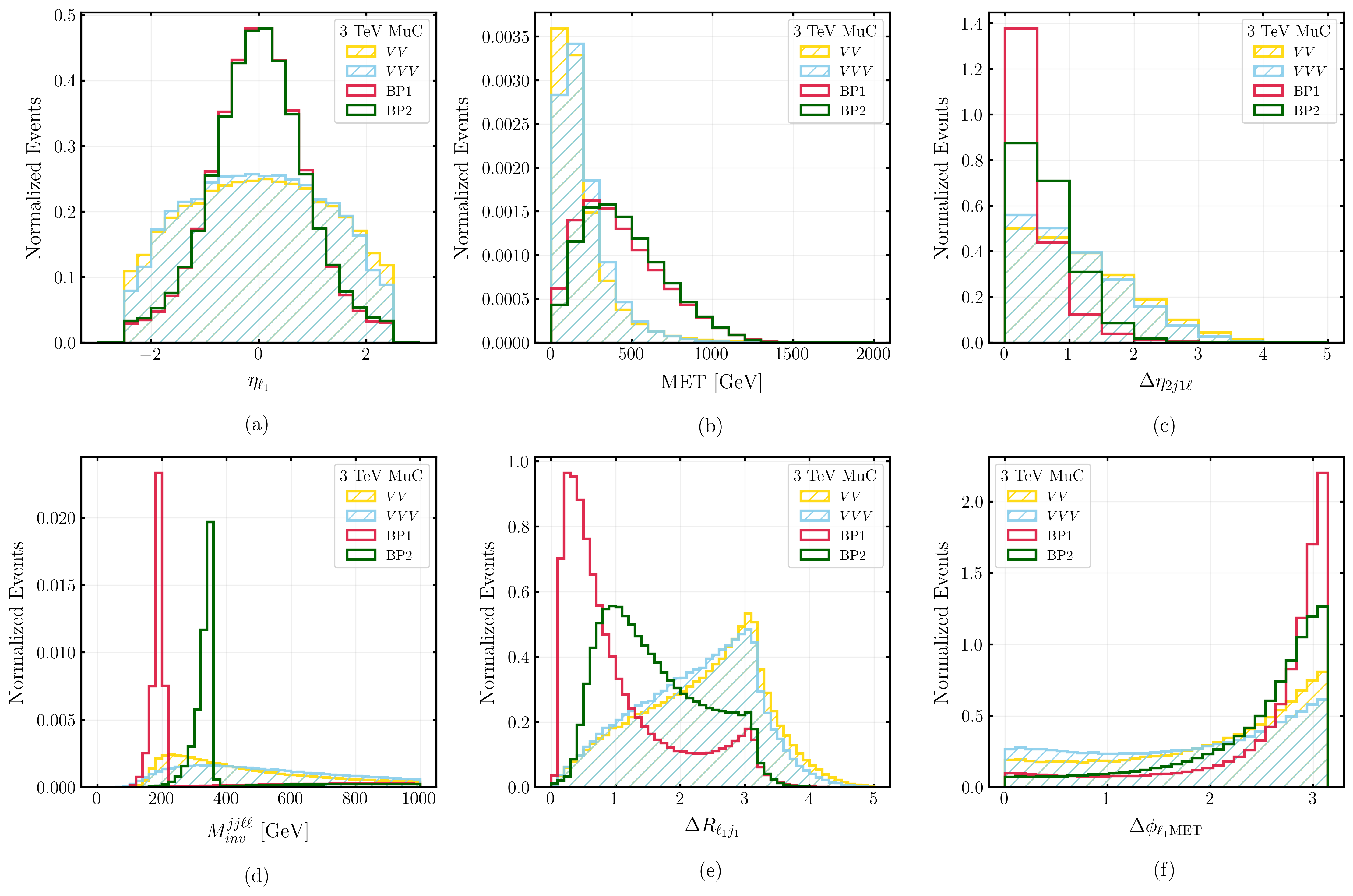}
	\caption{Distribution of discriminating kinematical variables for the $2j + 2\ell + {\rm MET} + 1\mu_F$ signal at a 3 TeV MuC. The backgrounds of $VV$ and $VVV$ are being taken with the specific final state at gen-level. The event histograms are normalized to one, for better visibility and comparison. The shaded histograms are for backgrounds, and the unshaded ones are for signals.}
	\label{fig:fs3kin}
\end{figure}

As we did for the previous two final states, we identify a few discriminating kinematic variables, whose distributions for both signal and background processes are shown in \autoref{fig:fs3kin}. The fiducial cross-section of BP3 remains too low, and hence at the 3 TeV MuC, we choose to present the distributions and results for BP1 (red solid) and BP2 (green solid) only. The $VV$ and $VVV$ backgrounds are shown in yellow and light blue shaded histograms. Right off the bat, we observe that the $2j+2\ell$ invariant mass, denoted as $M_{inv}^{jj\ell\ell}$, emerges as the variable with the best odds of distinguishing the signal (peaking at their respective masses) from the backgrounds (flat distributions). The other variables behave more-or-less the same way as they did for FS2. We find the optimal cut flow for the variables, and present the results in \autoref{tab:fs3_3tev}.

\begin{table}[h]
	\renewcommand{\arraystretch}{1.2}
	\centering
	\begin{tabular}{|c|c||c||c|c||c|c|}
		\hline 
		\multirow{2}{*}{BP}	& \multirow{2}{*}{Cut flow} & \multicolumn{3}{c||}{FS3 counts at 3 TeV MuC} &  \multicolumn{2}{c|}{{\makecell{Significance \\ at $ \int \mathcal{L} dt $ = 1 \abi}}} \\
		\cline{3-7}
		&& Signal& $VVV$ (BG) & $VV$ (BG) & $\eucal{S}$ & AMS \\
		\hline\hline
		\multirow{6}{*}{BP1} & \makecell{$\sigma_{\rm fiducial} \times \int \mathcal{L} dt $  } &93.52&1700.15&6377.30& -- &-- \\
		\cline{2-7}
		&S0:  $n_j = 2$ + $n_\ell =$ 2 + $n_{F\mu}$ = 1 &35.67&324.50&1783.05&0.77$\sigma$&0.77$\sigma$\\
		\cline{2-7}
		&S1:  S0 +  MET $\geq$ 200 GeV &30.01&143.41&689.38&1.02$\sigma$&1.03$\sigma$\\
		\cline{2-7}
		&S2:  S1 +  
		$\Delta R_{\ell_1 j_1} \leq$ 2  &25.70&61.31&319.30&1.27$\sigma$&1.29$\sigma$\\
		\cline{2-7} 
		&S3: S2 + $\Delta\eta_{2j1\ell} \leq 1$ &24.60&44.53&229.63&1.42$\sigma$&1.45$\sigma$\\
		\cline{2-7}
		&S4:  S3 + 180 GeV $\leq M_{inv}^{jj\ell\ell} \leq$ 210 GeV &13.87&1.29&24.09&2.22$\sigma$&2.36$\sigma$\\
		\hline\hline
		\multirow{6}{*}{BP2} & \makecell{$\sigma_{\rm fiducial}  \times \int \mathcal{L} dt$ } &23.53&1700.15&6377.30&--&--\\
		\cline{2-7}
		&S0:  $n_j = 2$ + $n_\ell =$ 2 + $n_{F\mu}$ = 1 &10.43&324.50&1783.05&0.23$\sigma$&0.23$\sigma$\\
		\cline{2-7}
		&S1:  S0 +  MET $\geq$ 200 GeV &8.86&143.41&689.38&0.31$\sigma$&0.31$\sigma$\\
		\cline{2-7}
		&S2:  S1 +  $\Delta R_{\ell_1 j_1} \leq$ 2 &6.78&61.39&319.30&0.34$\sigma$&0.34$\sigma$\\
		\cline{2-7} 
		&S3: S2 +  $\Delta\eta_{2j1\ell} \leq 1$  &5.67&44.53&229.63&0.34$\sigma$&0.34$\sigma$\\
		\cline{2-7}
		&S4:  S3 + 330 GeV $\leq M_{inv}^{jj\ell\ell} \leq$ 360 GeV&2.84&1.68&11.24&0.72$\sigma$&0.65$\sigma$\\
		\hline
	\end{tabular}
	\caption{Cut flow table of FS3 event counts for BP1 and BP2 at the 3 TeV MuC, against the $VV$ and $VVV$ backgrounds. Signal significance is evaluated for events after each cut.}
	\label{tab:fs3_3tev}
\end{table}

\autoref{tab:fs3_3tev} shows the cut flow for the FS3 for each BP under consideration as well as the $VV$ and $VVV$ backgrounds, with the corresponding values of the significance $\eucal{S}$ and the AMS score evaluated after each cut. It is important to note that, we are trying to follow a benchmark-independent cut flow till the S3 cut, and we apply the highly effective invariant mass window cut at the very end for each BP. As in the previous cases, S0 is the preselection cut, demanding exactly two light jets and two charged leptons, with one Forward muon in the dedicated detectors. This also includes the aforementioned $b$-jet veto to filter out top quark backgrounds. The next cut, S1, demands MET $\geq$ 200 GeV, ensuring we deal specifically with the $H^0 \to SS$ decay mode. S2 and S3 cuts are based on the separation between the jets and the leptons, which are expected to be closer for the signal compared to the backgrounds, as seen from \autoref{fig:fs3kin}(e) and (c) respectively.  The invariant mass window cut, denoted as S4, counts events within three 10 GeV bins of $M_{inv}^{jj\ell\ell}$ around the mass of the triplet scalar in each BP. Unsurprisingly, this cut yields the maximum significance for the signal, with $\eucal{S}$ (AMS) values of 2.22$\sigma$ (2.36$\sigma$) for BP1 at the 3 TeV MuC with 1 \abi luminosity. For BP2 however, both the  $\eucal{S}$  and AMS values fail to reach 1$\sigma$. 

\begin{figure}[h]
	\centering
	\subfigure[]{\includegraphics[width=0.45\linewidth]{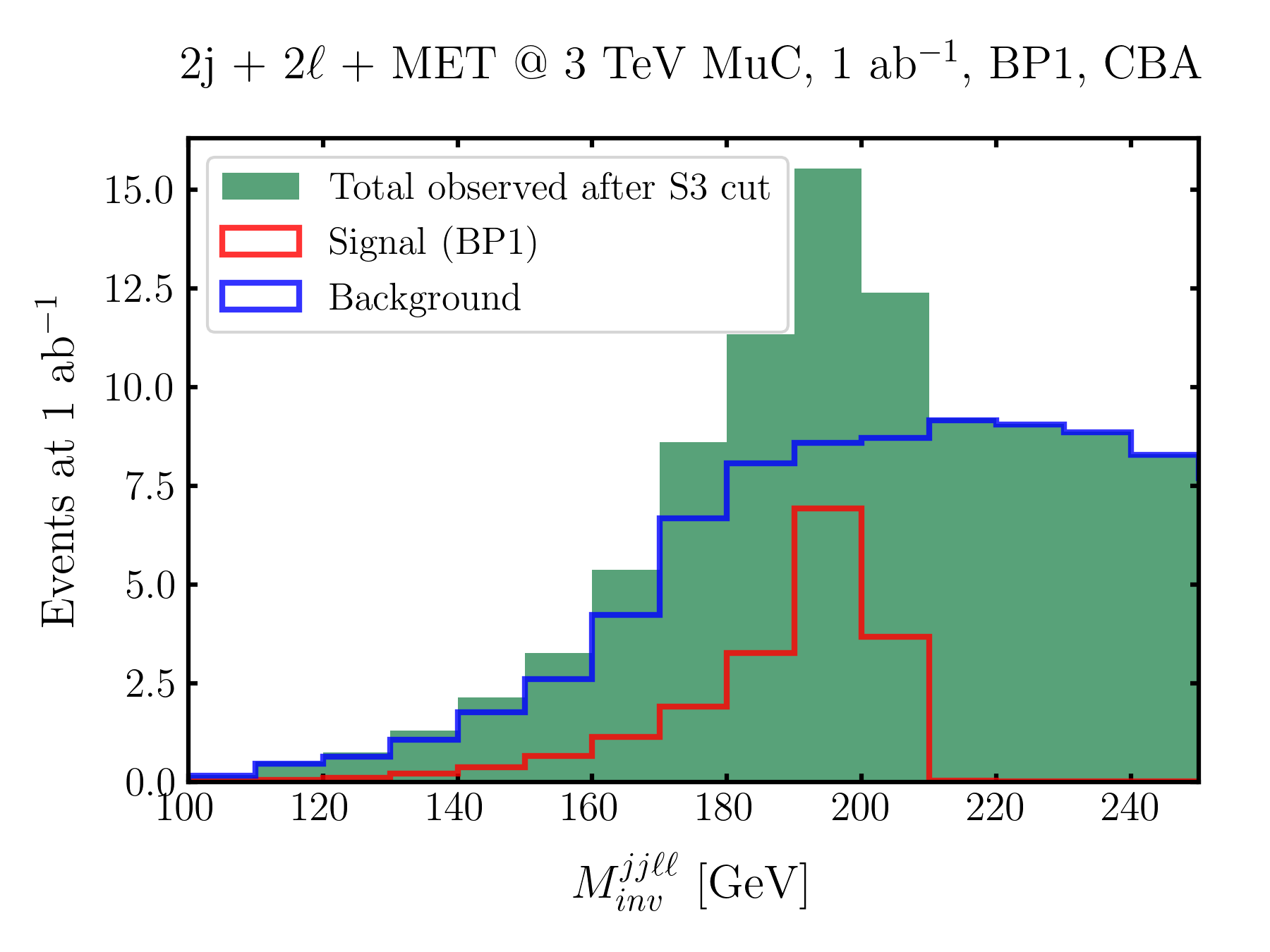}}
	\subfigure[]{\includegraphics[width=0.45\linewidth]{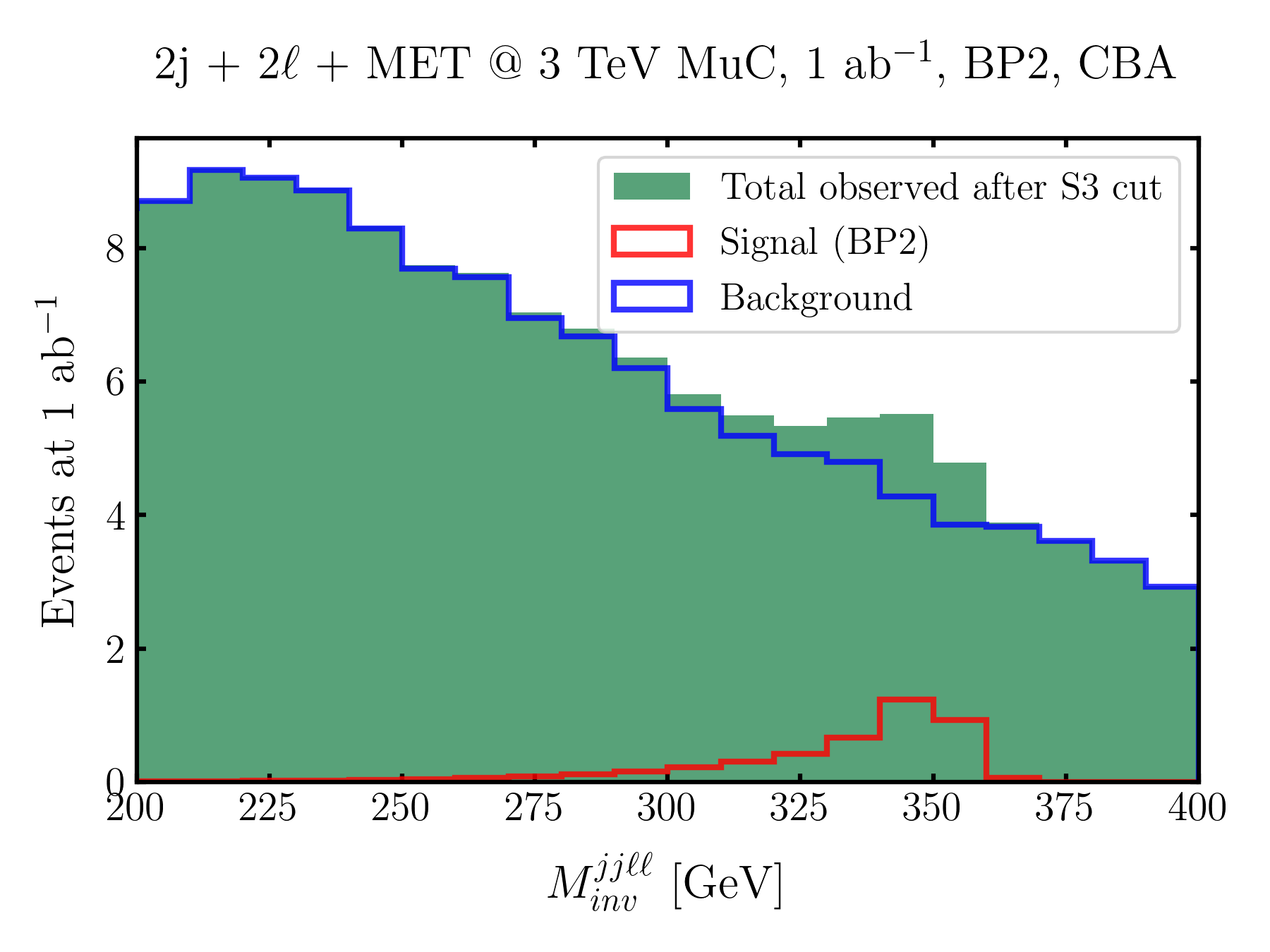}}
	\caption{$M_{inv}^{jj\ell\ell}$ distributions in GeV, for (a) BP1 and (b) BP2, keeping only the events that survive the S3 selection cut. The green filled histograms represent the total observable events at the 3 TeV MuC with 1 \abi luminosity, while the red and blue unfilled histograms stand for the underlying signal and background events.}
	\label{fig:mjjll_fs3_cba}
\end{figure}

To properly emphasize the peak significance calculation, we plot the $M_{inv}^{jj\ell\ell}$ distributions of all events that are filtered through the S3 selection cut, in \autoref{fig:mjjll_fs3_cba}(a) and (b), for BP1 and BP2 respectively. In each plot, the green filled histograms show the total events that are observed in the 3 TeV MuC with 1 \abi luminosity, and the underlying exact events for signal and background are shown in the unfilled red and blue histograms, respectively. We see that for BP1, the total observable event distribution shows a somewhat discernible peak at 200 GeV correspoding to the $H^\pm$ mass, while for BP2, the low number of signal events lead to no distinct peak at 350 GeV. This is reflected in the results from \autoref{tab:fs3_3tev}, where after the S4 cut, BP1 reach above 2$\sigma$ significance, while BP2 still cannot obtain a significance of 1$\sigma$. To somewhat enhance these values, we again turn our attention towards the XGB classifier, as discussed below.

\subsubsection{FS3 with a BDT}

For the purpose of training our \texttt{xgboost} classifier optimized to this final state, we identify the following set of kinematical variables:

\begin{itemize}
	\item $p_T$, $\eta$, $\phi$ of the two leptons and the two jets.
	\item Pseudorapidity separations between jet-lepton pairs, $\Delta\eta_{\ell_i j_i}$.
	\item Azimuthal angle separations between a lepton/jet and the MET, $\Delta\phi_{\ell/j_i MET}$.
	\item $\eta-\phi$ plane separation between jet and lepton combinations: $\Delta R_{\ell_i j_i}$.
	\item Missing transverse energy (MET).
	\item Invariant masses of lepton pairs ($M_{inv}^{\ell\ell}$), jet pairs ($M_{inv}^{jj}$), and $2j+2l$ combinations ($M_{inv}^{jj\ell\ell}$).
	\item Transverse mass of a jet or a lepton with the MET, defined as $M_T^{\ell/j_i MET}$.
\end{itemize}

\begin{figure}[h]
	\centering
	\includegraphics[width=\linewidth]{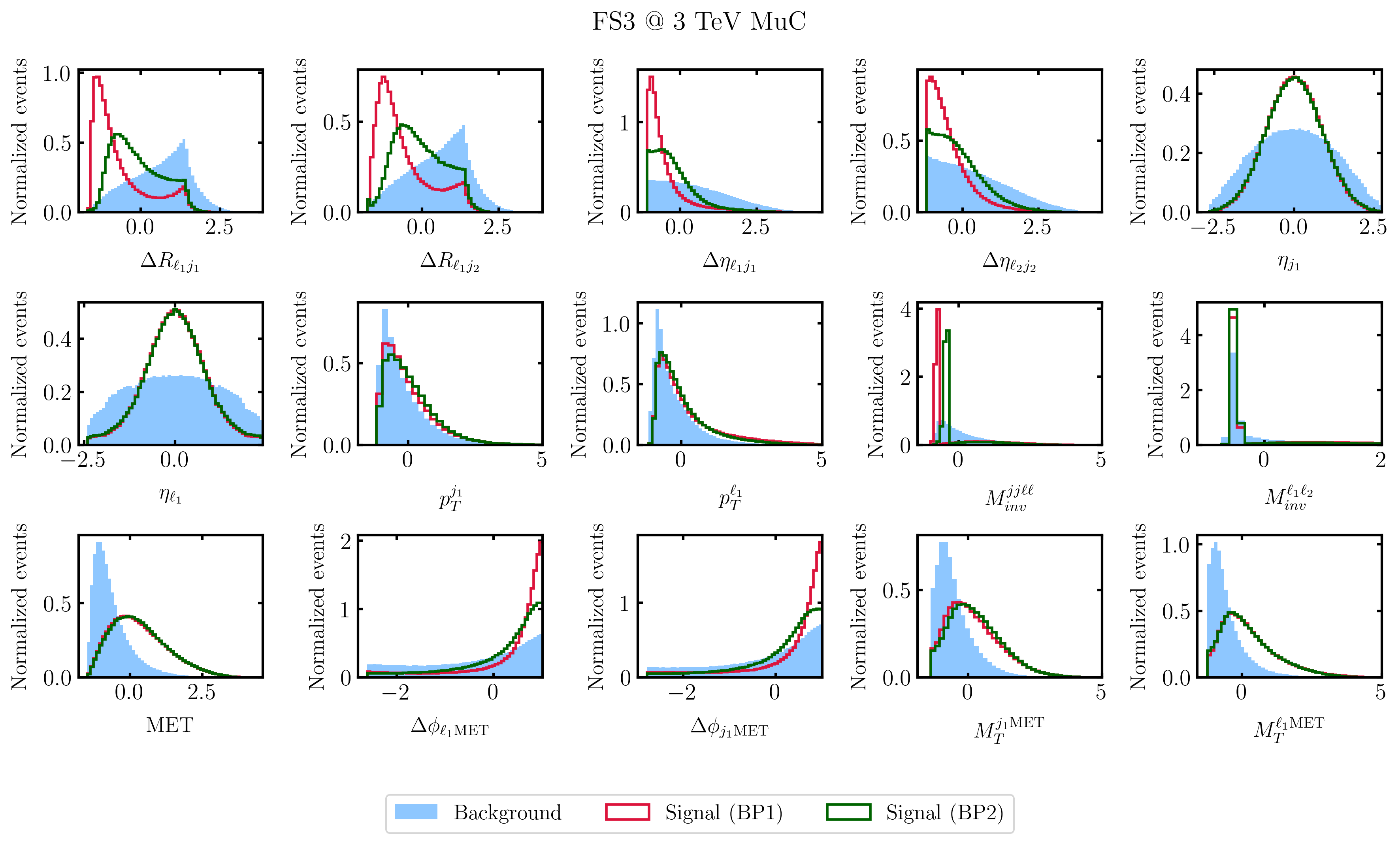}
	\caption{Feature distributions for the two signals and the combined $VV+VVV$ backgrounds, for \texttt{xgboost} training with FS3. The features are standardized to have mean =  0 and standard deviation = 1.}
	\label{fig:feat3tev_fs3}
\end{figure}

In \autoref{fig:feat3tev_fs3} we display the distributions for a few of these features, which are scaled so that they have mean = 0 and standard deviation = 1. The combined $VV+VVV$ background distributions are shown in the blue filled histograms, while the BP1 and BP2 signals are shown in red and green histograms. Using these features, we train the XGB classifier defined with the following parameters:

\begin{verbatim}
	cls=xgb.XGBClassifier(base_score=0.5, booster='gbtree', colsample_bylevel=0.8,
	colsample_bynode=0.8, colsample_bytree=0.8, gamma=1.5, 
	learning_rate=0.08, max_delta_step=0.4, max_depth=10, objective='binary:logistic',
	min_child_weight=8, n_estimators=1000, reg_alpha=10, reg_lambda=20, 
	scale_pos_weight=1, subsample=0.9, tree_method='exact')
\end{verbatim}

\begin{figure}[h]
	\centering
		\hspace{-0.7cm}
	\subfigure[]{\includegraphics[width = 0.32\linewidth]{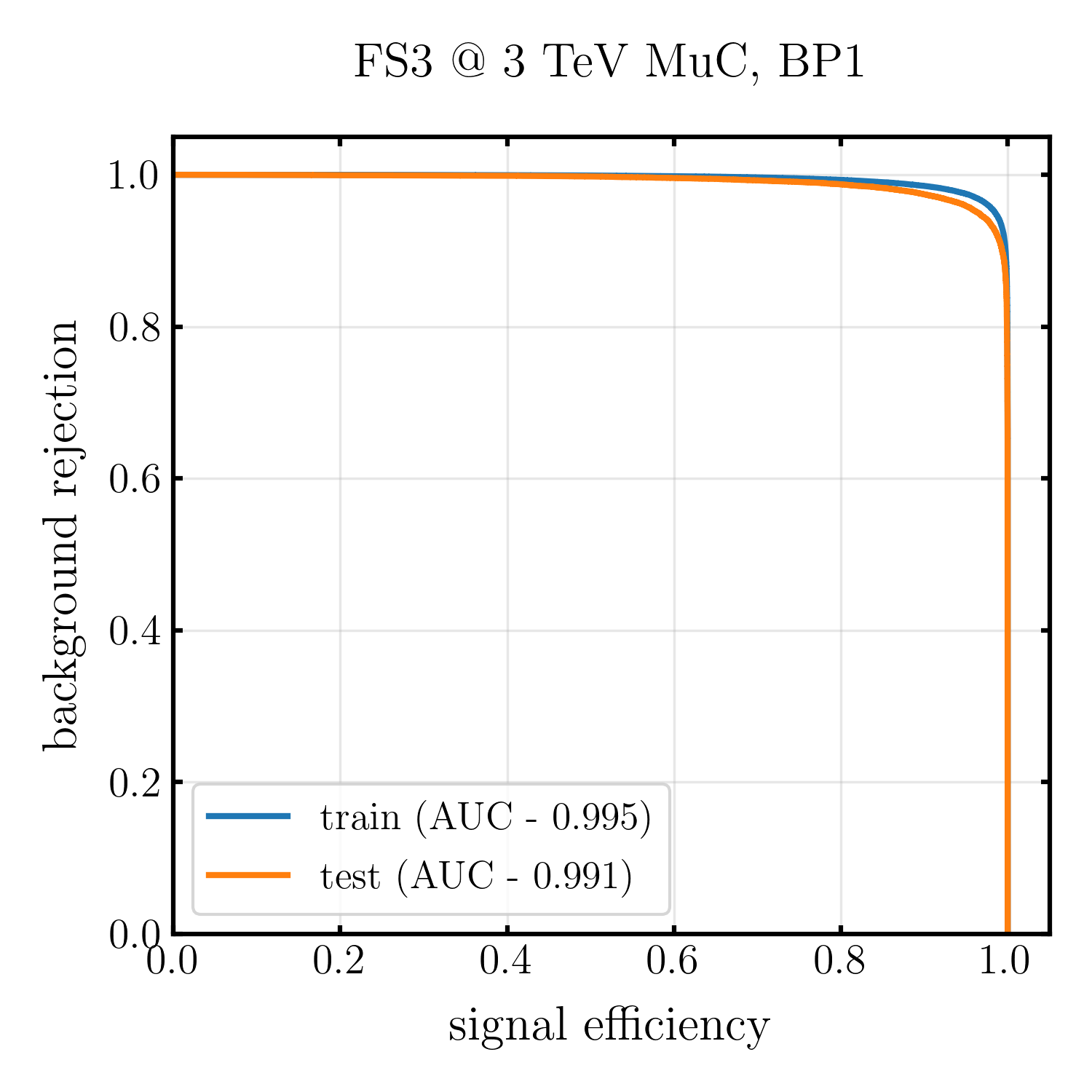}}
	\hspace{0.7cm}
	\subfigure[]{\includegraphics[width = 0.32\linewidth]{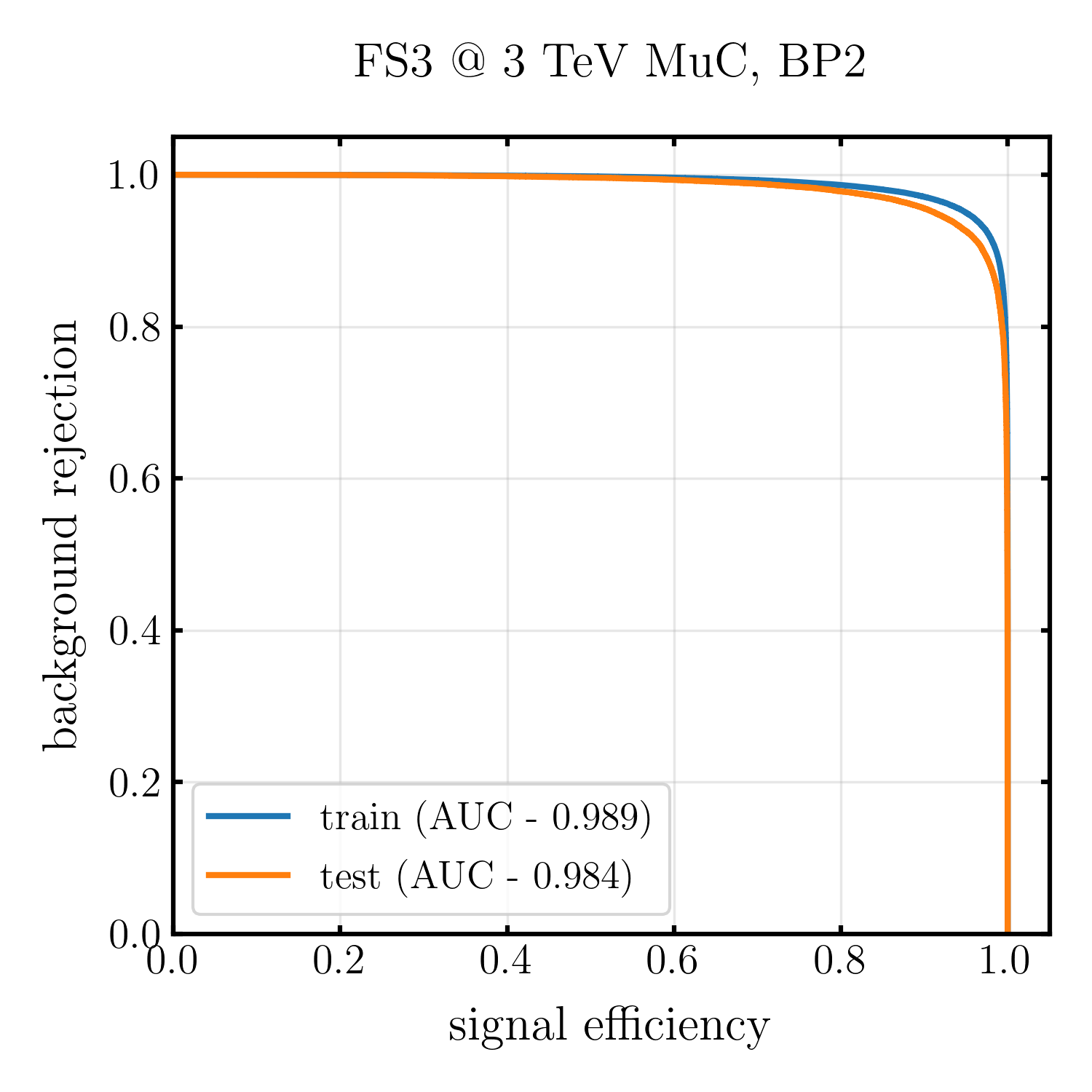}}
	\subfigure[]{\includegraphics[width = 0.37\linewidth]{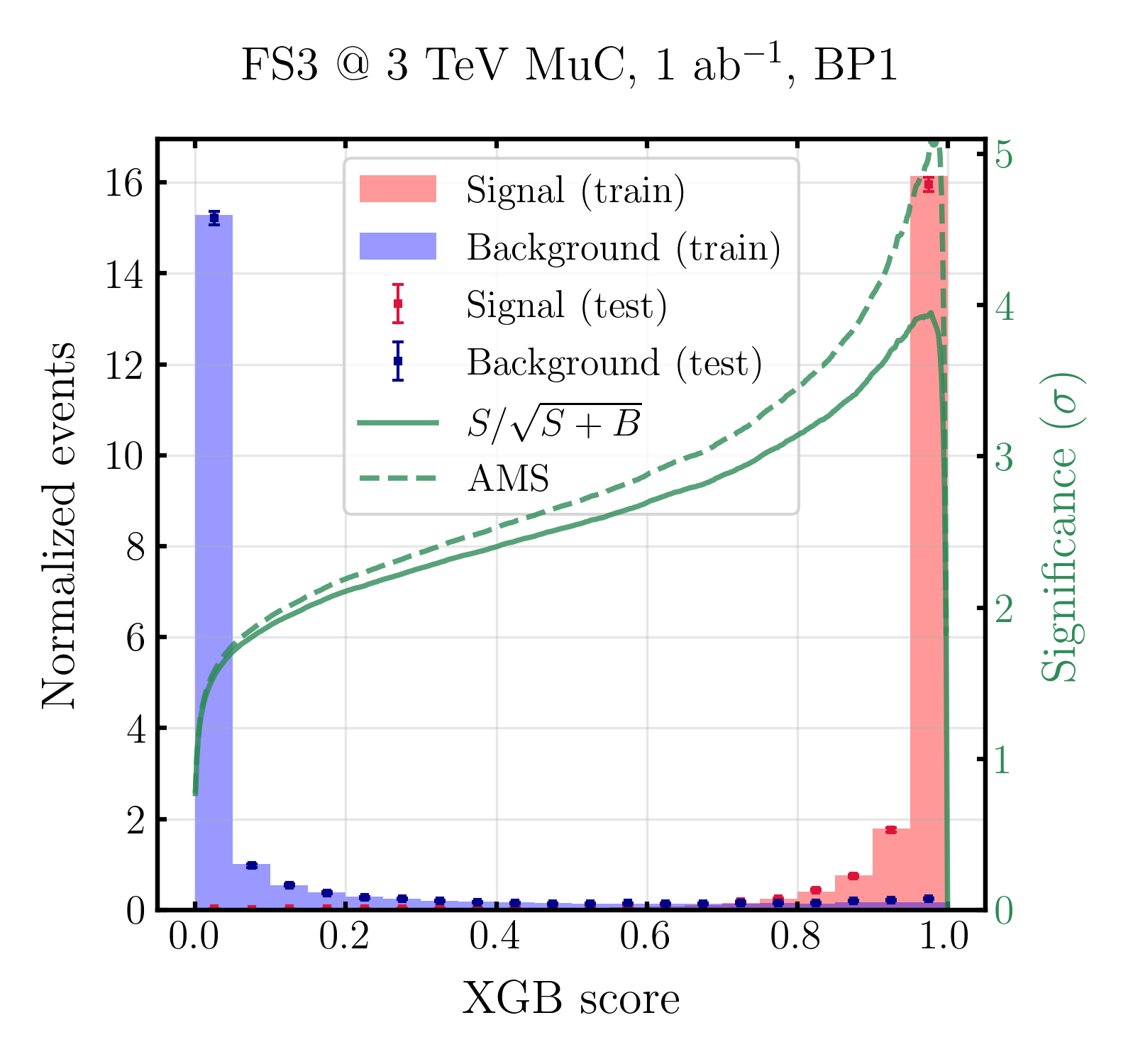}}
	\subfigure[]{\includegraphics[width = 0.37\linewidth]{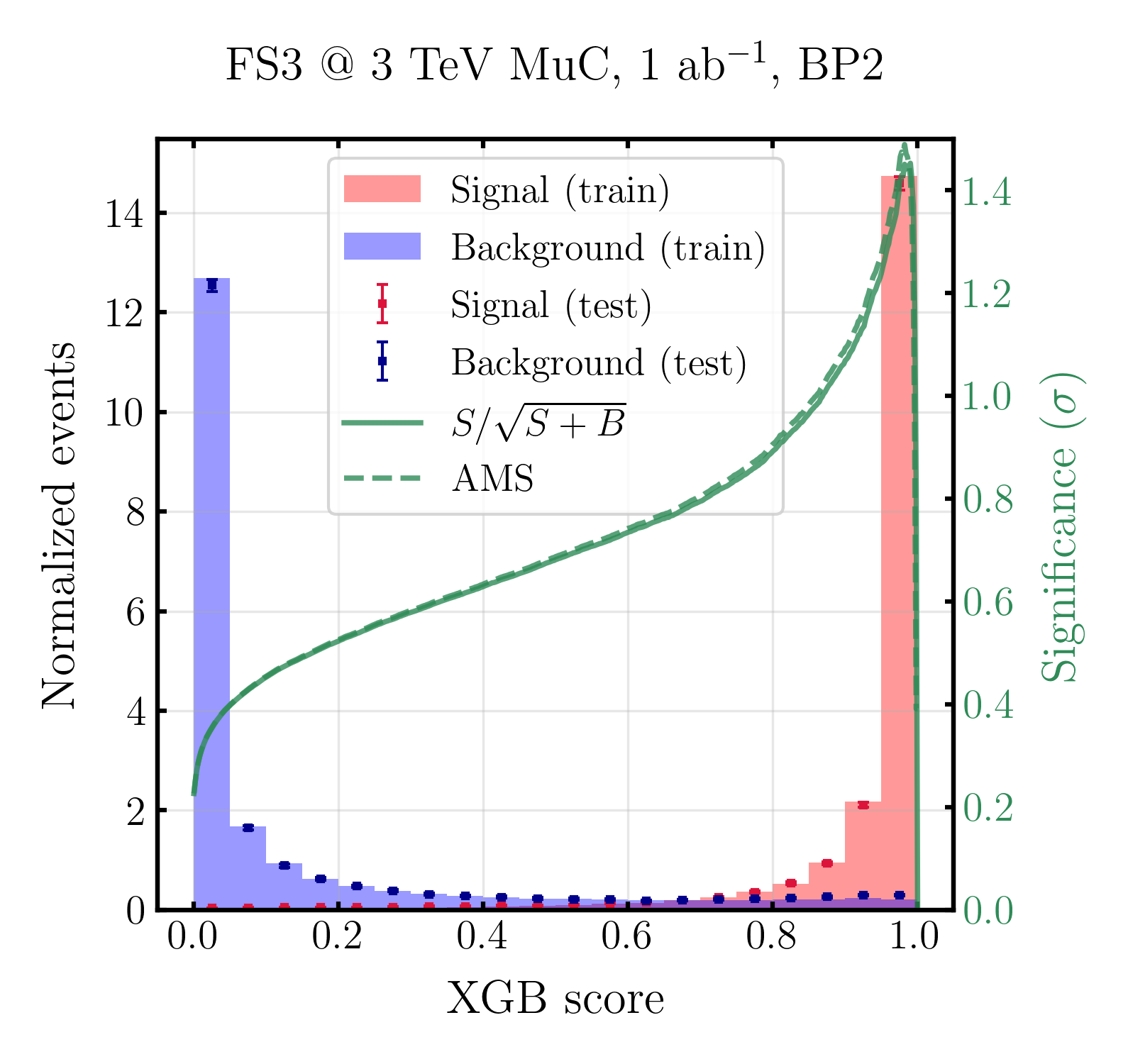}}
	
	\caption{Panels (a) and (b) show the ROC curves for BP1 and BP2 respectively, resulting from both the training data (blue) and the test set (orange), using features from FS3, with the AUC in brackets. Panels (c) and (d) show the signal (red) and background (blue) separation based on the XGB classifier score, for BP1 and BP2 respectively. The shaded histogram is for the training set, and the points with error bars are for the test set. The green curves represent the signal significance for the threshold cut on the XGB score. The y-axis on the right hand side with green ticks show the value of the significance in $\sigma$.}
	\label{fig:xgbres3tev_fs3}
\end{figure}

 The ROC AUC metric is utilized again to evaluate the performance of the classifier, which are graphically presented in \autoref{fig:xgbres3tev_fs3}(a) and (b), for BP1 and BP2, respectively. We notice that the FS3 features perform the best out of the three final states, with test AUC values of 0.991 and 0.984 for BP1 and BP2, respectively. The signal-to-background separation with respect to the XGB scores for BP1 and BP2 are shown in \autoref{fig:xgbres3tev_fs3}(c) and (d), respectively. The red and blue filled histograms correspond to the signal and background events in the training set, while the red and blue points with error bars represent the same for the test set. The solid and dashed green curves represent the signal significance $\eucal{S}$ and AMS scores respectively, for the corresponding XGB score cut. This results into the optimal XGB score thresholds of 0.974 and 0.990 for BP1 and BP2, with maximum achievable $\eucal{S}$(AMS) values of 3.95$\sigma$(5.02$\sigma$) and 1.45$\sigma$(1.43$\sigma$), respectively, at the 3 TeV MuC with 1 \abi of integrated luminosity. These outcomes are tabulated in \autoref{tab:fs3_3tev_bdt}.

\begin{table}[h]
	\renewcommand{\arraystretch}{1.2}
	\centering
	\begin{tabular}{|c|c||c||c|c||c|c|}
		\hline 
		\multirow{2}{*}{BP}	& \multirow{2}{*}{Cut flow} & \multicolumn{3}{c||}{FS3 counts at 3 TeV MuC with BDT} & \multicolumn{2}{c|}{{\makecell{Significance \\ at $ \int \mathcal{L} dt $ = 1 \abi}}} \\
		\cline{3-7}
		&& Signal& $VVV$ (BG) & $VV$ (BG) &  $\eucal{S}$ & AMS \\
		\hline\hline
		\multirow{3}{*}{BP1} & \makecell{$\sigma_{\rm fiducial}$ (fb) } &93.52&1700.15&6377.30& --&--\\
		\cline{2-7}
		&S0:  $n_\ell =$ 3 + $n_{F\mu}$ = 1 &35.67&324.50&1783.05&0.77$\sigma$&$\sigma$\\
		\cline{2-7}
		&S5:  S0 +  XGB score $\geq 0.974$ &24.32&2.04&11.54&3.95$\sigma$&5.02$\sigma$\\
		\cline{2-7}
		&S6: S5 + 180 GeV $\leq M_{inv}^{jj\ell\ell} \leq$ 210 GeV &14.69&1.00&8.00&3.08$\sigma$&3.75$\sigma$\\
		\hline\hline
		\multirow{3}{*}{BP2} & \makecell{$\sigma_{\rm fiducial}$ (fb) } &23.53&1700.15&6377.30& --&--\\
		\cline{2-7}
		&S0:  $n_\ell =$ 3 + $n_{F\mu}$ = 1 &10.43&324.50&1783.05&0.23$\sigma$&$\sigma$\\
		\cline{2-7}
		&S5:  S0 +  XGB score $\geq 0.990$ &3.93&0.57&2.81&1.45$\sigma$&1.43$\sigma$\\
		\cline{2-7}
		&S6: S5 + 330 GeV $\leq M_{inv}^{jj\ell\ell} \leq$ 360 GeV &2.81&0.08&1.76&1.30$\sigma$&1.18$\sigma$\\
		\hline
	\end{tabular}
	\caption{Cut flow table for FS3 event counts at the 3 TeV MuC with 1 \abi luminosity, for BP1 and BP2 against the $VV$ and $VVV$ backgrounds, with the XGB score threshold cut. Signal significance is evaluated for events after each cut.}
	\label{tab:fs3_3tev_bdt}
\end{table}

\begin{figure}[h]
	\centering
	\subfigure[]{\includegraphics[width=0.45\linewidth]{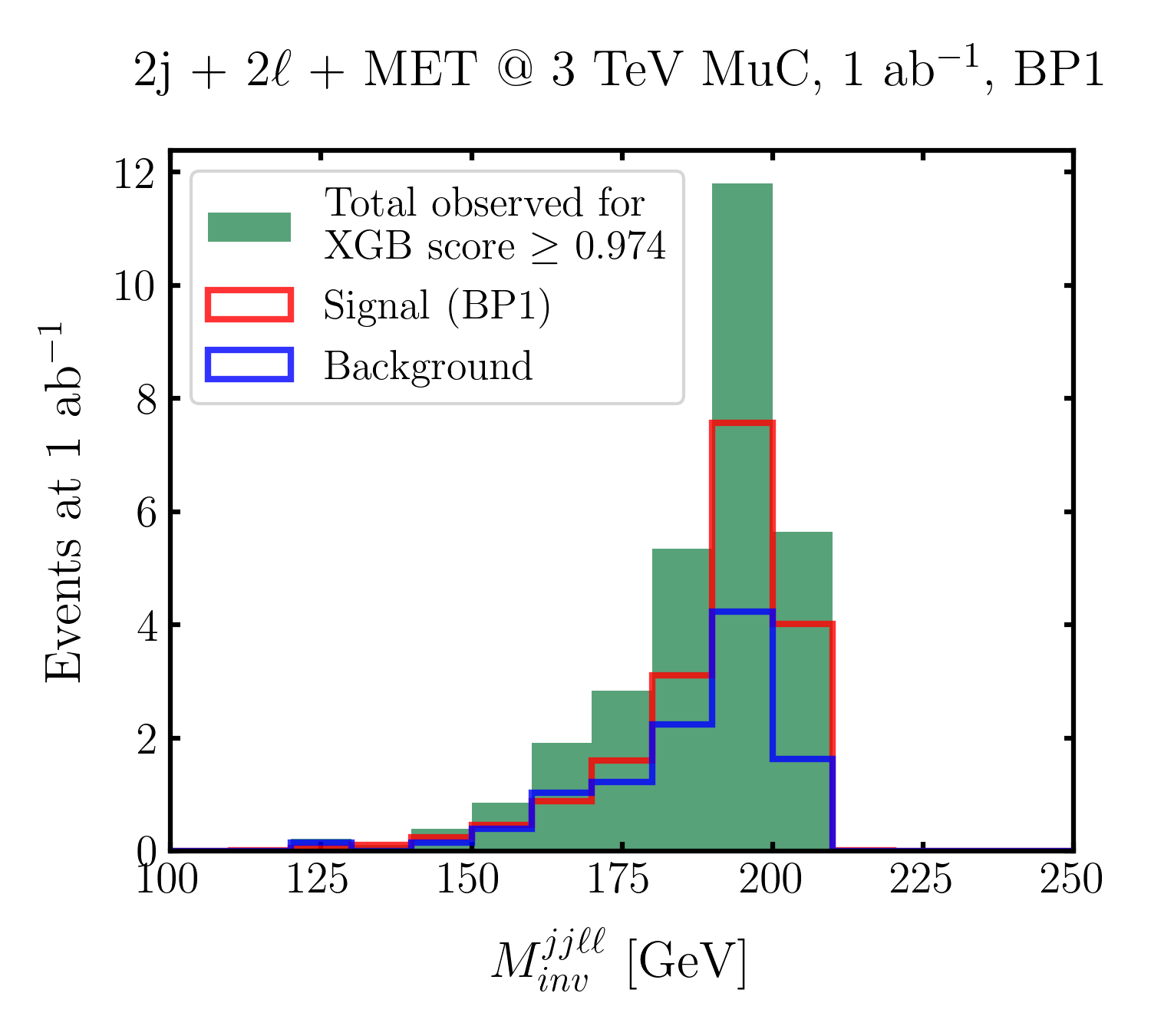}}
	\subfigure[]{\includegraphics[width=0.45\linewidth]{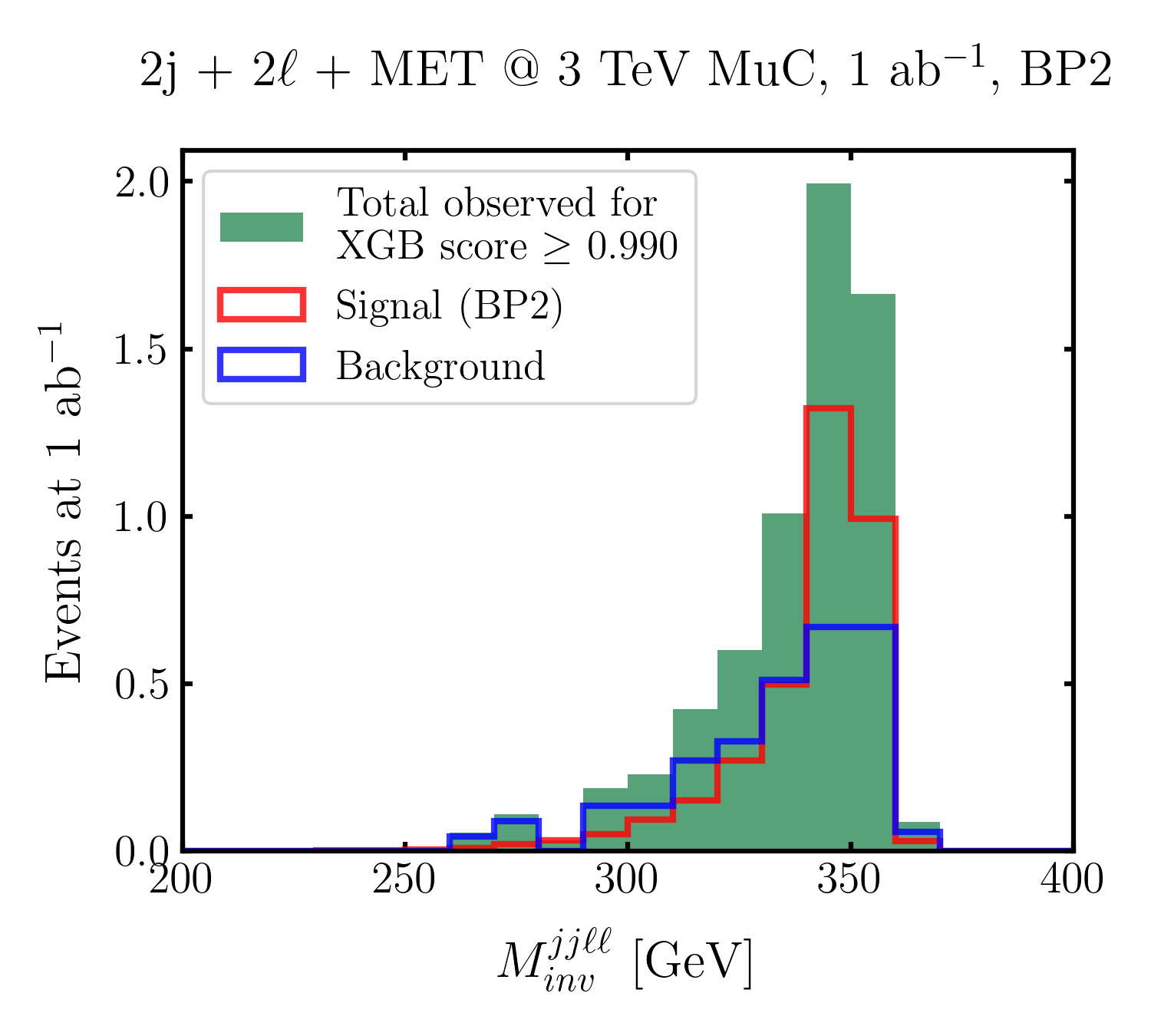}}
	\caption{$M_{inv}^{jj\ell\ell}$ distributions in GeV, for (a) BP1 and (b) BP2, keeping only the events that survive the XGB threshold cut. The green filled histograms represent the total observable events at the 3 TeV MuC with 1 \abi luminosity, while the red and blue unfilled histograms stand for the true positive (signal) and true negative (background) events.}
	\label{fig:mjjll_fs3_bdt}
\end{figure}

It is also important to note that, in the training of the BDT, the $M_{inv}^{jj\ell\ell}$ distributions are used as features, without any cuts on their values. After training and evaluation of the classifiers, we plot the $M_{inv}^{jj\ell\ell}$ distributions for the events that pass the XGB score cuts, for BP1 and BP2 in \autoref{fig:mjjll_fs3_bdt}(a) and (b) respectively. In these figures, the green filled histograms depict the invariant mass distributions for the total number of observed events after the particular XGB score cut. The true positive (signal) and true negative (background) distributions are marked with red and blue unfilled histograms, respectively. In each BP we witness the sharp peaks around 200 and 350 GeV, corresponding to the respective $H^\pm$ masses. Now, similar to the cut-based analysis, we opt to put the same 30 GeV invariant mass window cut on these events, and evaluate the signal significance for the peaks. In \autoref{tab:fs3_3tev_bdt}, we also entail the event counts and signal significance in these peaks, denoted as the S6 cut, with $\eucal{S}$(AMS) values of 3.08$\sigma$(3.75$\sigma$) and 1.30$\sigma$(1.18$\sigma$) for BP1 and BP2, respectively. While this cut apparently decreases the signal significance obtained from the S5 cut in  \autoref{tab:fs3_3tev_bdt}, one actually gains an advantage via the reconstruction of the triplet charged Higgs from a fully visible final state, a necessary trade-off to be made while hunting for a BSM resonance this way.

In \autoref{fig:mjjll_fs3_bdt}, we notice that the background distributions also follow the same shape as those of the signal, because of them being classified as false positives, as we have used the $M_{inv}^{jj\ell\ell}$ distributions in the training process. Training the classifier on the entire distribution need not necessarily induce bias, however, we still wish to compare it to a more generalized training approach, where we omit this invariant mass variable from the feature list, and retrain the same classifier.

\subsubsection{FS3 with a BDT, without using $M_{inv}^{jj\ell\ell}$ for training}

As mentioned above, we also choose to perform the XGB-based classification by excluding the $M_{inv}^{jj\ell\ell}$ feature from training. To compensate for this and to obtain the optimal performance, we slightly alter the classifier parameters as follows:

\begin{verbatim}
	cls=xgb.XGBClassifier(base_score=0.5, booster='gbtree', colsample_bylevel=0.8,
	colsample_bynode=0.8, colsample_bytree=0.8, gamma=1.5, 
	learning_rate=0.08, max_delta_step=0.3, max_depth=11, objective='binary:logistic',
	min_child_weight=8, n_estimators=1000, reg_alpha=15, reg_lambda=20, 
	scale_pos_weight=1, subsample=0.9, tree_method='exact')
\end{verbatim}

\begin{figure}[h]
	\centering
	\hspace{-0.7cm}
	\subfigure[]{\includegraphics[width = 0.32\linewidth]{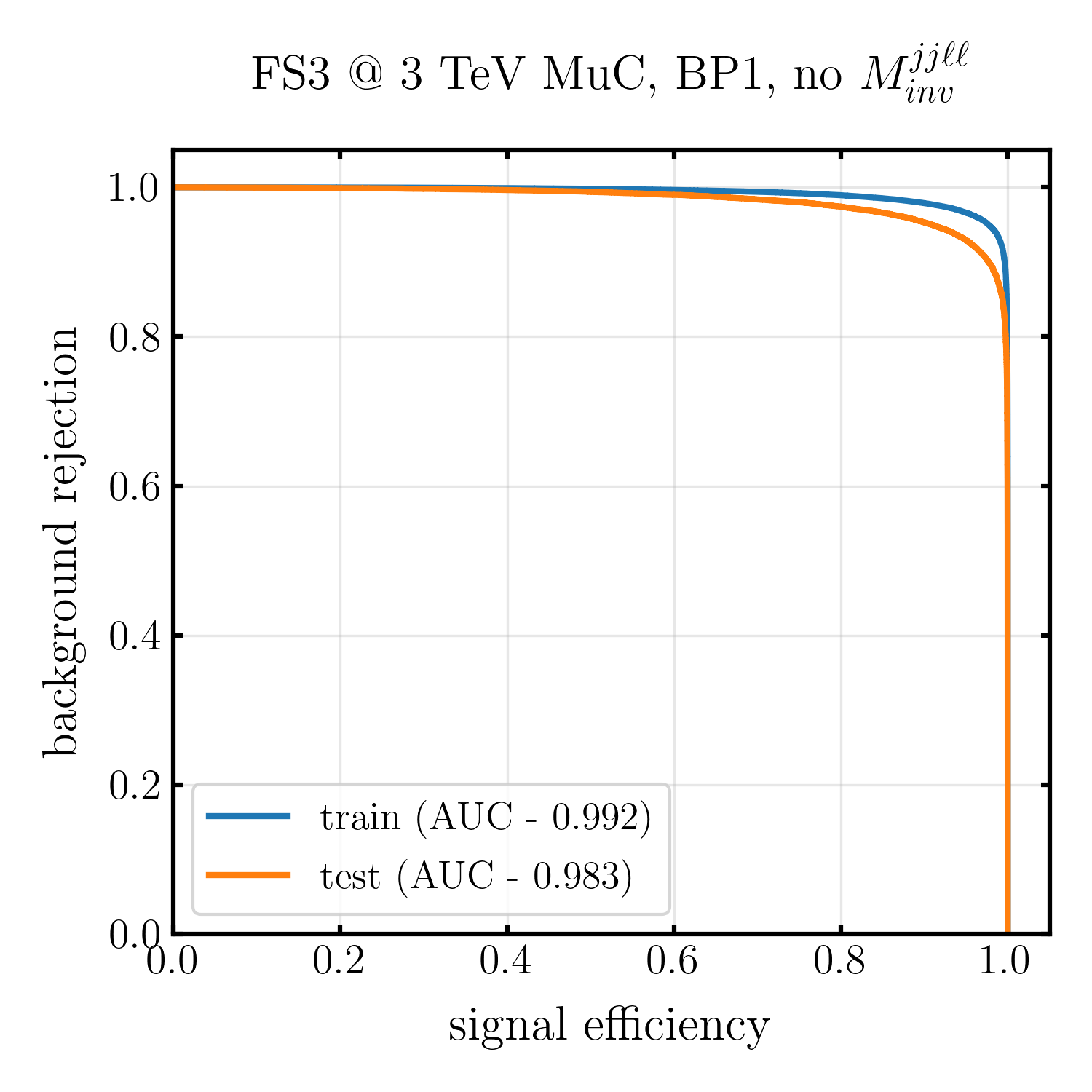}}
	\hspace{0.7cm}
	\subfigure[]{\includegraphics[width = 0.32\linewidth]{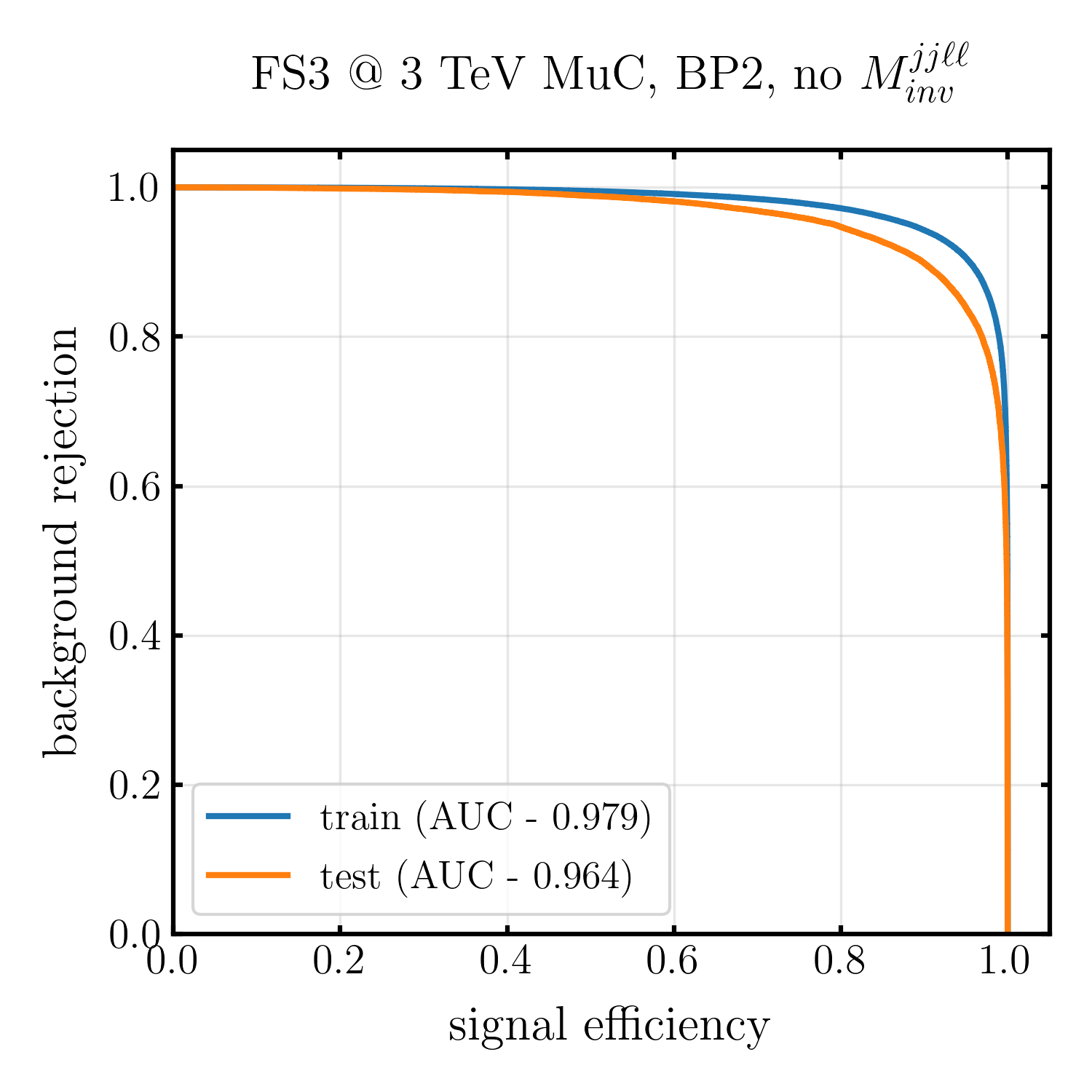}}
	\subfigure[]{\includegraphics[width = 0.37\linewidth]{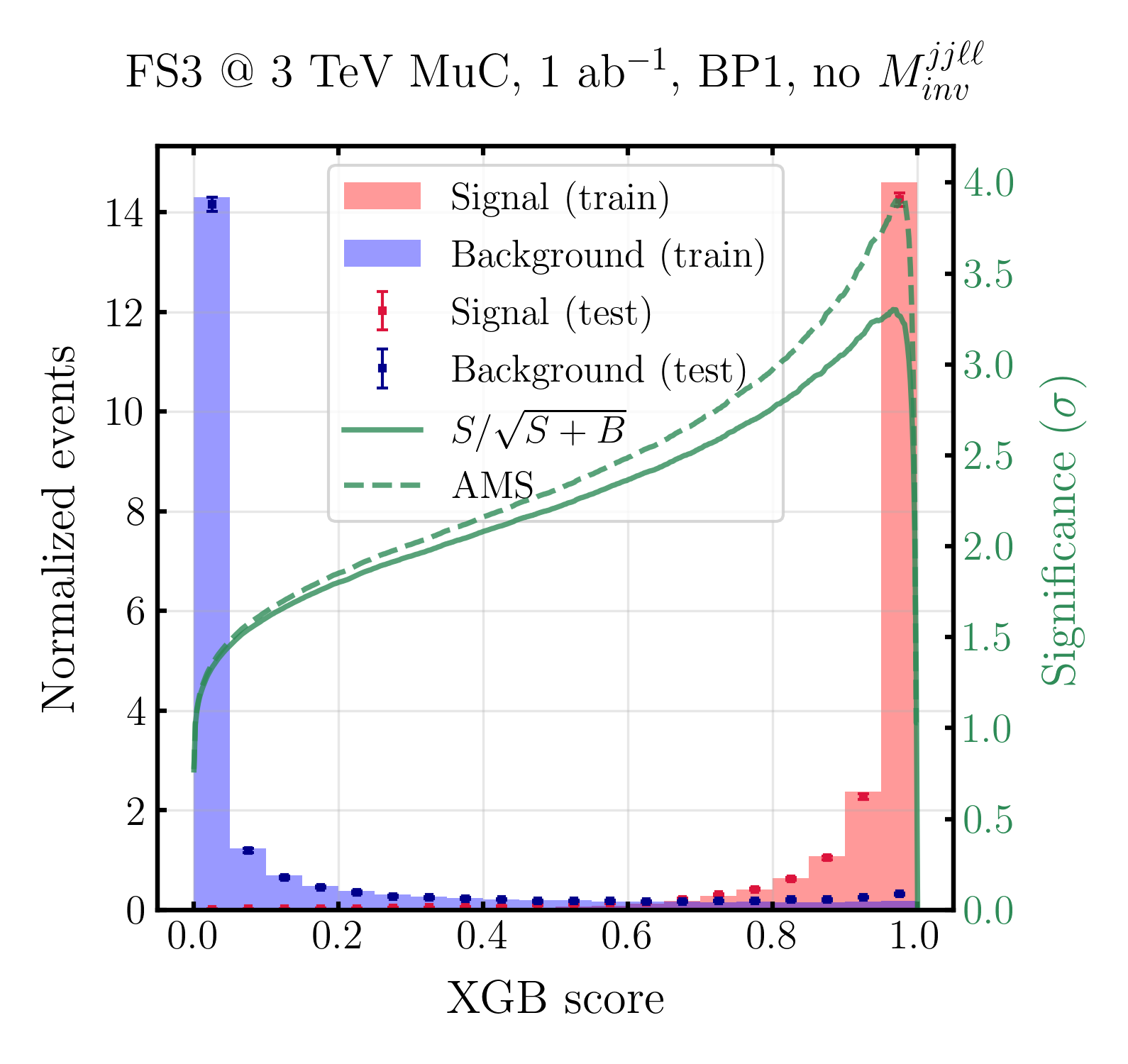}}
	\subfigure[]{\includegraphics[width = 0.37\linewidth]{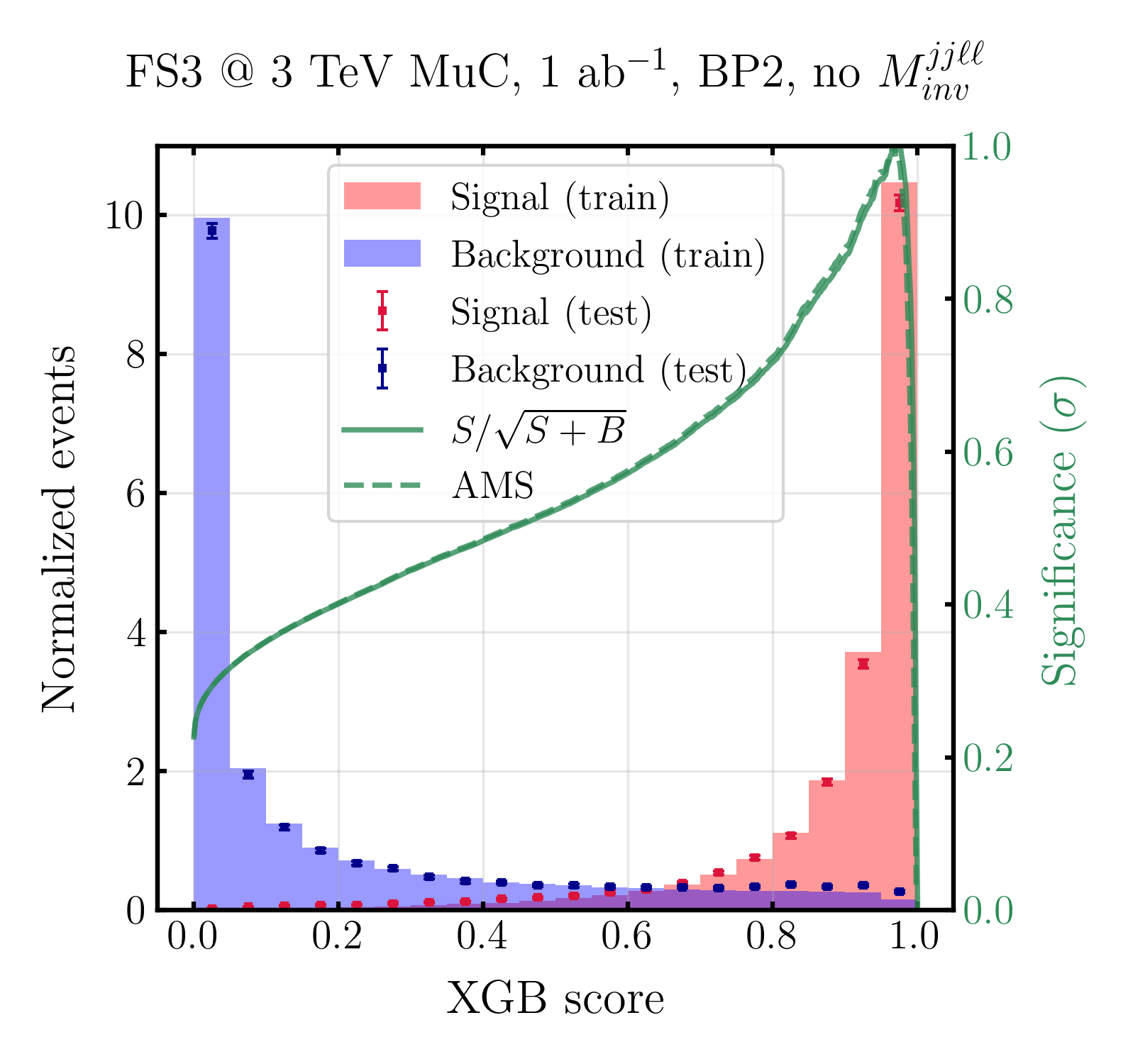}}
	
	\caption{Panels (a) and (b) show the ROC curves for BP1 and BP2 respectively, resulting from both the training data (blue) and the test set (orange), using features from FS3 excluding $M_{inv}^{jj\ell\ell}$, with the AUC in brackets. Panels (c) and (d) show the signal (red) and background (blue) separation based on the XGB classifier score, for BP1 and BP2 respectively. The shaded histogram is for the training set, and the points with error bars are for the test set. The green curves represent the signal significance for the threshold cut on the XGB score. The y-axis on the right hand side with green ticks show the value of the significance in $\sigma$.}
	\label{fig:xgbres3tev_fs3_noim}
\end{figure}

After training the same set as the previous case, and then testing it and validating it, we obtain the results shown in \autoref{fig:xgbres3tev_fs3_noim}. Clearly, from the ROC curves in  \autoref{fig:xgbres3tev_fs3_noim}(a) and (b), we notice that the performance of the classifier drops as compared to the previous case that included  $M_{inv}^{jj\ell\ell}$ in the feature set. The same effect is reflected in the signal-to-background separation and corresponding signal significance values based on the XGB score thresholds, as shown in \autoref{fig:xgbres3tev_fs3_noim}(c) and (d) for BP1 and BP2, respectively. The maximum achievable AMS value for BP1 comes down from $\sim5\sigma$ to $\sim4\sigma$, whereas for BP2, the significance values barely reach $1\sigma$. Compared to the CBA however, the XGB classifier still performs better. The outcomes of this particular analysis are displayed in \autoref{tab:fs3_3tev_bdt_noim}.

\begin{table}[h]
	\renewcommand{\arraystretch}{1.2}
	\centering
	\begin{tabular}{|c|c||c||c|c||c|c|}
		\hline 
		\multirow{2}{*}{BP}	& \multirow{2}{*}{Cut flow} & \multicolumn{3}{c||}{\makecell{FS3 counts at 3 TeV MuC with BDT, \\ no $M_{inv}^{jj\ell\ell}$}} & \multicolumn{2}{c|}{{\makecell{Significance \\ at $ \int \mathcal{L} dt $ = 1 \abi}}} \\
		\cline{3-7}
		&& Signal& $VVV$ (BG) & $VV$ (BG) &  $\eucal{S}$ & AMS \\
		\hline\hline
		\multirow{3}{*}{BP1} & \makecell{$\sigma_{\rm fiducial}$ (fb) } &93.52&1700.15&6377.30& --&--\\
		\cline{2-7}
		&S0:  $n_\ell =$ 3 + $n_{F\mu}$ = 1 &35.67&324.50&1783.05&0.77$\sigma$&$\sigma$\\
		\cline{2-7}
		&S5:  S0 +  XGB score $\geq 0.978$ &20.00&2.30&14.27&3.31$\sigma$&3.97$\sigma$\\
		\cline{2-7}
		&S6: S5 + 180 GeV $\leq M_{inv}^{jj\ell\ell} \leq$ 210 GeV &11.54&1.00&8.43&2.57$\sigma$&3.00$\sigma$\\
		\hline\hline
		\multirow{3}{*}{BP2} & \makecell{$\sigma_{\rm fiducial}$ (fb) } &23.53&1700.15&6377.30& --&--\\
		\cline{2-7}
		&S0:  $n_\ell =$ 3 + $n_{F\mu}$ = 1 &10.43&324.50&1783.05&0.23$\sigma$&$\sigma$\\
		\cline{2-7}
		&S5:  S0 +  XGB score $\geq 0.965$ &4.29&2.97&10.97&1.01$\sigma$&1.00$\sigma$\\
		\cline{2-7}
		&S6: S5 + 330 GeV $\leq M_{inv}^{jj\ell\ell} \leq$ 360 GeV &2.10&0.03&1.95&1.03$\sigma$&0.90$\sigma$\\
		\hline
	\end{tabular}
	\caption{Cut flow table for FS3 event counts at the 3 TeV MuC with 1 \abi luminosity, for BP1 and BP2 against the $VV$ and $VVV$ backgrounds, with the XGB score threshold cut. Signal significance is evaluated for events after each cut.}
	\label{tab:fs3_3tev_bdt_noim}
\end{table}

\begin{figure}[h]
	\centering
	\subfigure[]{\includegraphics[width=0.45\linewidth]{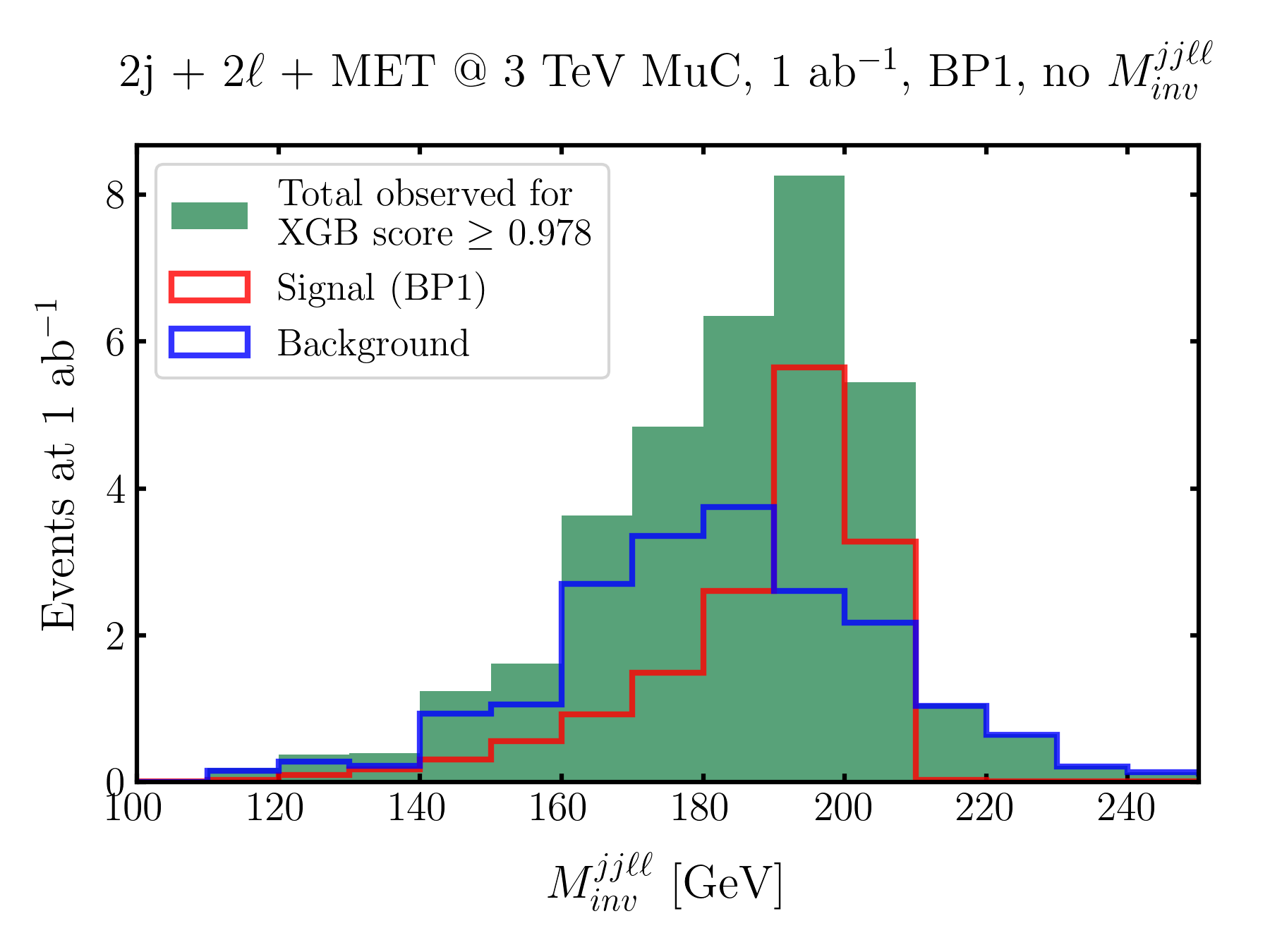}}
	\subfigure[]{\includegraphics[width=0.45\linewidth]{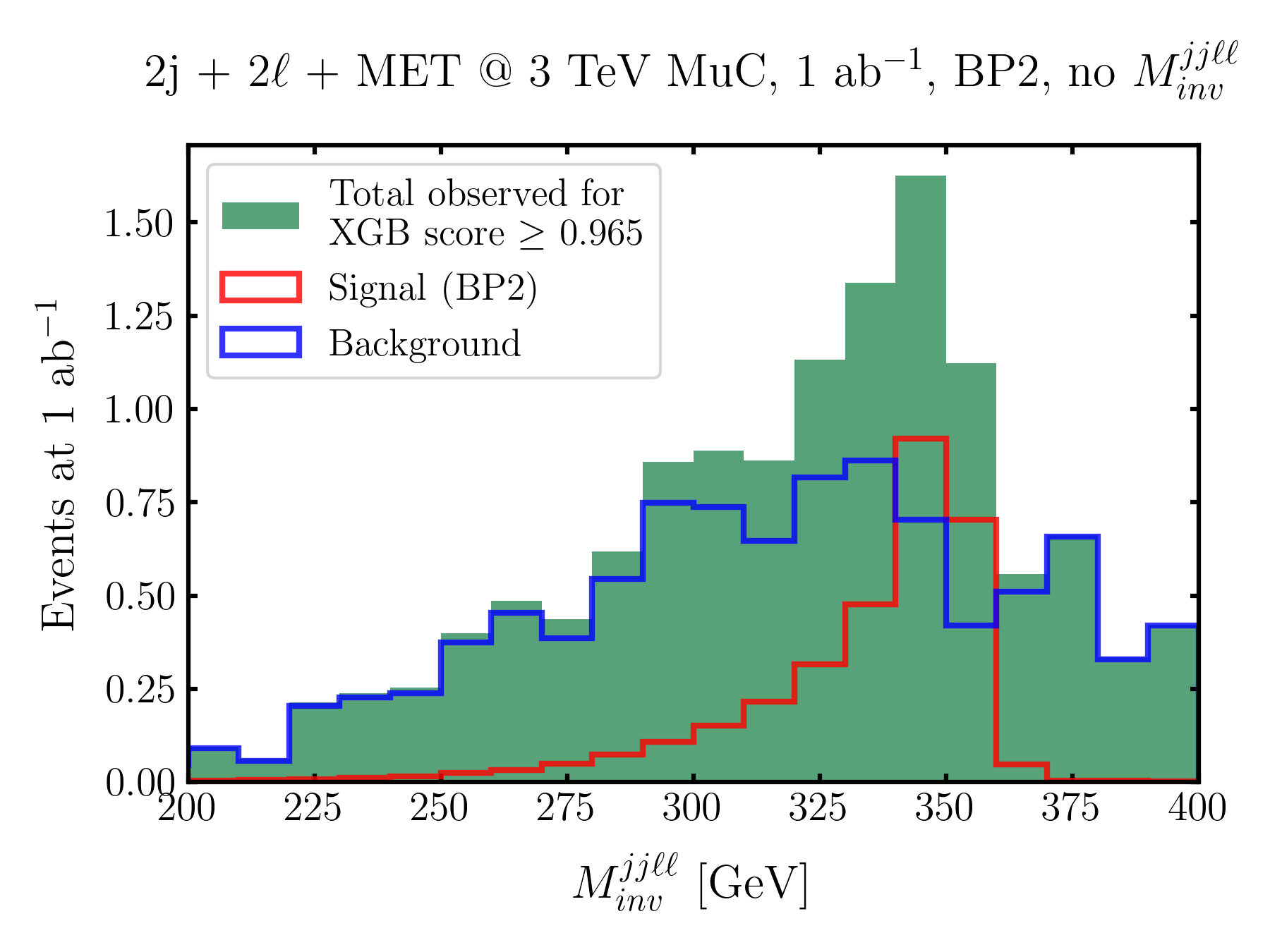}}
	\caption{$M_{inv}^{jj\ell\ell}$ distributions in GeV, for (a) BP1 and (b) BP2, keeping only the events that survive the XGB threshold cut. The green filled histograms represent the total observable events at the 3 TeV MuC with 1 \abi luminosity, while the red and blue unfilled histograms stand for the true positive (signal) and true negative (background) events.}
	\label{fig:mjjll_fs3_bdt_noim}
\end{figure}

From  \autoref{tab:fs3_3tev_bdt_noim}, we see that, the XGB score cuts of 0.978 and 0.965 for BP1 and BP2 respectively yield $\eucal{S}$(AMS) values of 3.31$\sigma$(3.97$\sigma$) and 1.01$\sigma$(1.00$\sigma$), which are lower than the classifier that includes $M_{inv}^{jj\ell\ell}$, but are still better than what we had from the simple CBA. Similar to the previous case, we take the events that survive this cut, and plot the $M_{inv}^{jj\ell\ell}$ distributions for them in \autoref{fig:mjjll_fs3_bdt_noim}.

The major difference that we notice between the $M_{inv}^{jj\ell\ell}$ distributions of \autoref{fig:mjjll_fs3_bdt} and \autoref{fig:mjjll_fs3_bdt_noim} is that, the distributions for the underlying true signal and true background events have different shapes, in the latter case. This is evidently because of the omission of this particular variable from the feature list. We do see however, that more background events can contaminate the signal peaks here, and hence the $\eucal{S}$(AMS)  values for the BP1 and BP2 peaks turn out to be 2.57$\sigma$(3.00$\sigma$) and 1.03$\sigma$(0.90$\sigma$), respectively. While not as good as the outcomes of \autoref{tab:fs3_3tev_bdt}, this approach still pinpoints the signal better compared to the CBA in \autoref{tab:fs3_3tev}, which is what we desire from this study.

With this elaborate discussion of three distinct final states, we witnessed that, from the 3 TeV MuC with 1 \abi of luminosity, we can achieve $\geq5\sigma$ significance for BP1 via FS2 and FS3, and also for BP2 via FS2, utilizing the XGB classifier. However, prospects for discovering BP3 at the 3 TeV MuC looks quite grim. In order to throughly assess the discovery probability of the model, we now turn to the next proposed stage of the muon collider, with the collision energy enhanced to 10 TeV. 

\section{Analysis at a 10 TeV Muon Collider}
\label{sec:an10}

The 10 TeV muon collider stage is proposed as the pinnacle of a discovery machine, coupled with unprecedented precision measurement of physics objects and couplings. For the purpose of our model, the logarithmic growth of the VBF cross-sections lays out a tantalising prospect for discovering all our benchmarks in the respective final states, with much less backgrounds compared to what a hadronic collider like the 100 TeV FCC-hh will bring. The kinematics of the final states, including the multiplicities of the leptons and jets, remain more-or-less similar as the 3 TeV stage, which again aides us to translate the same cuts from the 3 TeV analysis, The target luminosity offered by the 10 TeV MuC is a remarkable 10 \abi , which is also expected to elevate the prospects for the BP3, which produced low fiducial cross-sections for the final states at the 3 TeV MuC. In the following subsections, we present the analysis for the same three final states at the 10 TeV MuC for our model. Owing to the expectedly large statistics for both signal and background yields, we rely on $\eucal{S} = S/\sqrt{S+B}$ as the approximation for signal significance throughout the 10 TeV analysis.

\subsection{FS1: 3 leptons + MET + 1 Forward Muon}

For FS1, we keep the preselection criteria as well as the optimal cut-flow same as the case for the 3 TeV MuC, and present the outcomes of the analysis in \autoref{tab:fs1_10tev}. As mentioned above, the distributions of our kinematical variables preserve their shapes for both signal and backgrounds when we move to the 10 TeV collider, and hence we do not plot them again here.

\begin{table}[h]
	\renewcommand{\arraystretch}{1.2}
	\centering
	\begin{tabular}{|c|c||c||c|c||c|}
		\hline 
		\multirow{2}{*}{BP}	& \multirow{2}{*}{Cut flow} & \multicolumn{3}{c||}{FS1 counts at 10 TeV MuC} &  \multicolumn{1}{c|}{{\makecell{Significance \\ at $ \int \mathcal{L} dt $ = 10 \abi}}} \\
		\cline{3-6}
		&& Signal& $VVV$ (BG) & $VV$ (BG) & $\eucal{S}$  \\
		\hline\hline
		\multirow{6}{*}{BP1} & \makecell{$\sigma_{\rm fiducial} \times \int \mathcal{L} dt $  } &9122.27&13338.42&83385.10& --  \\
		\cline{2-6}
		&S0:  $n_\ell =$ 3 + $n_{F\mu}$ = 1 &4770.45&5562.41&30127.62&23.72$\sigma$\\
		\cline{2-6}
		&S1:  S0 +  MET $\geq$ 200 GeV &4442.36&3397.81&17420.40&27.95$\sigma$\\
		\cline{2-6}
		&S2:  S1 +  $\Delta R_{\ell_1 \ell_2} \leq$ 2 &4401.37&750.86&9013.10&36.98$\sigma$\\
		\cline{2-6} 
		&S3: S2 + $M_{inv}^{3\ell} \leq$ 200 GeV & 4400.36 &41.22&1741.50&55.96$\sigma$\\
		\cline{2-6} 
		&S4: S3 + $\abs{M_{inv}^{\ell_i\ell_j} - M_Z} \leq 10$ GeV & 4272.06 &20.58&1673.70&55.30$\sigma$\\
		\hline\hline
		\multirow{6}{*}{BP2} & \makecell{$\sigma_{\rm fiducial}  \times \int \mathcal{L} dt$ } &4360.49&13338.42&83385.10& -- \\
		\cline{2-6}
		&S0:  $n_\ell =$ 3 + $n_{F\mu}$ = 1 &2728.73&5562.41&30127.62&13.92$\sigma$\\
		\cline{2-6}
		&S1:  S0 +  MET $\geq$ 200 GeV &2578.65&3397.81&17420.40&16.85$\sigma$\\
		\cline{2-6}
		&S2:  S1 +  $\Delta R_{\ell_1 \ell_2} \leq$ 2 &2509.72&750.86&9013.10&22.65$\sigma$\\
		\cline{2-6} 
		&S3: S2 +  $M_{inv}^{3\ell} \leq$ 350 GeV &2509.02&127.45&3770.42&31.35$\sigma$\\
		\cline{2-6} 
		&S4: S3 + $\abs{M_{inv}^{\ell_i\ell_j} - M_Z} \leq 10$ GeV & 2410.49 &52.52&3619.83&30.91$\sigma$\\
		\hline\hline 
		\multirow{6}{*}{BP3} & \makecell{$\sigma_{\rm fiducial}  \times \int \mathcal{L} dt$ } &1296.16&13338.42&83385.10& -- \\
		\cline{2-6}
		&S0:  $n_\ell =$ 3 + $n_{F\mu}$ = 1 &813.01&5562.41&30127.62&4.26$\sigma$\\
		\cline{2-6}
		&S1:  S0 +  MET $\geq$ 200 GeV &772.37&3397.81&17420.40&5.26$\sigma$\\
		\cline{2-6}
		&S2:  S1 +  $\Delta R_{\ell_1 \ell_2} \leq$ 2 &728.17&750.86&9013.10&7.10$\sigma$\\
		\cline{2-6} 
		&S3: S2 +  $M_{inv}^{3\ell} \leq$ 500 GeV &727.98&196.72&4843.84&9.58$\sigma$\\
		\cline{2-6} 
		&S4: S3 + $\abs{M_{inv}^{\ell_i\ell_j} - M_Z} \leq 10$ GeV & 694.00 &67.77&4640.21&9.44$\sigma$\\
		\hline
	\end{tabular}
	\caption{Cut flow table of FS1 event counts for BP1 and BP2 at the 10 TeV MuC with 10 \abi luminosity, against the $VV$ and $VVV$ backgrounds. Signal significance is evaluated for events after each cut.}
	\label{tab:fs1_10tev}
\end{table}

Evidently, with the significant enhancement in the VBF cross-sections at the 10 TeV MuC, coupled with the 10 \abi of integrated luminosity, the event yields for the signal processes experience an overwhelming increment as compared to the 3 TeV MuC with 1 \abi luminosity. With the exact same cut-flow as \autoref{tab:fs1_3tev} up until the S3 selection, we see that, for BP1 and BP2, the maximum achievable signal significance in terms of $\eucal{S}$ become $\sim56\sigma$ and $\sim31\sigma$ respectively. While for BP3 the fiducial cross-section was too small to yield any events at the 3 TeV MuC in this final state, we witness not only a respectable event yield, but also a strong signal significance value of $\sim9.5\sigma$. To allow ourselves a more pinpoint probe of the $H^\pm \to ZW^\pm \to 3\ell$ channel, we can now afford to introduce another cut, denoted as S4, where we demand exactly one pair of leptons out of the three to have an invariant mass within the $Z$-boson peak. This weeds out $WWW$-background events further, thereby reducing the count of $VVV$ backgrounds significantly. While the slight reduction of signal event counts drop the overall significance by a tiny amount as compared to S3, it helps us establish the triplet nature of the charged scalar, with the $v_t$-aided decay mode as the source of this final state signal. 

This result indicates two things: firstly a probe of the custodial symmetry-breaking signature of the $Y=0$ triplet scalar with a singlet DM can be performed with very early data-taking runs of a future 10 TeV MuC, and secondly, for signals like this that does not drown in a sea of QCD backgrounds, we do not require additional ML-based classification, thus saving precious computation time. If this analysis was to be performed at the 100 TeV FCC-hh, the ever-present large hadronic background would still compel us to resort to ML methods, which is an analysis that we leave for a future study. Keeping this discussion in mind, for this final state as well as the remaining two final states, we will not be requiring the \texttt{xgboost} analysis.

\subsection{FS2: 2 $b$-jets + 1 lepton + MET + 1 Forward Muon}

\begin{table}[h]
	\renewcommand{\arraystretch}{1.2}
	\centering
	\begin{tabular}{|c|c||c||c|c|c||c|}
		\hline 
		\multirow{2}{*}{BP}	& \multirow{2}{*}{Cut flow} & \multicolumn{4}{c||}{FS2 counts at 10 TeV MuC} &   \multicolumn{1}{c|}{{\makecell{Significance \\ at $ \int \mathcal{L} dt $ = 10 \abi}}} \\
		\cline{3-7}
		&& Signal& \makecell{$ZW$ \\ (BG)} & \makecell{$hW$ \\ (BG)} &  \makecell{$t\bar{b}$ \\ (BG)} & $\eucal{S}$\\
		\hline\hline
		\multirow{6}{*}{BP1} & \makecell{$\sigma_{\rm fiducial} \times \int \mathcal{L} dt $  } &64901.53&184282.00&175240.76&38931.36& -- \\
		\cline{2-7}
		&S0:  $n_b = 2 + n_\ell =$ 1 + $n_{F\mu}$ = 1 &17070.86&38700.33&41905.67&12345.51&51.46$\sigma$\\
		\cline{2-7}
		&S1:  S0 + MET $\geq$ 200 GeV &16013.93&22298.86&21897.21&6410.55&62.04$\sigma$\\
		\cline{2-7}
		&S2:  S1 + $\Delta R_{\ell_1 b_1} \leq$ 2.5  &14297.81&8000.97&4620.22&4412.44&80.77$\sigma$\\
		\cline{2-7} 
		&S3: S2 + $\Delta\eta_{2b1\ell} \leq 1$&13883.87&4959.95&2618.79&3521.46&87.84$\sigma$\\
		\cline{2-7}
		&S4:  S3 + $M_{inv}^{bb\ell} \leq$ 200 GeV&13702.33&2795.37&1634.29&1309.65&98.27$\sigma$\\
		\hline\hline
		\multirow{6}{*}{BP2} & \makecell{$\sigma_{\rm fiducial}  \times \int \mathcal{L} dt$ } &39029.93&184282.00&175240.76&38931.36&--\\
		\cline{2-7}
		&S0:  $n_b = 2 + n_\ell =$ 1 + $n_{F\mu}$ = 1 &18171.32&38700.33&41905.67&12345.51&54.51$\sigma$\\
		\cline{2-7}
		&S1:  S0 + MET $\geq$ 200 GeV &17351.22&22298.86&21897.21&6410.55&66.56$\sigma$\\
		\cline{2-7}
		&S2:  S1 + $\Delta R_{\ell_1 b_1} \leq$ 2.5  &16121.39&8000.97&4620.22&4412.44&88.53$\sigma$\\
		\cline{2-7} 
		&S3: S2 + $\Delta\eta_{2b1\ell} \leq 1$&15158.09&4959.95&2618.79&3521.46&93.54$\sigma$\\
		\cline{2-7}
		&S4:  S3 +  $M_{inv}^{bb\ell} \leq$ 350 GeV &15010.48&4093.82&2287.24&2448.16&97.21$\sigma$\\
		\hline\hline
		\multirow{6}{*}{BP3} & \makecell{$\sigma_{\rm fiducial}  \times \int \mathcal{L} dt$ } &10223.16&184282.00&175240.76&38931.36&--\\
		\cline{2-7}
		&S0:  $n_b = 2 + n_\ell =$ 1 + $n_{F\mu}$ = 1 &5019.00&38700.33&41905.67&12345.51&16.03$\sigma$\\
		\cline{2-7}
		&S1:  S0 + MET $\geq$ 200 GeV &4806.60&22298.86&21897.21&6410.55&20.42$\sigma$\\
		\cline{2-7}
		&S2:  S1 + $\Delta R_{\ell_1 b_1} \leq$ 2.5  &4438.30&8000.97&4620.22&4412.44&30.28$\sigma$\\
		\cline{2-7} 
		&S3: S2 + $\Delta\eta_{2b1\ell} \leq 1$&3976.74&4959.95&2618.79&3521.46&32.39$\sigma$\\
		\cline{2-7}
		&S4:  S3 +   $M_{inv}^{bb\ell} \leq$ 500 GeV &3942.73&4545.50&2482.81&2959.37&33.41$\sigma$\\
		\hline
	\end{tabular}
	\caption{Cut flow table of FS2 event counts for BP1-BP3 at the 10 TeV MuC with 10 \abi luminosity, against the backgrounds. Signal significance is evaluated for events after each cut.}
	\label{tab:fs2_10tev}
\end{table}

For FS2, once again we expect higher event counts compared to FS1, for all the three BPs. The corresponding cut flow at 10 \abi of luminosity is presented below in \autoref{tab:fs2_10tev}.

One interesting behaviour that we notice here at 10 TeV is that, for BP2, despite having almost half the fiducial cross-section, about a thousand more events pass the pre-selection cut of S0 $\equiv n_b = 2 + n_\ell =$ 1 + $n_{F\mu}$ = 1 for BP2 than BP1. This is owing to that fact that, at high energies, the comparatively lower mass of $H^\pm$ in BP1 allows the subsequent $b$-jets to be more forward or collimated, dropping the number of events with two $b$-jets even lower than what we saw at 3 TeV in \autoref{fig:fs2kin}(a). Hence, up until the S3 cut, BP2 yields higher signal significance than BP1, despite being heavier and being produced less. Nonetheless, with the S4 cut corresponding to $M_{inv}^{bb\ell} \leq M_{H^\pm}$, the order is somewhat restored again, with maximum values of $\eucal{S}$ of 98.27$\sigma$ for BP1, and  97.21$\sigma$ for BP2. In case of BP3, we again achieve very optimistic results, with the S4 cut yielding a maximum significance of 33.41$\sigma$ with the 10 \abi luminosity. 

\subsection{FS3: 2 jets + 2$\ell$ + MET + 1 Forward Muon}

Seeking the opportunity to reconstruct the charged scalar from $M_{inv}^{jj\ell\ell}$ with $\geq 5\sigma$ significance for all the three BPs, we analyse the FS3 events at the 10 TeV MuC with 10 \abi luminosity, and present the results in \autoref{tab:fs3_10tev} below:

\begin{table}[h]
	\renewcommand{\arraystretch}{1.2}
	\centering
	\begin{tabular}{|c|c||c||c|c||c|}
		\hline 
		\multirow{2}{*}{BP}	& \multirow{2}{*}{Cut flow} & \multicolumn{3}{c||}{FS1 counts at 10 TeV MuC} &  \multicolumn{1}{c|}{{\makecell{Significance \\ at $ \int \mathcal{L} dt $ = 10 \abi}}} \\
		\cline{3-6}
		&& Signal& $VVV$ (BG) & $VV$ (BG) & $\eucal{S}$  \\
		\hline\hline
		\multirow{6}{*}{BP1} & \makecell{$\sigma_{\rm fiducial} \times \int \mathcal{L} dt $  } &12600.20&186696.01&253607.19& -- \\
		\cline{2-6}
		&S0:  $n_j = 2$ + $n_\ell =$ 2 + $n_{F\mu}$ = 1 &2634.25&20929.37&41079.54&10.36$\sigma$\\
		\cline{2-6}
		&S1:  S0 +  MET $\geq$ 200 GeV &2513.17&14363.27 &20102.43&13.07$\sigma$\\
		\cline{2-6}
		&S2:  S1 +  
		$\Delta R_{\ell_1 j_1} \leq$ 2  &2260.46&5109.87&8392.87&18.00$\sigma$\\
		\cline{2-6} 
		&S3: S2 + $\Delta\eta_{2j1\ell} \leq 1$ &2205.35&3785.26&5956.47&20.18$\sigma$\\
		\cline{2-6
		}
		&S4:  S3 + 180 GeV $\leq M_{inv}^{jj\ell\ell} \leq$ 210 GeV &1561.33&47.79&637.51&32.94$\sigma$\\
		\hline\hline
	\multirow{6}{*}{BP2} & \makecell{$\sigma_{\rm fiducial} \times \int \mathcal{L} dt $  } &4959.50&186696.01&253607.19& --  \\
	\cline{2-6}
	&S0:  $n_j = 2$ + $n_\ell =$ 2 + $n_{F\mu}$ = 1 &1260.81&20929.37&41079.54&5.01$\sigma$\\
	\cline{2-6}
	&S1:  S0 +  MET $\geq$ 200 GeV &1203.68&14363.27 &20102.43&6.37$\sigma$\\
	\cline{2-6}
	&S2:  S1 +  
	$\Delta R_{\ell_1 j_1} \leq$ 2  &1073.60&5109.87&8392.87&8.89$\sigma$\\
	\cline{2-6} 
	&S3: S2 + $\Delta\eta_{2j1\ell} \leq 1$ &990.05&3785.26&5956.47&9.55$\sigma$\\
	\cline{2-6}
	&S4:  S3 + 330 GeV $\leq M_{inv}^{jj\ell\ell} \leq$ 360 GeV &499.35&53.76&296.21&17.13$\sigma$\\
		\hline\hline 
		\multirow{6}{*}{BP3} & \makecell{$\sigma_{\rm fiducial} \times \int \mathcal{L} dt $  } &1372.23&186696.01&253607.19& --  \\
		\cline{2-6}
		&S0:  $n_j = 2$ + $n_\ell =$ 2 + $n_{F\mu}$ = 1 &449.85&20929.37&41079.54&1.79$\sigma$\\
		\cline{2-6}
		&S1:  S0 +  MET $\geq$ 200 GeV &434.16&14363.27 &20102.43&2.32$\sigma$\\
		\cline{2-6}
		&S2:  S1 +  
		$\Delta R_{\ell_1 j_1} \leq$ 2  &367.42&5109.87&8392.87&3.12$\sigma$\\
		\cline{2-6} 
		&S3: S2 + $\Delta\eta_{2j1\ell} \leq 1$ &319.54&3785.26&5956.47&3.18$\sigma$\\
		\cline{2-6}
		&S4:  S3 + 480 GeV $\leq M_{inv}^{jj\ell\ell} \leq$ 510 GeV &158.54&58.62&133.14&8.47$\sigma$\\
		\hline
	\end{tabular}
	\caption{Cut flow table of FS1 event counts for BP1 and BP2 at the 10 TeV MuC with 10 \abi luminosity, against the $VV$ and $VVV$ backgrounds. Signal significance is evaluated for events after each cut.}
	\label{tab:fs3_10tev}
\end{table}

\begin{figure}[h]
	\centering
	\subfigure[]{\includegraphics[width=0.32\linewidth]{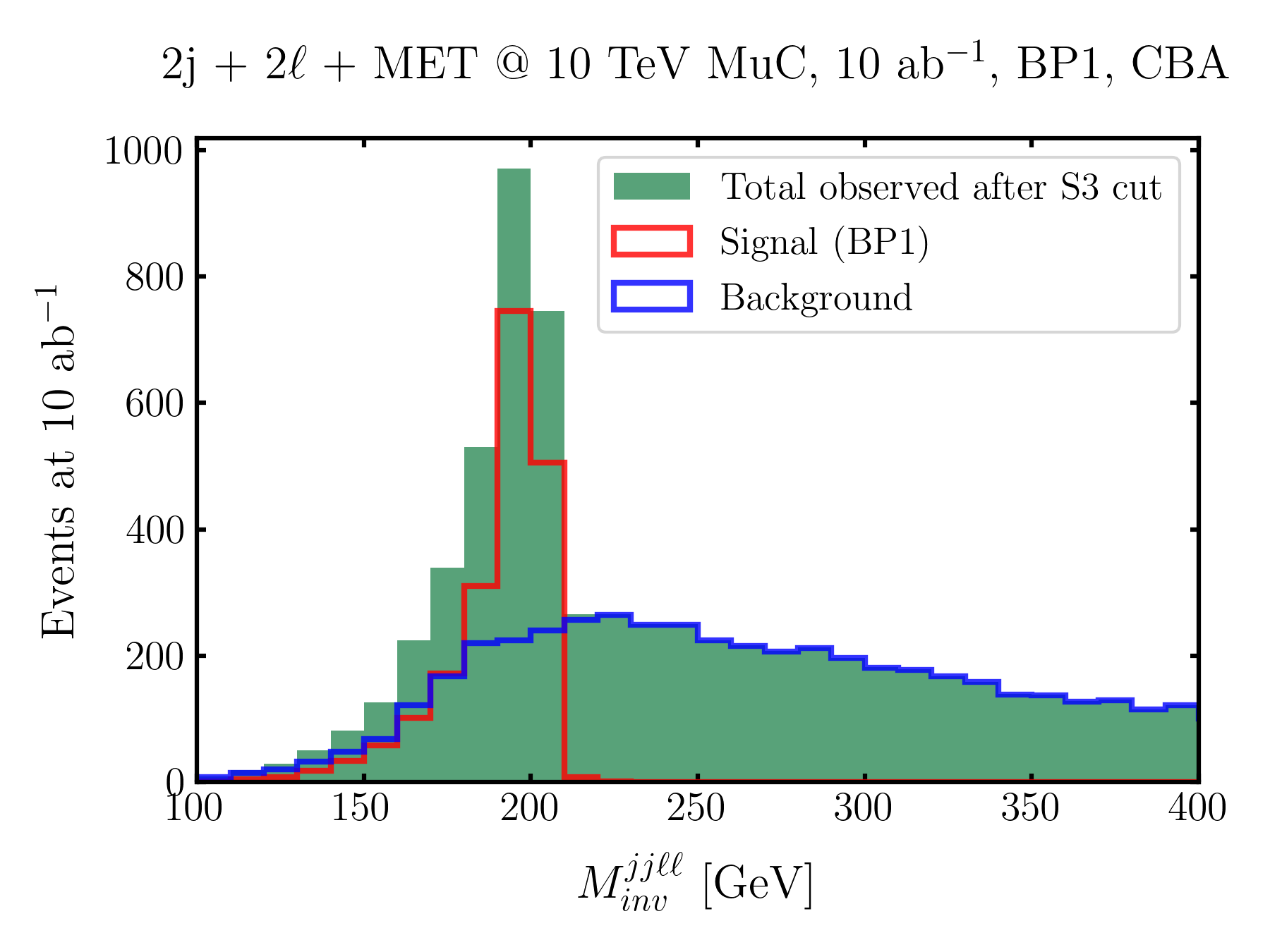}}
	\subfigure[]{\includegraphics[width=0.32\linewidth]{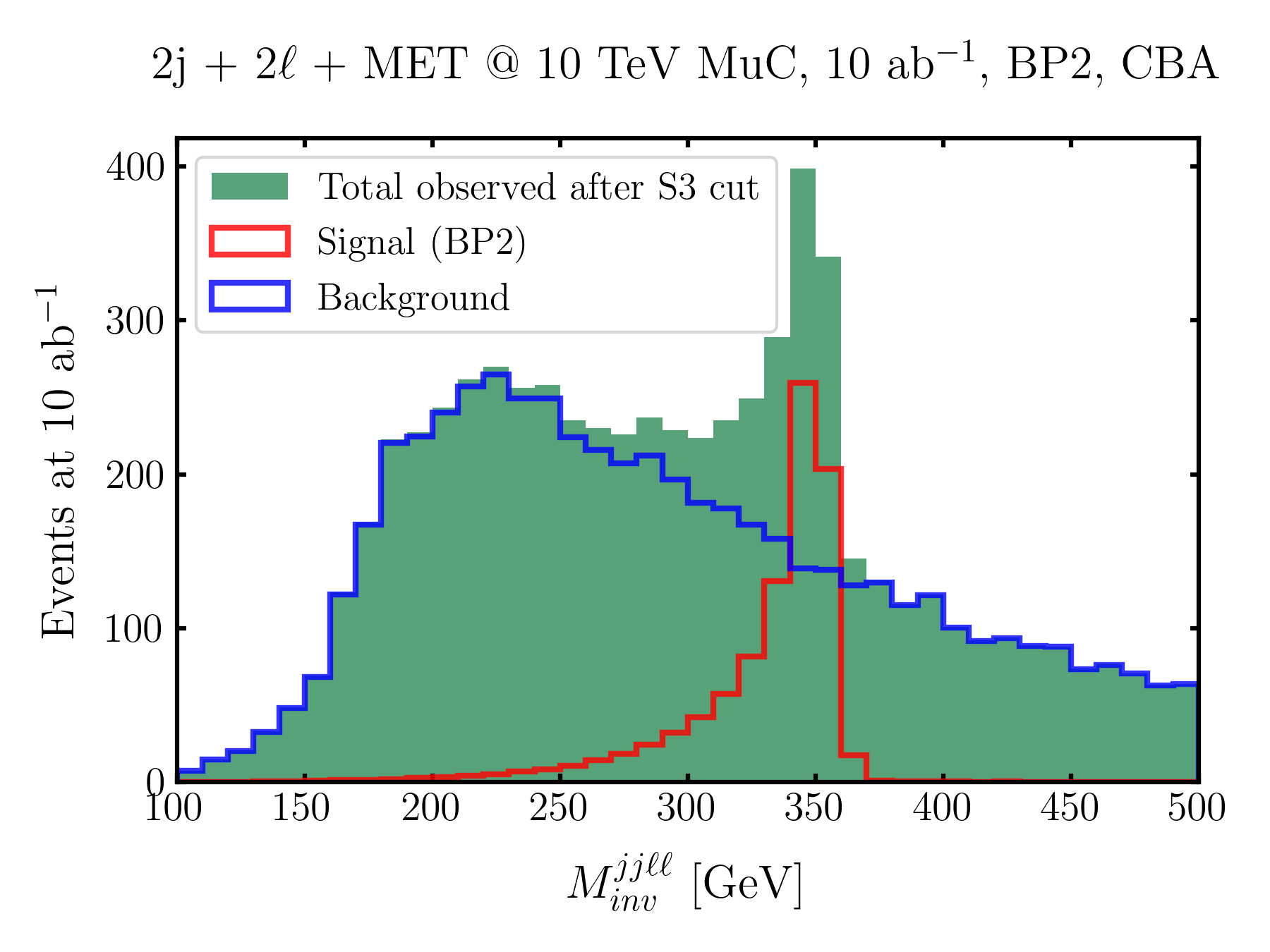}}
	\subfigure[]{\includegraphics[width=0.32\linewidth]{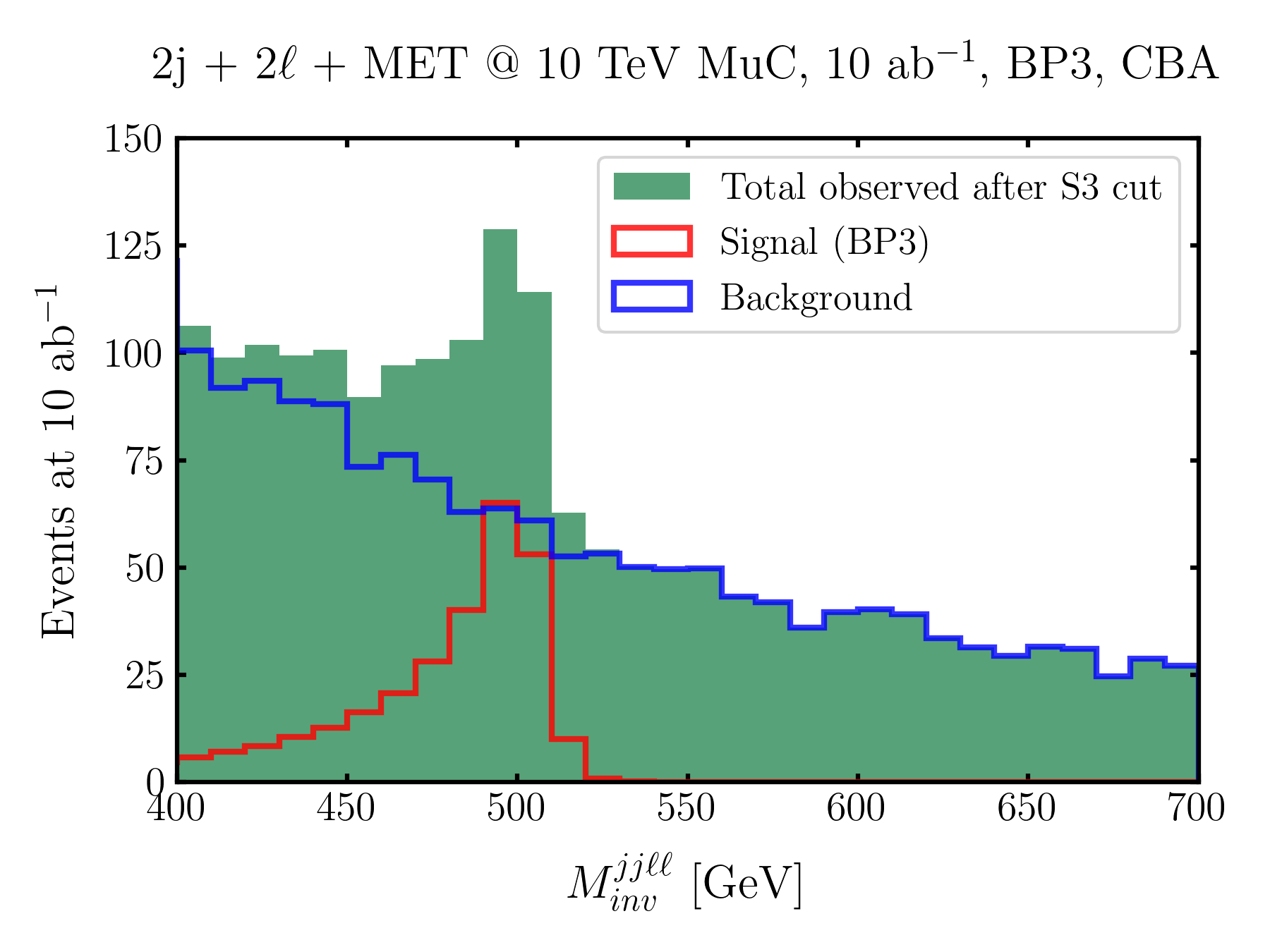}}
	\caption{$M_{inv}^{jj\ell\ell}$ distributions in GeV, for (a) BP1 and (b) BP2, keeping only the events that survive the S3 selection cut. The green filled histograms represent the total observable events at the 10 TeV MuC with 10 \abi luminosity, while the red and blue unfilled histograms stand for the underlying signal and background events.}
	\label{fig:mjjll_fs3_cba_10}
\end{figure}

We observe straight away that for BP1 and BP2, even the pre-selection cut yields $\geq 5\sigma$ significance. We again keep the cut flow consistent for all the BPs, and put the invariant mass window cut of three 10 GeV bins around the BP-specific $M_{H^\pm}$ value as the final cut, designated as S4. After the S4 cut, BP1 and BP2 yield significance values of 32.94$\sigma$ and 17.13$\sigma$ respectively. For BP3, due to the comparatively lower fiducial cross-sections, up until the S3 cut we reach $\sim3\sigma$ significance only. However, with the inclusion of the S4 cut on the $M_{inv}^{jj\ell\ell}$, the maximum achievable value of $\eucal{S}$ becomes 8.47$\sigma$, thus confirming the discovery potential of this channel for all the three BPs. To witness how the mass peaks look like over the background, we plot the $M_{inv}^{jj\ell\ell}$ distributions for the total observed events for all three BPs after the S3 cut, in \autoref{fig:mjjll_fs3_cba_10}(a)-(c) respectively. 

In all the panels of \autoref{fig:mjjll_fs3_cba_10}, the green filled histograms represent the distributions of the total observed events that survive the S3 cut, while the underlying signal and total background events are overlaid with red and blue unfilled histograms. The distributions show the true event counts with 10 \abi luminosity. The strongest peak, expectedly, is observed for BP1, and the BP2 peak is also quite healthy. While at first glance the BP3 peak does not look as appealing, the underlying background estimation shows, as seen in \autoref{tab:fs3_10tev}, that we still achieve more than 5$\sigma$ significance. 

After successfully analysing the results for three definitive collider benchmarks, we now intend to project our outcomes over a range of parameters that influence the final states, to estimate the discovery reach of both the 3 TeV and 10 TeV muon colliders. In the next section, detailed discussions on our approach and the subsequent results are presented.

\section{Parameter Space Reach at the Muon Collider}
\label{sec:reach}

As mentioned previously, our goal now is to extend our analysis, in order to provide a benchmark-independent overview of how much of the parameter space that the muon collider can cover. Looking at the final states in our consideration, a significant role in the discernibility of the model lies in the large MET yield from the $H^0 \to SS$ decay, the rate of which depends on two parameters- $\lambda_{st}$ and $M_S$- for a fixed value of $M_{H^\pm}$ and $M_{H^0}$. As we do not have much splitting between $M_{H^\pm}$ and $M_{H^0}$, the reach of the collider for a specific target luminosity can hence be presented in terms of a three-way parameter space, consisting of $M_{H^\pm}$, $M_S$, and $\lambda_{st}$. It is important to note that, these discovery projections are presented purely from a collider search perspective, and not all of the projected parameter space may be allowed by DM experiments. Additionally, for $\lambda_{st}$, we consider the upper limit from perturbative unitarity i.e. $\lambda_{st} \leq 4\pi$, to present the results. It is noteworthy that, a collider experiment cannot directly measure the coupling and the DM mass separately from any of these final states, as different combinations of them can lead to the same branching ratio for the $H^0 \to SS$, yielding the same amount of MET. Hence, these projections should be considered mainly as probes for the triplet scalar, which is our primary focus nonetheless. 

\subsection{Discovery projections at a 3 TeV muon collider}
\label{sec:r3}

\begin{figure}[h]
	\centering
	\includegraphics[width=\linewidth]{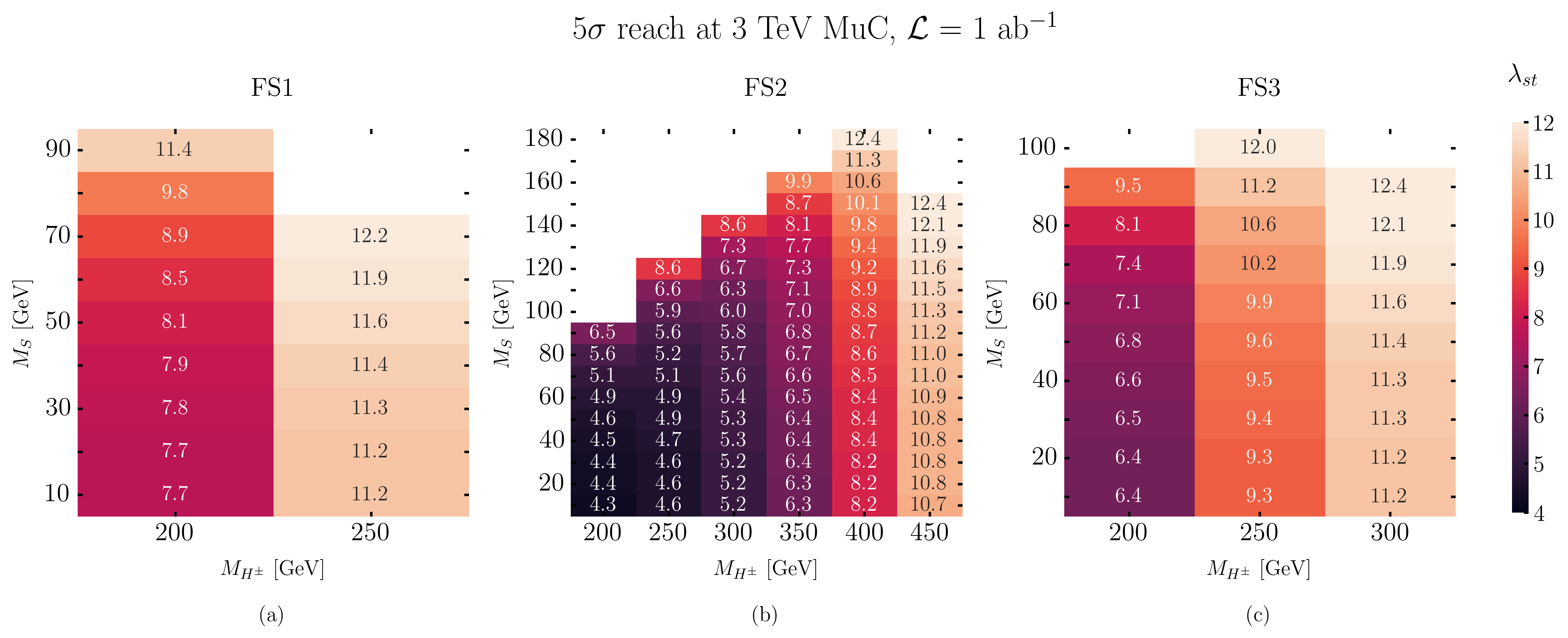}
	\caption{5$\sigma$ discovery reach in the $M_{H^\pm}$-$M_S$ plane, with a colour map over the smallest $\lambda_{st}$ required, at the 3 TeV muon collider with 1 \abi luminosity, for (a) FS1, (b) FS2, (c) FS3.}
	\label{fig:reach3}
\end{figure}

For the 3 TeV muon collider, our analysis in \autoref{sec:an3} reveals that, even with the aide of the BDT-based classifier, only FS2 is capable of providing more than 5$\sigma$ significance for the first two benchmark points. However, higher values of $\lambda_{st}$, or lower values of $M_S$, can lead to enhancement of the $H^0 \to SS$ branching ratios, which in turn allows more fiducial cross-sections with large MET contributions. Keeping this in mind, our approach for this case is to train the XGB classifier for each final state with triplet scalar masses from 200 GeV onwards with 50 GeV intervals, keeping the $\lambda_{st}$ and $M_S$ values fixed across the points. After evaluating the maximum achievable AMS score for each of the cases with 1 \abi of luminosity, we can then scale the fiducial cross-sections with the combinations of $M_S$ and $\lambda_{st}$ that can provide a $5\sigma$ value of the AMS. With the accumulated results, in \autoref{fig:reach3}, we present heatmap plots for each final state, keeping $M_{H^\pm}$ values along the x-axis in 50 GeV intervals, and $M_S$ values in the y-axis with 10 GeV intervals, colour-coding the minimum $\lambda_{st}$ required to achieve the 5$\sigma$ AMS with 1 \abi luminosity at the 3 TeV MuC.

In \autoref{fig:reach3}, panels (a), (b), and (c) correspond to FS1, FS2, and FS3, respectively, with the colour bar showing the $\lambda_{st}$ range from a minimum of 4 to a maximum of $4\pi$. The plots are only presented up to the $M_{H^\pm}$ values that can accommodate $\lambda_{st} \leq 4\pi$ for the required 5$\sigma$ AMS score. Clearly, FS2 with its large fiducial cross-sections provide the broadest range of discovery for $M_{H^\pm}$. With different combinations of $M_S$ and $\lst$ within the perturbative unitarity limit yielding enough events with large MET, one can probe a maximum of  $M_{H^\pm} \sim 450$ GeV. In stark contrast, FS1 has the least favourable projection, owing to the low rates of obtaining three-lepton events, with no more than 250 GeV of triplet charged scalar mass being in the discoverable range. FS2 provides a comparatively better mass reach over the considered three-parameter space, due to the highly discriminating $M_{inv}^{jj\ell\ell}$ feature, enhancing the discovery potential of $M_{H^\pm}$ up to 300 GeV. With the hope of achieving a larger range of $M_{H^\pm}$ reach, we turn our attention to the 10 TeV MuC.

\subsection{Discovery projections at a 10 TeV muon collider}
\label{sec:r10}

\begin{figure}[h]
	\centering
	\includegraphics[width=\linewidth]{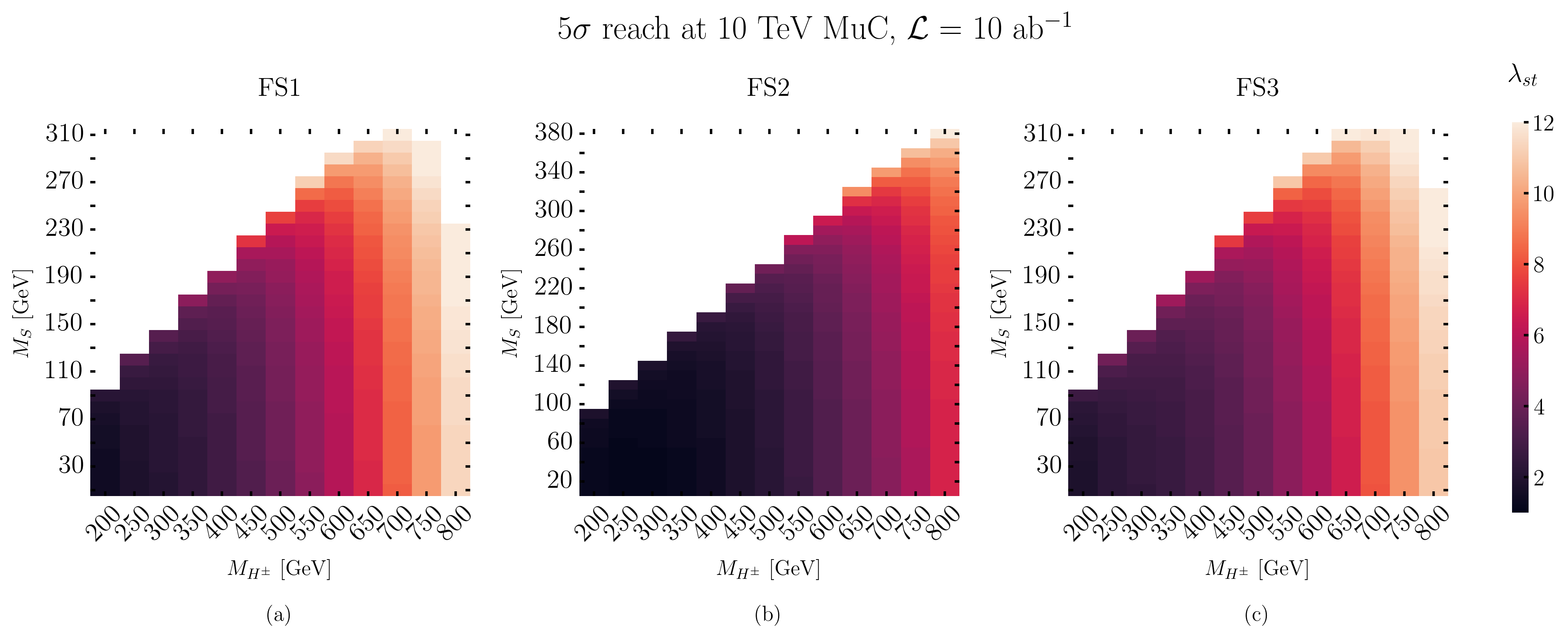}
	\caption{5$\sigma$ discovery reach in the $M_{H^\pm}$-$M_S$ plane, with a colour map over the smallest $\lambda_{st}$ required, at the 10 TeV muon collider with 10 \abi luminosity, for (a) FS1, (b) FS2, (c) FS3.}
	\label{fig:reach10}
\end{figure}

For the 10 TeV muon collider, we do not resort to the usage of the XGB classifier, as discussed in the benchmark analysis part in \autoref{sec:an10}. Hence, for the discovery projections, we adapt the same approach as the 3 TeV case, but replacing the BDT training with just the regular cut-flow evaluation for each of the $M_{H^\pm}$ value. As our statistics are large enough, we present the results for the combinations of parameter space that can obtain $S/\sqrt{S+B}$ = 5$\sigma$, instead of the AMS. The three-way reach plots similar to the previous case are shown in \autoref{fig:reach10}, with the target luminosity of 10 \abi.

Here also, \autoref{fig:reach10}(a)-(c) represent the results for FS1-FS3 at the 10 TeV MuC with 10 \abi of luminosity. In each case, the colour map runs over the minimum value of $\lambda_{st}$ that can yield the required number of events that pass the cut flow to give a 5$\sigma$ significance, for a particular combination of $M_{H^\pm}$ and $M_S$ values in 50 GeV and 10 GeV intervals, respectively. Here, the colour bar for $\lambda_{st}$ has a minimum of 1 and a maximum of $4\pi$, the aforementioned perturbative unitarity limit. With the enhanced event rates at the 10 TeV MuC, the model remains discoverable for all three final states upto the considered triplet scalar mass range of 800 GeV, a significant improvement over its 3 TeV counterpart. Depending on the $M_{H^\pm}$ value, The maximum possible value of $M_S$ that can yield enough events for 5$\sigma$ also increase compared to the 3 TeV case, with FS2 evidently providing the most optimistic projection of $M_S \lsim \frac{M_{H^\pm}}{2}$ for each $M_{H^\pm}$ bin. For each of the three final states, $\lst \lsim 2.0$ is also adequate for a 5$\sigma$ probe for lower-mass triplet scalars. For this analysis, we terminate the triplet masses at 800 GeV, as for heavier $H^0$, the branching ratio of $H^0 \to SS$ drops to $< 5\%$ even with high couplings and low DM masses, which is not a significant enough branching ratio for our consideration. Nonetheless, the 10 TeV MuC with 10 \abi luminosity turns out to be an overwhelmingly optimistic aide in discovering the HTM+S scenario.

\subsection{Discussion on signal similarity}

Before we conclude the work, we wish to mention how this model's signatures differ from other triplet scalar scenarios that can yield the same final states if extended with the additional DM. In $Y=0$ triplet-extended supersymmetric models \cite{Bandyopadhyay:2014vma, Bandyopadhyay:2017klv}, as well as in a complex $Y=0$ scalar triplet extension \cite{Bandyopadhyay:2020otm}, one can encounter multi-lepton signatures similar to the HTM+S case. These models are however not the most minimal, as already discussed in the introduction, and they come with additional particles whose signatures, if detected in tandem with the specific asymmetric production as mentioned here, may set them apart from our case.Considering the custodial symmetry-preserving Georgi-Machacek (GM) model, the larger triplet vev there leads to very high VBF production rates at the LHC, from which the GM triplet charged scalar decaying to the $ZW^\pm$ mode is excluded for masses upto $\sim1.2$ TeV from recent CMS results \cite{CMS:2021wlt}. Another model that can mimic this signal is the type-II seesaw with a $Y=1$ scalar triplet, which has a similar vev constraint from custodial symmetry breaking. While the event rates for type-II seesaw scalar productions can be similar (usually larger by a factor) to the HTM+S scenario, the presence of the doubly-charged scalar decaying to a pair of same-signed $W^\pm$-bosons in the model can distinguish it from our case. Provided that doubly-charged scalars are easier to produce than singly-charged or neutral scalars, a discovery for the type-II seesaw model, if it exists, is expected before a $Y=0$ triplet scalar. Such doubly-charged scalars with triplet vev of $\mathcal{O}(1)$ GeV are already excluded by combined ATLAS and CMS searches up to masses of $\sim420$ GeV  \cite{Ashanujjaman:2021txz}. Additionally, favoured values of $\rho$-parameter preserving vevs for the type-II seesaw are $\leq 10^{-4}$ GeV, for which the $ZW^\pm$ decay mode is highly suppressed in favour of purely leptonic decay $H^\pm \to \ell \nu$, thus vastly differing from our model's signatures, as well as being already excluded by CMS searches for masses below $\sim700$ GeV \cite{CMS:2017pet}. Heavy fermions from the type-III seesaw triplet can also mimic our multi-lepton signal, whose masses below $\sim 700$ GeV are ruled out by CMS searches \cite{CMS:2019lwf}. Our model successfully evades these exclusion limits in all possible final states for the $\lsim 500$ GeV mass range of triplet scalars under consideration, and hence we push the case for analysis of its signals at the future muon collider through this article. 
\vspace{-0.2cm}
\section{Conclusion}
\label{sec:conc}

In this work, we have presented the most minimal extension of the SM that contains a triplet charged scalar with a custodial symmetry-violating signature, as well as a stable DM candidate. A real $SU(2)_L$ triplet scalar with $Y=0$, which obtains a non-zero vev $v_t$, yields a pair of triplet-like charged scalars $H^\pm$ alongside a neutral scalar $H^0$ after EWSB, with little mass splitting between them at the tree-level. The charged scalars can decay into $ZW^\pm$, facilitated purely by the presence of $v_t$, which contributes only to the $W^\pm$-boson mass, enabling tree-level violation of the custodial symmetry. The DM comes in the form of a singlet scalar $S$, which transforms as odd under a discrete $Z_2$ symmetry, as opposed to rest of the fields in the model. The contribution of $v_t$ to only $M_{W^\pm}$ and not $M_Z$ imposes a constraint of $v_t \lsim 3$ GeV, from the $\rho$-parameter measurement. Additional constraints are imposed on the model parameters, especially the portal couplings $\lhs, \lht$ and $\lst$, from observations such as $h\to\gamma \gamma$ and $h\to$invisible decay branching rations. Having a WIMP-like DM candidate also brings limits the parameter space further, owing mainly to the bounds from the Planck data for observed relic, as well as the DM direct detection experiments. Considering all these constraints, we carefully choose a set of three benchmark points with triplet scalar masses of 200 GeV, 350 GeV, and 500 GeV for BP1-BP3 respectively, on which we perform a detailed collider analysis. The final state that we want to focus on at the colliders involve the three possible leptons from the $H^\pm \to ZW\pm$ decay, as well as the hefty amount of missing energy from the $H^0 \to SS$ decay, facilitated by the $\lst$ coupling. For the optimal feasibility of the latter, we keep $M_S \sim 59$ GeV in all our three BPs and $\lst \sim \mathcal{O}(1)$, both of which are allowed by the DM constraints.

The production mode that we wish to look for involve the asymmetric production of the triplet scalars, i.e. one $H^\pm$ and one $H^0$, with the $H^0$ decaying fully invisibly. At the LHC, due to the small value of $v_t$, the $pp\to H^\pm H^0$ production cross-sections drop sharply with the increase in $M_{H^\pm}$. Additionally, as we primarily focus on a hadronically quiet three-lepton final state with large MET, the FCC-hh at 100 TeV is not deemed an efficient option, owing to large QCD backgrounds. We find the proposed multi-TeV muon collider to be the optimal way to proceed with the production, especially in the VBF mode $\mu^+ \mu^- \to H^\pm H^0 \mu^\mp \nu_\mu$ that yield comparatively higher cross-sections. Another advantage of the VBF modes at the muon collider is the presence of exactly one Forward muon from the VBF, with high pseudorapidity, carrying large enough energy to pass through the tungsten nozzles that will absorb the soft BIB particles at the MuC, getting detected at a dedicated Forward muon detector. The inclusion of a Forward muon can successfully isolate this VBF process from non-VBF SM backgrounds. Hence, we construct three final states for the model signatures, each of which include exactly one Forward muon and large MET, alongside three charged leptons for FS1, two $b$-jets and one charged lepton for FS2, and two light jets and two charged leptons in FS3. Out of these three, FS1 is dominantly contributed by the custodial symmetry violating $H^\pm \to ZW^\pm$ process, and while the other two final states are not unique to this kind of a triplet, they can work together to enhance the discovery potential of the model.

We start our analysis at the 3 TeV MuC with a target luminosity of 1 \abi. For FS1, the most desirable final state of the model, the fiducial cross-section is low enough to enable a meaningful analysis for BP1 and BP2 only. The VBF multilepton background still proves to be quite high, and a traditional cut-based analysis fails to achieve a 3$\sigma$ significance for BP1, and not even 1$\sigma$ for BP2. Hence, we choose to employ a BDT-based classifier using the \texttt{xgboost} algorithm to enhance our chances of finding hints for this final state. Conceivably, the XGB classifier enhances the BP1 significance to $\geq 3\sigma (1\sigma)$ for BP1 and BP2 in this final state. FS2, with more fiducial cross-section due to contributions from all the three dominant decay modes of $H^\pm$, performs the best out of the three, with the cut-based analysis itself yielding $>5\sigma$ significance for BP1. With the deployment of the XGB classifier, this enhances to $\sim10\sigma$ for BP1, and even BP2 reaches the $5\sigma$ discovery threshold. While BP3 has an analysable fiducial cross-section here, it fails to achieve $1\sigma$ significance even with the XGB classifier. FS3 is particularly designed to reconstruct the charged triplet scalar resonance via $M_{inv}^{jj\ell\ell}$, and from the cut-based traditional analysis we obtain a maximum mass peak significance of $\sim 2.4\sigma$ for BP1, but $<1\sigma$ for BP2. Training an XGB classifier keeping the $M_{inv}^{jj\ell\ell}$ among the training variables, the mass peak significances are enhanced to $\sim 3.8\sigma (1.2\sigma)$ for BP1 (BP2). $3\sigma$ significance for BP1 can also be achieved if we do not use the $M_{inv}^{jj\ell\ell}$ in the training feature set to make the approach more generalized. Concluding the 3 TeV MuC analysis with 1 \abi of luminosity, we realise that while $3\sigma$ hint is achievable for at least one BP across all the three final states using an XGB classifier, we need to rely on the next stage of the MuC, with a collision energy of 10 TeV and a remarkable 10 \abi of luminosity.

The 10 TeV MuC predictably proves to be hugely fruitful, with $\gsim 10\sigma$ significance being achievable for all three BPs, in each of the final state under consideration, without needing to employ the XGB classifier. The healthy number of events in FS1 for all three BPs allow us to pinpoint the $H^\pm \to ZW^\pm$ decay mode by demanding a lepton pair to have invariant mass within the $Z$-mass peak. With the final event yields of FS3, we are also successful in reconstructing the $H^\pm$ mass peak with $>5\sigma$ significance for all three benchmarks, from the $M_{inv}^{jj\ell\ell}$ distributions.

For an article dedicated to the collider search of the model, it is also imperative to present a discovery projection of the parameter space that directly contribute to the event counts in the final states under consideration, independent of benchmark choices. Subsequently, we present $5\sigma$ discovery reach plots for each final state in the $M_{H^\pm}-M_{S}$ plane, with a colour map over the lowest possible $\lst$ that can yield enough fiducial cross-section with sufficiently large MET corresponding to $H^0 \to SS$ decay, for both 3 TeV and 10 TeV MuC energies, in \autoref{fig:reach3} and \autoref{fig:reach10} respectively. Due to the fact that collider experiments cannot measure exact values of $\lst$ and $M_S$, these projections are focused on the reach for $M_{H^\pm}$, which at the 3 TeV MuC with 1 \abi luminosity is restricted to a maximum of 450 GeV via FS2 despite using the XGB classifier. The 10 TeV MuC with its hefty 10 \abi luminosity can successfully distinguish this model for $M_{H^\pm} \lsim 800$ GeV across all the three final states just from the CBA, with $\lst \lsim 2$ being enough for low-mass triplet scalars. The combined presence of the DM mass and $\lst$, that contribute to the large MET required to isolate the model signatures from the background, is treated here as an aid in finding information of the triplet charged scalar that comes with it. Another point of discussion here is that, we are considering the fully invisible decay of $H^0 \to SS$ while calculating the fiducial production cross-sections of the benchmark points in each final state, to reduce the cost of computation especially for the XGB classifier training, while sacrificing some amount of event yield. In reality, the $H^0 \to ZZ \to 4\nu$ channel can also contribute to the fully invisible mode especially for BP3, albeit with very low probability of $<1\%$. Hence, especially for the events at the tail i.e. $\geq 200$ GeV of the MET distribution from inclusive event generation, the $H^0 \to SS$ contribution will dominate nonetheless. 

To summarize, a simultaneous extension of the SM with a $Y=0$ scalar triplet and a $Z_2$-odd scalar singlet $S$ is the most minimal framework within which one can hunt for signatures of custodial symmetry breaking, as well as have a stable DM candidate. The low triplet vev of $v_t \lsim 3$ GeV from $\rho$-parameter constraints help the model evade the bounds from current LHC searches even with $\lsim 500$ GeV masses. Such triplet scalars can be copiously produced from VBF at a future 3 TeV muon collider, where fully invisible decay of the neutral triplet scalar into a DM pair enhances the detectability of the multi-lepton signatures that ensue from the charged triplet scalar, especially with the implementation of a BDT-based classifier. For the 10 TeV muon collider upgrade, one does not require such classifiers, and with traditional cut-and-count methods, 5$\sigma$ discovery of the custodial symmetry-violating $H^\pm ZW^\pm$ vertex can be achieved. 
\vspace{-0.5cm}
\section*{Acknowledgements}

SP thanks the Council of Scientific and Industrial Research (CSIR), India for funding his research (File no: 09/1001(0082)/2020-EMR-I). SP acknowledges valuable discussions with Saranya Samik Ghosh and Prabhat Solanki, as well as the lectures and tutorials by Aruna K. Nayak and Sanu Varghese in the DML@LHC-2022 workshop held at IIT Hyderabad, regarding the BDT implementation. SP also thanks Shilpa Jangid for initial discussions regarding the model. PB wants to thank the support of DST and SERB with the grants  AV/KAR/2022/0167 for the DML@LHC-2022 workshop and  SSY/2023/001078 for the Phoenix-2023 conference  that played a crucial part in this project. 
\bibliography{HTMS_refs}

\appendix

\section{Decay widths of triplet scalars}
\label{sec:appa}

The partial decay widths of the dominant decay modes of $H^\pm$ and $H^0$ are provided here, with the couplings defined in \autoref{sec:decay}. The triangle function $\lambda(x,y,z) = x^2 + y^2 +z^2 -2xy -2yz -2zx$ is used in the phase space factors.

\textbf{Charged triplet scalar decay:}

\begin{align}
	\Gamma(H^\pm \to Z W^\pm) &= \dfrac{\sqrt{\lambda(M_{H^\pm}^2, M_W^2, M_Z^2)}}{16\pi M_{H^\pm}^3} g_{H^\pm ZW^\pm}^2 \left[2+ \dfrac{(M_W^2 + M_Z^2 -M_{H^\pm}^2)^2}{4 M_W^2 M_Z^2}  \right] , \\
	\Gamma(H^+ \to t \bar{b}) &= \dfrac{3\sqrt{\lambda(M_{H^\pm}^2, M_t^2, M_b^2)}}{16\pi M_{H^\pm}^3}\,	g_{H^+ t \bar{b}}^2 \left[M_{H^\pm}^2 - M_t^2 - M_b^2 -2M_t M_b\right], \\
	\Gamma(H^\pm \to h W^\pm) &= \begin{aligned} \dfrac{\sqrt{\lambda(M_{H^\pm}^2, M_W^2, M_h^2)}}{8\pi M_{H^\pm}^3} g_{H^\pm W^\pm h}^2 \dfrac{1}{M_W^2}&{}[M_{H^\pm}^4 + (M_{h}^2 - M_W^2)^2 \\
	&-2M_{H^\pm}^2(M_{h}^2 + M_W^2)].
	\end{aligned}
\end{align}

\textbf{Neutral triplet scalar decay:}

\begin{align}
	\Gamma(H^0 \to SS) = {}& \frac{g_{T^0 SS}^2}{16\pi M_{T^0}}\sqrt{1-\frac{4M_S^2}{M_{T^0}^2}}\\
	\Gamma(H^0 \to W^+ W^-) = {}& \frac{g_{H^0 W^+ W^-}^2}{16\pi M_{H^0}} \sqrt{1-\frac{4M_W^2}{M_{H^0}^2}}	 \left(3+\frac{M_{H^0}^4}{4M_W^4}-\frac{M_{H^0}^2}{M_W^2}\right)\\
	\Gamma(H^0 \to ZZ) = {}& \frac{g_{T^0 ZZ}^2}{32\pi M_{T^0}} \sqrt{1-\frac{4M_Z^2}{M_{H^0}^2}}	 \left(3+\frac{M_{T^0}^4}{4M_Z^4}-\frac{M_{H^0}^2}{M_Z^2}\right)\\
	\Gamma(H^0 \to hh) = {}& \frac{g_{H^0 hh}^2}{16\pi M_{H^0}}\sqrt{1-\frac{4M_h^2}{M_{H^0}^2}}
\end{align}

\end{document}